\pdfoutput=1
\documentclass[aps,pre,preprint,eqsecnum,showpacs]{revtex4-1}
\usepackage[final]{changes}
\usepackage{cancel}
\usepackage{graphicx}
\usepackage{url}
\usepackage{amsmath}
\usepackage{amssymb}
\usepackage{multirow}

\def\transpose{\mathrm{T}}
\DeclareMathOperator*{\argmin}{arg\,min}
\DeclareMathOperator*{\argmax}{arg\,max}
\def\diff{d}

\def\VALgL{2.2}
\def\VALgLr{1.0}
\def\VALgO{0.82g_{L}}
\def\VALgV{0.75}

\begin{document}
\title{
Performance Optimization in Two-dimensional
Brownian Rotary Ratchet Models
}
\author{Hiroki \surname{Tutu}}
\affiliation{Graduate School of Informatics, Kyoto University,
Kyoto 606-8501, Japan}
\author{Katsuya \surname{Ouchi}}
\affiliation{Kobe Design University, Kobe 651-2196, Japan}
\author{Takehiko \surname{Horita}}
\affiliation{
Department of Mathematical Sciences, Osaka Prefecture University,
Sakai 599-8531, Japan}
\date{\today}
%
%
\begin{abstract}
With a model for two-dimensional (2D) Brownian rotary ratchets being capable of
producing a net torque under athermal random forces, its optimization for
mean angular momentum ($L$), mean angular velocity ($\omega$), 
and efficiency ($\eta$) is considered.
In the model, supposing that such a small ratchet system is placed in a thermal bath,
the motion of the rotor in the stator is described
by the Langevin dynamics of a particle in a 2D
ratchet potential, which consists of a static and a time-dependent
interaction between rotor
and stator; for the latter, we examine a force [randomly directed d.c. field (RDDF)]
for which only the direction is instantaneously updated
in a sequence of events in a Poisson process.
Because of the chirality of the static part of the potential,
it is found that the RDDF causes net rotation while coupling with the thermal fluctuations.
Then, to maximize the efficiency of the power
consumption of the net rotation,
we consider optimizing the static part of the
ratchet potential.
A crucial point is that the newly designed form of ratchet potential
enables us to capture
the essential feature of 2D ratchet potentials 
with two closed curves and allows us
to systematically construct an optimization strategy.
In this paper, we show a method for maximizing
$L$, $\omega$, and $\eta$, its outcome in 2D two-tooth ratchet systems,
and a direction of optimization for a three-tooth ratchet system.
\end{abstract}

\maketitle

\section{\label{Intro} Introduction}

A ratchet is a mechanical device that combines a pawl and a wheel such that
the former limits the rotation of the latter to only one direction.
Also, a ratchet mechanism can refer to dynamism among objects 
that rectifies incoming stimulative actions into directed movement.
The mechanism of a ratchet is attributed to the nature of
a nonequilibrium (or macroscopic) system.
If the size of the ratchet is reduced to nanoscale,
the rectifying action of the ratchet becomes unreliable or probabilistic because
the influence of the surrounding molecules is comparable to the input stimuli
to the ratchet; the pawl moves erroneously and allows the wheel to rotate in the opposite 
(i.e., undesired) direction.
Such a very small ratchet system is
called a Brownian ratchet (BR) or Smoluchowski--Feynman ratchet from
Smoluchowski's (and Feynman's) thought experiment \cite{Smoluchowski1912,FeynmannLecI}.
To be consistent with the Second Law of Thermodynamics,
if the temperature of the ``agents'' causing
the input stimuli to the ratchet equals the temperature of the ratchet,
there can be no net rotation of the wheel.
This contraposition implies that if net rotation does appear, the statistical property of
the input agents differs from that in thermal equilibrium, or
that the temperature of the pawl is lower than that of the input 
agents \cite{FeynmannLecI,PhysRevLett.71.1477,PhysRevE.81.061104}.
The problem of how net rotation or unidirectional motion results
from unbiased stimuli in the thermal environment has been analyzed by numerous studies
with various types of ratchet model \cite{Reimann200257,RevModPhys.81.387}.
Because of its universal nature in nonequilibrium phenomena,
the concept of a ratchet mechanism has attracted a great deal of attention 
from various perspectives, e.g., biological \cite{VALE199097,Wang2002,Kawaguchi20142450} 
and artificial molecular motors 
\cite{ANIE:ANIE200504313,RevModPhys.81.387,Kuhne14122010,C5NR08768F}, 
optical thermal ratchets \cite{PhysRevLett.74.1504}, 
dielectrophoretic ratchets \cite{PhysRevE.86.041106}, 
and granular ratchet systems\cite{1742-5468-2008-10-P10011,PhysRevLett.104.248001,
PhysRevLett.107.138001,PhysRevE.86.061310,PhysRevLett.110.120601,
PhysRevE.87.040101,PhysRevE.87.052209,PhysRevE.94.032910}.

In this study, we consider the rectification behavior of
two-dimensional (2D) BR models for a rotating thin rod inside
a cylinder, and its optimization for the rotational performance.
Firstly, we outline our dynamical model,
in which we suppose that 
the thin rod (rotor) contacts diagonally with the cylinder (stator) at the upper and lower
 edges, and rotates inside the cylinder through mutual ratchet interaction
under temporally varying fields
\cite{PhysRevE.84.061119,PhysRevE.87.022144,doi:10.7566/JPSJ.84.044004}.
Real systems that are relevant to such Brownian rotary ratchets
may be found in microscopic light-driven 
rotors \cite{:/content/aip/journal/apl/78/2/10.1063/1.1339258},
the artificial molecular rotor of
caged supramolecules \cite{Kuhne14122010}, 
or synthetic molecular systems, e.g.,
\cite{ANIE:ANIE200603618,TCR:TCR201402007}.

As in \cite{PhysRevE.84.061119,PhysRevE.87.022144,doi:10.7566/JPSJ.84.044004},
we describe the state of rotation as a trajectory on a 2D plane.
Representing the state of the rotor tip at time $t$ as
$\boldsymbol{X}\equiv (X_t,Y_t)^{\transpose}$ (hereinafter,
$\transpose$ denotes the transpose of a vector or matrix, and
bold face represents a 2D vector),
we assume that $\boldsymbol{X}$ obeys Langevin dynamics:
\begin{equation}
\gamma\dot{\boldsymbol{X}} = -\partial_{\boldsymbol{X}}
 V_{0}(\boldsymbol{X})  -\partial_{\boldsymbol{X}}
 V_h(\boldsymbol{X}, t) +\boldsymbol{f}_{I}(\boldsymbol{x})
 + \boldsymbol{R}_t. 
\label{LEQ}
\end{equation}
Here,
$\partial_{\boldsymbol{x}} \equiv 
(\frac{\partial}{\partial x}, \frac{\partial}{\partial y})^{\transpose}$,
$\gamma$ ($=1$) denotes a viscosity coefficient,
and $\boldsymbol{R}_t$ is a random force
with properties $\langle \boldsymbol{R}_t\rangle=\boldsymbol{0}$ and
$
\langle \boldsymbol{R}_{t}\boldsymbol{R}_{t'}^{\transpose} \rangle = 
2D \delta(t-t')\Hat{1}
$, 
where $\Hat{1}$ and $\langle A_t\rangle$ denote a $2\times 2$ unit matrix and
the average of $A_t$ over all possible process of
$\boldsymbol{R}_t$, respectively.
Here, $\boldsymbol{R}_t$ corresponds to the thermal fluctuation, and
 the noise intensity $D$ is assumed to satisfy
$D=k_{\mathrm{B}}T$ with a temperature $T$ and
the Boltzmann constant $k_{\mathrm{B}}$.
In addition, $V_0(\boldsymbol{x})$ represents the 2D ratchet potential
for the static part of the rotor--stator interaction (one can imagine 
the interaction between the pawl and the wheel for this).
The function $V_h(\boldsymbol{x}, t)$ is the temporally varying part
of the interaction:
\begin{align}
V_h(\boldsymbol{x}, t)
&=
- h\boldsymbol{N}_t\cdot\boldsymbol{x}
,
\quad\boldsymbol{N}_t=(\cos\Phi_t,\sin\Phi_t)^{\transpose}
,
\label{LEQ:Poth}
\end{align}
where $h\boldsymbol{N}_t$ represents a force on the rotor.
The angle $\Phi_t$ switches successively to independent values in
$[0,2\pi)$ in a sequence of events described as a Poisson process
with a mean interval $\Omega^{-1}$.
In other words, the mean and auto-correlation function of $\boldsymbol{N}_t$ obey
\begin{equation}
\left\langle\boldsymbol{N}_t\right\rangle_{\Phi}=\boldsymbol{0},
\quad
\left\langle
\boldsymbol{N}_{t}\boldsymbol{N}_{0}^{\transpose}
\right\rangle_{\Phi}
= \frac{e^{-\Omega t}}{2}\Hat{1},
\label{cor_NN}
\end{equation}
where $\langle A_t\rangle_{\Phi}$ denotes
the average of $A_t$ over all possible process of
$\Phi_t$ (Appendix.~\ref{OnNoise}).
We can regard $h\boldsymbol{N}_t$ as an external field or
a force due to
a temporal deformation of the stator, and call this a randomly directed d.c. field (RDDF).
For simplicity, we consider the load
$\boldsymbol{f}_I(\boldsymbol{x})$ for the rotation as
\begin{equation}
\boldsymbol{f}_{I}(\boldsymbol{x})
 = \frac{I}{2\pi} 
\left(
\frac{y}{\lvert\boldsymbol{x}\rvert^2},
-\frac{x}{\lvert\boldsymbol{x}\rvert^2}\right)^{\transpose}
\equiv
-\frac{I}{2\pi} \partial_{\boldsymbol{x}}
\theta(\boldsymbol{x}),
\label{LEQ:PotI} 
\end{equation}
where $\theta(\boldsymbol{x})\equiv\tan^{-1}\frac{y}{x}$ and
$I$ denotes the load torque.

In the absence of an external field and load ($h=I=0$) in Eq.~(\ref{LEQ}),
we have an equilibrium state that corresponds to thermal equilibrium;
there is no net circulation of $\boldsymbol{X}$ about the origin, so there is
no net rotation. As mentioned above,
net rotation requires the (agents of) external field to be 
athermal \cite{PhysRevLett.71.1477,Reimann200257}.
Here, as in Eq.~(\ref{cor_NN}), the RDDF can have a sufficiently long correlation time and
be athermal.
In general, there are two basic types of 2D field:
either one in which only the field angle varies but the magnitude is constant,
or a uni-axially polarized field.
The RDDF is classed as the former type because
the force angle varies randomly without bias.
An example of the latter field type is reported in \cite{PhysRevE.84.061119};
with dynamics in a two-tooth ratchet potential under
a uni-axially polarized sinusoidal field, it is shown that a net rotation appears
with a rotational direction that depends on the polarization angle. The
ranges of angle for the clockwise and counterclockwise rotations are 
asymmetric, reflecting the chirality of the ratchet potential
(cf. \cite{PhysRevE.77.061111}, which reports on 
the occurrence of unidirectional rotation
with a symmetric (achiral) two-well hindered-rotation potential).

An aim of the present study is to show that a combination of the
two-tooth ratchet potential
and the RDDF (as a basic example of an athermal unbiased field)
can support net rotation in a constant direction that is
determined by only the chirality of the ratchet potential.
Such a net rotational state is also capable of producing a positive power against
the load in Eq.~(\ref{LEQ:PotI}) for a sufficiently small $I$.
Another aim is to formulate a method of optimizing the 2D ratchet potential 
to maximize the efficiency of rotational output.
In previous papers by some or all of the authors 
\cite{PhysRevE.84.061119,PhysRevE.87.022144,doi:10.7566/JPSJ.84.044004},
analyses of the two- and three-tooth models
with the four- and six-state approaches have been shown
\cite{PhysRevE.84.061119,PhysRevE.87.022144},
and the analytical framework for estimating energetic efficiency
\cite{doi:10.7566/JPSJ.84.044004} has been developed,
in which any optimization has been disregarded.

Here, 
we define the efficiency of the rotational output.
The balance between the input power of the external field
($\boldsymbol{f}_{h}\equiv h\boldsymbol{N}_t$) and
the combined power consumed by the load
[$\boldsymbol{f}_{I}\equiv \boldsymbol{f}_{I}(\boldsymbol{X})$]
and the other resistive forces is
\begin{equation}
\overline{
\dot{\boldsymbol{X}}\cdot \boldsymbol{f}_{h}
}
=
\overline{
(-\dot{\boldsymbol{X}}\cdot\boldsymbol{f}_{I})
}
+
\gamma
\left(
\overline{\lvert\langle\Dot{\boldsymbol{X}}\rangle\rvert^2}
+
\overline{L_t'}\,\overline{\omega_t'}
\right)
+Q_T,
\label{EnergyBalanceEq}
\end{equation}
where
\begin{align}
L_t'\equiv &\;
X_t(\dot{Y}_t-\langle\dot{Y}_t\rangle ) -Y_t
(\dot{X}_t-\langle \dot{X}_t\rangle)
,
\label{def:L'}
\\
\omega_t'\equiv &\; 
\frac{
X_t(\dot{Y}_t-\langle\dot{Y}_t\rangle ) -Y_t
(\dot{X}_t-\langle \dot{X}_t\rangle)
}{X_t^2+Y_t^2}
,
\label{def:omega'}
\\
Q_T
\equiv
&\;
\frac{k_{\mathrm B}T}{\gamma}(\overline{\partial_xF_x +\partial_yF_y})
+
\gamma
\overline{\left(
L_t' -
 \overline{L_t'}
\right)
\left(
\omega_t'- \overline{\omega_t'}\right)}
\nonumber\\
&+
\frac{1}{\gamma}
\overline{\left(\frac{X_t\Tilde F_x+Y_t\Tilde F_y}{\sqrt{X_t^2+Y_t^2}}\right)^2},
\label{def:Qt}
\end{align}
with $\boldsymbol{F}\equiv (F_x,F_y)^{\top}\equiv
\boldsymbol{f}_{h}+\boldsymbol{f}_{I}$ and
$
\Tilde{\boldsymbol{F}}\equiv \boldsymbol{F}-\gamma
\langle \dot{\boldsymbol{X}} \rangle$.
The equality (\ref{EnergyBalanceEq}) is derived in \cite{doi:10.7566/JPSJ.84.044004}
based on
\cite{PhysRevE.75.061115,PhysRevLett.83.903,PhysRevE.68.021906,PhysRevE.70.061105}.
Here, we define the long time average of $A$ as
$
\overline{A}
\equiv \overline{A(\boldsymbol{X},\Phi_t)} \equiv \int_{0}^{T_{\mathrm{tot}}}
d t\, A(\boldsymbol{X},\Phi_t)/T_{\mathrm{tot}}$ for $T_{\mathrm{tot}} \gg \Omega^{-1}$, 
and assume 
$
\overline{A}
= 
\left\langle\! \langle
A(\boldsymbol{X},\Phi_t)\rangle\!
\right\rangle_{\Phi}
$ (ergodic hypothesis), where
$ \left\langle\! \langle  A \rangle\! \right\rangle_{\Phi}$ means 
doubly averaging over all possible realization of the stochastic processes
$\{\boldsymbol{R}_t\}_{t=0}^{T_{\mathrm{tot}}}$ and $\{\Phi_t\}_{t=0}^{T_{\mathrm{tot}}}$.
The products of the dynamical variables are considered in
the Stratonovich sense \cite{Sekimoto2010}.

The left-hand side (LHS) in Eq.~(\ref{EnergyBalanceEq}) is the input power.
The first term on the right-hand side (RHS) is the power consumed by the load.
The second and third terms are the dissipation rates
associated with the mean translational and rotational motions, respectively
(these can be interpreted as the power consumed while 
drawing in the surrounding molecules).
Here, $L_t'$ and $\omega_t'$ denote the angular momentum and angular velocity, respectively,
defined in coordinates fixed to the mean translational motion.
The final term $Q_T$ in Eq.~(\ref{EnergyBalanceEq})
can be regarded as an excess dissipation rate resulting from the difference
between the dissipation due to velocity fluctuations---consisting of the second (rotational component) and third (radial component) terms in 
Eq.~(\ref{def:Qt})---and the input power from the thermal bath (the first term multiplied by minus one).
Using the input power and the output powers associated with the rotation
in the RHS of Eq.~(\ref{EnergyBalanceEq}),
the rectification efficiency (or generalized efficiency)
\cite{PhysRevE.75.061115,PhysRevLett.83.903,PhysRevE.68.021906,PhysRevE.70.061105}
is defined as
\begin{gather}
\eta =
\frac{
\gamma \overline{L_t'}\,\overline{\omega_t'}
-
\overline{
\dot{\boldsymbol{X}}\cdot\boldsymbol{f}_{I}
}
}{
\overline{\dot{\boldsymbol{X}}\cdot \boldsymbol{f}_h}
}.
\label{def:eta}
\end{gather}
This definition is usable even in the absence of a load ($I=0$).

There have been many studies of the rotation or transport efficiency of ratchet systems.
In one-dimensional ratchet systems in particular, proposals have been made for exact expressions
for the efficiency or for models that realize highly efficient performance, e.g.,
\cite{0295-5075-27-6-002,PhysRevE.60.2127,PhysRevE.69.021102,PhysRevE.75.061115,PhysRevE.90.032104}.
In the context of maximization of efficiency,
although there are various aspects to optimization \cite{PhysRevE.68.046125,PhysRevE.74.066602},
basic approaches may be classified into two types:
those that optimize the temporally
varying part of the ratchet potential 
\cite{Tarlie03031998,PhysRevE.79.031118,PhysRevE.87.032111},
and those that optimize the static part 
\cite{PhysRevE.75.061115,PhysRevE.69.021102}.
Experiments relevant to these optimization approaches can be found in 
\cite{PhysRevLett.74.1504,PhysRevE.86.041106}.
However, in the present context and to the best of our knowledge, 
there have been few theoretical studies on 2D ratchet models\cite{schmid2009}.

In considering the optimization of the static part of the ratchet potential,
a basic idea is to design the ratchet potential in the following form:
\begin{equation}
V_0(\boldsymbol{x}) =
\frac{1}{4}
\left[
1-\left\{
v_{0}(\boldsymbol{x})
\right\}^m
\right]^2
-
K v_{1}(\boldsymbol{x})
,
\label{LEQ:Pot0} 
\end{equation}
where $m \geq 1$.
For $m\gg 1$, the curve of
$v_{0}(\boldsymbol{x})=1$ 
approximates a potential valley that
mimics a constraint on the rotor--stator contact and 
along which the orbit of the rotational-motion concentrates.
The purpose of $v_{1}(\boldsymbol{x})$ is to create the local minima and saddles
in the valley.
The functions $v_{0}(\boldsymbol{x})$ and $v_{1}(\boldsymbol{x})$ are non-decreasing functions of
 $\lvert\boldsymbol{x}\rvert$,
 and the region specified by $v_{0}(\boldsymbol{x})\leq 1$ is a simply connected space.
These details are shown in Sec.~\ref{TT-Model}.
Here, an important point is that for $m\gg 1$ we can characterize a ratchet potential
with two curves specified by
$v_{0}(\boldsymbol{x})=1$ and $v_{1}(\boldsymbol{x})=E$ with a constant $E$
as shown later.
This allows us to easily design an optimized
ratchet potential that maximizes the rotational output.

In this study, we develop an optimization method by using
a 2D two-tooth ratchet potential.
Of course, our approach is applicable to more general 2D ratchet potentials
in the form of Eq.~(\ref{LEQ:Pot0}).
In Sec.~\ref{TT-Model}, for the two-tooth ratchet model,  we provide 
$v_{0}(\boldsymbol{x})$ and $v_{1}(\boldsymbol{x})$ and describe their details.
In Sec.~\ref{Quantities}, we define indexes with which to characterize the performance
of the ratchet model; we show analytical expressions for these, 
which are obtained using the same approach as in \cite{doi:10.7566/JPSJ.84.044004}.
In Sec.~\ref{Optimization}, we formulate the optimization problem.
In Sec.~\ref{NumericalResults}, we test the results of the optimization.
In Sec.~\ref{Discuss}, we suggest a way to optimize
three-tooth ratchet models.
In. Sec.~\ref{summary}, we summarize the whole study.

\section{\label{TT-Model} Two-tooth Ratchet Model}

\begin{figure}[t]
\def\Size{9.5cm}
\centering
\includegraphics[width=\Size,keepaspectratio,clip]{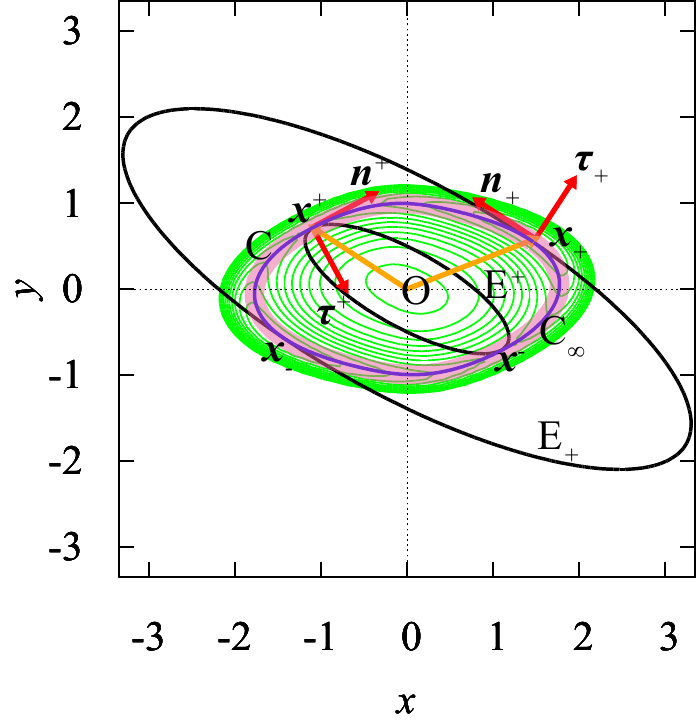}
\caption{
(Color online)
Contour plot of $V_0(\boldsymbol{x})$ with a skeleton of
curves
$\mathrm{C}_{\infty}:\{\boldsymbol{x}\mid v_0(\boldsymbol{x})=1\}$,
$\mathrm{E}_{+}:\{\boldsymbol{x}\mid v_1(\boldsymbol{x})=E_{+}\}$ and
$\mathrm{E}^{+}:\{\boldsymbol{x}\mid v_1(\boldsymbol{x})=E^{+}\}$.
The parameters of $V_0(\boldsymbol{x})$ are
$(m,a, b, K, d, e, f, \lambda, \alpha, \beta) = 
(2, 1.8, 1, 0.02396, 3, 8, 1, 0.27, 0.34\pi, 0.05\pi)$
(the ``$d=3$'' row of B1 in Table~\ref{TAB:B}), for which
$E_{+}=13.7888$ and $E^{+}=1.78493$.
The tangent points between $\mathrm{C}_{\infty}$ and $\mathrm{E}_{+}$
($\mathrm{E}^{+}$)
almost agree with the local minima (saddles) of $V_0(\boldsymbol{x})$,
i.e., $\boldsymbol{x}_{+}$ and $-\boldsymbol{x}_{+}$
($\boldsymbol{x}^{+}$ and $-\boldsymbol{x}^{+}$),
and so do the valleys $\mathrm{C}$ and $\mathrm{C}_{\infty}$.
$\{\boldsymbol{n}_{+}, \boldsymbol{\tau}_{+}\}$ 
($\{\boldsymbol{n}^{+}, \boldsymbol{\tau}^{+}\}$)
denote the eigenvectors
of $\Hat{G}_{0}(\boldsymbol{x})$ at the minima
(saddles), which also almost agree with
the common tangent and normal vectors, i.e.,
$\{\boldsymbol{n}_{v}, \boldsymbol{\tau}_{v}\}$,
 between $\mathrm{C}_{\infty}$ and $\mathrm{E}_{+}$
($\mathrm{C}_{\infty}$ and $\mathrm{E}^{+}$).
}
\label{fig:show_pot}
\end{figure}

For $V_0(\boldsymbol{x})$ in Eq.~(\ref{LEQ:Pot0}),
let us consider a ratchet potential with a two-fold symmetry as shown in
Fig.~\ref{fig:show_pot}, and call it the two-tooth ratchet model.
In such a case, $v_{0}(\boldsymbol{x})$ and 
$v_{1}(\boldsymbol{x})$ also have two-fold symmetry.
Here, we define them as
\begin{align}
 v_{0}(\boldsymbol{x})
  &=
  \lvert\boldsymbol{a}\cdot \boldsymbol{x}\lvert^2
  +
  \lambda
  \frac{\lvert\boldsymbol{e}\cdot \boldsymbol{x}\rvert^2
  \lvert\boldsymbol{e}_{\perp}\cdot \boldsymbol{x}\rvert^2}
{\lvert\boldsymbol{x}\rvert^2}
,
\label{v0:core}
\\
 v_{1}(\boldsymbol{x})
  &= \frac{1}{2}
\lvert\boldsymbol{d}\cdot\boldsymbol{x}\rvert^2
,
\label{v1:core}
\end{align}
where
$\boldsymbol{a}$, $\boldsymbol{d}$, $\boldsymbol{e}$, and $\boldsymbol{e}_{\perp}$ are
complex vector-valued parameters:
\begin{gather*}
\boldsymbol{a}=
\begin{pmatrix}
\frac{1}{a}\\ \frac{i}{b}
\end{pmatrix}
,
\quad
 \boldsymbol{e}=
\begin{pmatrix}
 \cos\beta&-\sin\beta\\
\sin\beta&  \cos\beta
\end{pmatrix}
\begin{pmatrix}
\frac{1}{e}\\\frac{i}{f}
\end{pmatrix}
,
\\
\boldsymbol{e}_{\perp}=
\begin{pmatrix}
0&-1\\
1& 0
\end{pmatrix}
\boldsymbol{e}
,
\quad
 \boldsymbol{d}\equiv
\begin{pmatrix}
 \cos\alpha &-\sin\alpha\\
 \sin\alpha & \cos\alpha
\end{pmatrix}
\begin{pmatrix}
d\\
i
\end{pmatrix},
\label{ComplexVecs}
\end{gather*}
with $i^2=-1$, $a > 0$, $b > 0$, $e \geq 0$, $f \geq 0$, and $0\leq\beta<\frac{\pi}{4}$.

We assume $m\gg 1$ and $0 < K \ll 1$ in Eq.~(\ref{LEQ:Pot0}), unless stated otherwise.
Then, the curve
$\mathrm{C}_{\infty}: \{\boldsymbol{x}\mid v_0(\boldsymbol{x})=1\}$
approximately indicates the potential valley.
If $\lambda = 0$, $\mathrm{C}_{\infty}$ is an ellipse, i.e.,
$
\lvert\boldsymbol{a}\cdot \boldsymbol{x}\rvert^2
=
(\boldsymbol{a}\cdot \boldsymbol{x})
(\boldsymbol{a}^{\ast}\cdot \boldsymbol{x})
=\left(\frac{x}{a}\right)^2+\left(\frac{y}{b}\right)^2=1$, 
otherwise, for $\lambda\ne 0$, it adds a fourth harmonic deformation, with
reference axes $(\cos\beta,\sin\beta)^{\transpose}$ and 
$(-\sin\beta,\cos\beta)^{\transpose}$.
The sharpness of
the potential profile normal to $\mathrm{C}_{\infty}$
is tuned by $m$
(as shown in Sec.~\ref{subsec:Hesse},  
the curvature is proportional to $m^2$ for $m\geq 1$).
Function $v_{1}(\boldsymbol{x})$ is a potential function with
an anisotropic axis $(\cos\alpha,\sin\alpha)^{\transpose}$.
The curve of
$\lvert\boldsymbol{d}\cdot\boldsymbol{x}\rvert^2 = \mathrm{constant}$
is an ellipse whose short axis is along $(\cos\alpha,\sin\alpha)^{\transpose}$
and whose eccentricity is $\sqrt{1-d^{-2}}$ ($d>1$).
If $\mathrm{C}_{\infty}$ does not have line symmetry with respect to the anisotropic axis,
 the pathway along the valley has a ratchet property.

\subsection{\label{subsec:v0v1} Features of the potential function}

Let $\mathrm{O}$, $\mathrm{C}$, $\boldsymbol{x}_{\sigma}$ ($\sigma\in \{-,+\}$),
and $\boldsymbol{x}^{\mu}$ ($\mu\in \{-,+\}$)  be the origin, the potential valley
of $V_{0}(\boldsymbol{x})$, the local minimum, and the saddle, respectively 
(Fig.~\ref{fig:show_pot}) [
$\boldsymbol{x}_{+}$ and $\boldsymbol{x}^{+}$ are placed in $x>0$ and $y>0$,
 and $\boldsymbol{x}_{-}= -\boldsymbol{x}_{+}$ and
$\boldsymbol{x}^{-}= -\boldsymbol{x}^{+}$].

The minima and saddles satisfy
$\partial_{ \boldsymbol{x}} V_{0}(\boldsymbol{x})= \boldsymbol{0}$, and
Eq.~(\ref{LEQ:Pot0}) leads to
\begin{equation}
\frac{m}{2}\left[
1-
\left\{
v_{0}(\boldsymbol{x})
\right\}^m
\right]
\left\{
v_{0}(\boldsymbol{x})
\right\}^{m-1}
\partial_{\boldsymbol{x}}v_{0}(\boldsymbol{x})
+K
\partial_{\boldsymbol{x}}v_{1}(\boldsymbol{x})
=
\boldsymbol{0}
.
\label{force} 
\end{equation}
Using the orthogonal vectors
\begin{equation}
\boldsymbol{\tau}_{v}
\equiv
\frac{ \partial_{\boldsymbol{x}}v_{0}(\boldsymbol{x})}
{\lvert\partial_{\boldsymbol{x}}v_{0}(\boldsymbol{x})\rvert},
\quad
\boldsymbol{n}_{v}
\equiv
\begin{pmatrix}
 0&-1\\ 1 &0
\end{pmatrix}
\boldsymbol{\tau}_{v}
,
\label{n_v}
\end{equation}
we decompose Eq.~(\ref{force})
in two directions as
\begin{gather}
\frac{m}{2}\left[
1-
\left\{
v_{0}(\boldsymbol{x})
\right\}^m
\right]
\left\{
v_{0}(\boldsymbol{x})
\right\}^{m-1}
=-K
\frac{ \boldsymbol{\tau}_{v}\cdot
\partial_{\boldsymbol{x}}v_{1}(\boldsymbol{x})}
{\lvert\partial_{\boldsymbol{x}}v_{0}(\boldsymbol{x})\rvert}
,
\label{force_1}
\\
\boldsymbol{n}_{v}\cdot \partial_{\boldsymbol{x}}v_{1}(\boldsymbol{x})
=0
.
\label{force_2}
\end{gather}

When taking the limit $m\rightarrow \infty$
in Eqs.~(\ref{force_1}) and (\ref{force_2}), 
the minima and the saddles, $\{\boldsymbol{x}_{\sigma},\boldsymbol{x}^{\mu}\}$, 
satisfy
\begin{gather}
\boldsymbol{n}_{v}
\cdot
\partial_{ \boldsymbol{x}}
v_1(\boldsymbol{x})
=
0, \quad \boldsymbol{x}\in \mathrm{C}_{\infty}.
\label{3th:force_2}
\end{gather}
For a geometrical interpretation of Eq.~(\ref{3th:force_2}),
let us define $\mathrm{E}:\{\boldsymbol{x}\mid v_1(\boldsymbol{x})=E\}$
as a family of curves specified by the parameter $E$.
Then, Eq.~(\ref{3th:force_2}) means that with certain values of $E$, the curves
$\mathrm{E}$ and $\mathrm{C}_{\infty}$ have tangent points at
$\boldsymbol{x} \in \{\boldsymbol{x}_{\sigma},\boldsymbol{x}^{\mu}\}$
 at which $\boldsymbol{n}_v$ is tangent to both curves.
As shown in Fig.~\ref{fig:show_pot}, there are two cases of tangency
depending on $E$;
let $\mathrm{E}_{+}:\{\boldsymbol{x}\mid v_1(\boldsymbol{x})=E_{+}\}$
[ $\mathrm{E}^{+}:\{\boldsymbol{x}\mid v_1(\boldsymbol{x})=E^{+}\}$]
be a curve that is tangent to
$\mathrm{C}_{\infty}$ at $\boldsymbol{x}=\boldsymbol{x}_{\sigma}$
[$\boldsymbol{x}=\boldsymbol{x}^{\mu}$]
as $E$ reaches $E_{+}$ [$E^{+}$].
Since we choose $K>0$, we have $E^{+}\leq E_{+}$. Therefore,
$\mathrm{E}_{+}$ is externally tangent to
$\mathrm{C}_{\infty}$, and
$\mathrm{E}^{+}$ is internally tangent to $\mathrm{C}_{\infty}$.
However, these describe only the local relationships between
$v_0(\boldsymbol{x})$ and $v_1(\boldsymbol{x})$ at
$\boldsymbol{x}=\boldsymbol{x}_{\sigma}$ ($E= E^{+}$) 
and $\boldsymbol{x}^{\mu}$ ($E^{+}$) as they contact;
the global relationships between them remain undefined.
As global conditions in which
 $\mathrm{E}_{+}$ ($\mathrm{E}^{+}$) contacts 
with $\mathrm{C}_{\infty}$  only at two points 
$\boldsymbol{x}=\boldsymbol{x}_{+}$
and $\boldsymbol{x}_{-}$
($\boldsymbol{x}=\boldsymbol{x}^{+}$ and $\boldsymbol{x}^{-}$),
we insist that all points
on $\mathrm{C}_{\infty}$ satisfy
\begin{equation}
E^{+} \leq
v_1(\boldsymbol{x})
\leq E_{+},
\label{ineq:E}
\end{equation}
where equal cases of the left and right sides hold
at $\boldsymbol{x}=\boldsymbol{x}^{\mu}$ and
$\boldsymbol{x}_{\sigma}$, respectively.
In this case, letting $\Delta V$ be the difference of $V_{0}(\boldsymbol{x})$ [Eq.~(\ref{LEQ:Pot0})]
between the saddle and the local minimum, we have
\begin{equation}
 \Delta V = K(E_{+}-E^{+}).
\label{DV}
\end{equation}

\subsection{\label{subsec:Hesse} Hessian matrix}

The Hessian matrix 
$\Hat{G}_{0}(\boldsymbol{x})\equiv
\partial_{\boldsymbol{x}}\partial_{\boldsymbol{x}}^{\transpose} 
V_{0}(\boldsymbol{x})$ 
is diagonalized approximately for $m\gg 1$.
We denote its eigenvectors by
$\boldsymbol{n}(\boldsymbol{x})$ and $\boldsymbol{\tau}(\boldsymbol{x})$, i.e.,
\begin{align}
\Hat{G}_{0}(\boldsymbol{x})
\boldsymbol{\tau}(\boldsymbol{x})
&=\Lambda_{\tau}(\boldsymbol{x})
\boldsymbol{\tau}(\boldsymbol{x}),
\label{eigen_tau}
\\
\Hat{G}_{0}(\boldsymbol{x})
\boldsymbol{n}(\boldsymbol{x})
&=\Lambda_{n}(\boldsymbol{x})
\boldsymbol{n}(\boldsymbol{x}),
\label{eigen_n}
\end{align}
where
$\Lambda_{n}(\boldsymbol{x})$ and $\Lambda_{\tau}(\boldsymbol{x})$
are the corresponding eigenvalues, respectively;
$\boldsymbol{n}(\boldsymbol{x})$ and $\boldsymbol{\tau}(\boldsymbol{x})$
are tangent and normal to $\mathrm{C}$ at 
$\boldsymbol{x}\in\{\boldsymbol{x}_{\sigma},\boldsymbol{x}^{\mu}\}$;
$\Lambda_{n}(\boldsymbol{x})$ and $\Lambda_{\tau}(\boldsymbol{x})$
are equivalent to the curvatures of $V_{0}(\boldsymbol{x})$ along the
$\boldsymbol{n}(\boldsymbol{x})$ and $\boldsymbol{\tau}(\boldsymbol{x})$ axes,
respectively.
Hereinafter, we denote these eigenvectors by
$\boldsymbol{n}(\boldsymbol{x}_{\sigma})\equiv  \boldsymbol{n}_{\sigma}$,
$\boldsymbol{\tau}(\boldsymbol{x}_{\sigma})\equiv \boldsymbol{\tau}_{\sigma}$,
$\boldsymbol{n}(\boldsymbol{x}^{\mu})\equiv \boldsymbol{n}^{\mu}$, and
$\boldsymbol{\tau}(\boldsymbol{x}^{\mu})\equiv \boldsymbol{\tau}^{\mu}$.
In addition, we define
the reference direction of $\boldsymbol{n}_{\sigma}$ ($\boldsymbol{n}^{\mu}$)
as directed in the counterclockwise (clockwise) 
pathway of $\mathrm{C}$, and
$\boldsymbol{\tau}_{\sigma}$ ($\boldsymbol{\tau}^{\mu}$) as directed in the right-hand side
of $\boldsymbol{n}_{\sigma}$ ($\boldsymbol{n}^{\mu}$) (see Fig.~\ref{fig:show_pot}).

From Eq.~(\ref{LEQ:Pot0}), we have 
\begin{align}
\Hat G_{0}(\boldsymbol{x})
=&
-\frac{m}{2}\left[
m-1-(2m-1)
v_{0}(\boldsymbol{x})^m
\right]
v_{0}(\boldsymbol{x})^{m-2}
\partial_{\boldsymbol{x}}v_{0}(\boldsymbol{x})
\partial_{\boldsymbol{x}}^{\transpose}v_{0}(\boldsymbol{x})
\nonumber
\\
&-
\frac{m}{2}\left[
1-
v_{0}(\boldsymbol{x})^m
\right]
v_{0}(\boldsymbol{x})^{m-1}
\partial_{\boldsymbol{x}}
\partial_{\boldsymbol{x}}^{\transpose}v_{0}(\boldsymbol{x})
-K
\partial_{ \boldsymbol{x}}
\partial_{ \boldsymbol{x}}^{\mathrm{T}}
v_1(\boldsymbol{x})
 .
 \label{Hessian}
\end{align}
Substituting $v_0(\boldsymbol{x})=1$
into the first two factors in the first term in Eq.~(\ref{Hessian}),
and $v_0(\boldsymbol{x})=1+\delta v_0$ into the second term,
we approximate $\Hat G_{0}(\boldsymbol{x})$ as
\begin{align}
\Hat G_{0}(\boldsymbol{x})
\approx &
\frac{m^2}{2}
\lvert\partial_{\boldsymbol{x}}v_{0}(\boldsymbol{x})\rvert^2
\boldsymbol{\tau}_{v}
\boldsymbol{\tau}_{v}^{\mathrm T}
+
\frac{m^2}{2}\delta
v_{0}
\partial_{\boldsymbol{x}}
\partial_{\boldsymbol{x}}^{\transpose}v_{0}(\boldsymbol{x})
 -K
\partial_{ \boldsymbol{x}}
\partial_{ \boldsymbol{x}}^{\mathrm{T}}
v_1(\boldsymbol{x})
 ,
\label{G0:pre}
\end{align}
where $\boldsymbol{\tau}_{v}$ is defined in Eq.~(\ref{n_v}), and,
from Eq.~(\ref{force_1}), $\delta v_0$ is estimated as
\begin{equation}
\delta v_{0}\approx
 \frac{2K\boldsymbol{\tau}_{v}\cdot
\partial_{\boldsymbol{x}}v_{1}(\boldsymbol{x})}
{m^2 \lvert\partial_{\boldsymbol{x}}v_{0}(\boldsymbol{x})\rvert}
.
\label{delta_v0}
\end{equation}

From Eqs.~(\ref{G0:pre}) and (\ref{delta_v0}),
neglecting the nondiagonal components (which are not essential),
we obtain
\begin{gather}
\Hat G_{0}(\boldsymbol{x})
\approx
 \frac{m^2}{2}
\lvert\partial_{\boldsymbol{x}}v_{0}(\boldsymbol{x})\rvert^2
\boldsymbol{\tau}_{v}
\boldsymbol{\tau}_{v}^{\mathrm T}
+K g(\boldsymbol{x})
 \boldsymbol{n}_{v}
 \boldsymbol{n}_{v}^{\mathrm T}
,
\label{Hessian:G}
\\
 g(\boldsymbol{x})\equiv
   \boldsymbol{n}_{v}^{\mathrm T} 
   \left[
    \left\{
 \boldsymbol{\tau}_{v}\cdot
\partial_{ \boldsymbol{x}}
v_1(\boldsymbol{x})
\right\}
 \frac{
\partial_{\boldsymbol{x}}
\partial_{\boldsymbol{x}}^{\transpose}v_{0}(\boldsymbol{x})
 }{ \lvert\partial_{\boldsymbol{x}}v_{0}(\boldsymbol{x})\rvert }
  -
\partial_{ \boldsymbol{x}}
\partial_{ \boldsymbol{x}}^{\mathrm{T}}
v_1(\boldsymbol{x})
\right]
\boldsymbol{n}_{v}
\label{Hessian:g}
\end{gather}
for $\boldsymbol{x} \in \{ \boldsymbol{x}_{\sigma}, \boldsymbol{x}^{\mu}\}$.
This is valid for $m \gg 1$, in which
the eigenvectors of the Hessian matrix at 
$\boldsymbol{x} \in \{ \boldsymbol{x}_{\sigma}, \boldsymbol{x}^{\mu}\}$, i.e.,
$\boldsymbol{\tau}_{\sigma}$, $\boldsymbol{n}_{\sigma}$,
$\boldsymbol{\tau}^{\mu}$ and $\boldsymbol{n}^{\mu}$,
are well approximated with $\boldsymbol{\tau}_{v}$ and $\boldsymbol{n}_{v}$
in Eq.~(\ref{n_v}).
We thus have
\begin{equation}
\Lambda_{\tau}(\boldsymbol{x})\approx 
 \frac{m^2}{2}
\lvert\partial_{\boldsymbol{x}}v_{0}(\boldsymbol{x})\rvert^2, 
\quad
\Lambda_{n}(\boldsymbol{x})\approx K g(\boldsymbol{x})
\label{eigenvals}
\end{equation}
at $\boldsymbol{x} \in \{ \boldsymbol{x}_{\sigma}, \boldsymbol{x}^{\mu}\}$
for $m\gg 1$.

\section{\label{Quantities} Performance indexes}

We characterize the rotational-motion performance
of the 2D ratchet using
the mean angular momentum (MAM)
\begin{equation}
L \equiv \overline{X_t\dot{Y}_t-Y_t\dot{X}_t},
\label{MAM} 
\end{equation}
the mean angular velocity (MAV) $\omega \equiv \overline{\Dot{\Theta}_t}$, and the
efficiency
\begin{gather}
\eta =
 \frac{
\gamma L\omega
 +
P_I
}{P_h}
,
\label{def:eta1}
\end{gather}
where
\begin{gather}
\Theta_t 
\equiv 
\int_{0}^{t} d s
\left(
\frac{X_s\dot Y_s-Y_s\dot X_s
}{\lvert\boldsymbol{X}\rvert^2}
\right)
\equiv
\theta(\boldsymbol{X})
-
\theta(\boldsymbol{X}_0)
,
\label{def:theta}
\\
P_I \equiv
-
\overline{
\dot{\boldsymbol{X}}\cdot \boldsymbol{f}_{I}(\boldsymbol{X})
}
=\frac{I}{2\pi}\overline{\dot \theta(\boldsymbol{X})}
=
\frac{I \omega}{2\pi}
,
\label{def:P_I}
\\ 
  P_h \equiv
 \overline{h\boldsymbol{N}_t
\cdot \Dot{\boldsymbol{X}}(t)}
,
\label{def:P_h}
\end{gather}
i.e., the counterclockwise displacement angle about the origin,
the power consumed by the load, and the input power of the external field
(which is equivalent to the total power consumption), respectively.
We have replaced Eq.~(\ref{def:eta}) with Eq.~(\ref{def:eta1}) because
the long-time averages of
the relative angular momentum 
$L_t'$ [Eq.~(\ref{def:L'})] and
the relative angular velocity $\omega_t'$ [Eq.~(\ref{def:omega'})]
agree with $L$ and $\omega$, respectively, to $o(h^2)$
(see Appendix~\ref{App:XxX}).
Hereinafter,
$O(\cdot)$ and $o(\cdot)$ denote the Landau symbols (Big- and Little-O).

In Eq.~(\ref{MAM}), the direction $L>0$ corresponds to counterclockwise
rotation. 
The direction of the ratchet (chirality) is defined as 
the direction in which one goes around
a circular pathway along $\mathrm{C}$ through each of the minima 
from the side of steeper gradient to the more gentle one. Hence, the ratchet in Fig.~\ref{fig:show_pot}
has counterclockwise chirality.
In the following analytical and numerical simulation results,
under the RDDF,
the net rotation of the ratchet tends to be the same as the chirality.
In the numerical simulations, we examine only the case of $I=0$ and
we treat the efficiency as
\begin{gather}
\eta =
 \frac{
\gamma L\omega}{P_h}.
\label{def:eta0}
\end{gather}

In this paper, we consider a ratchet system in a thermal bath
under a weak and slow external field, and we impose the following requirements: 
1) the typical magnitudes of $V_{h}(\boldsymbol{x}, t)$ and $I$
(which are denoted by $O(h)$ and $O(I)$, respectively, in an energetic dimension)
are smaller than the energy barrier 
$\Delta V$ [see Eq.~(\ref{DV})] to a sufficient extent, it being assumed hereinafter that $O(I)\sim O(h)$;
2) the mean switching time of the RDDF ($T_{p}\equiv \frac{2\pi}{\Omega}$)
is longer than the typical relaxation time $T_r$ of a trajectory to a sufficient extent,
i.e., $\Omega T_r \ll 1$, where $T_r^{-1}$ is related to
the curvature of $V_{0}(\boldsymbol{x})$ 
at the minima [or more likely is governed by the smallest eigenvalue
of $\Hat{G}_{0}(\boldsymbol{x}_{\sigma})$].

In a previous paper \cite{doi:10.7566/JPSJ.84.044004},
we proposed a framework for obtaining approximate expressions for
the performance indexes 
($L$, $\omega$, and $P_h$) using a master equation
for coarse-grained states under the assumptions mentioned above.
For a self-contained description, we briefly introduce 
the basic construction of the master equation and 
its applications to the computation of $L$, $\omega$, and $P_h$
in Secs.~\ref{DefStates} and \ref{CGK-theory}. 
In Sec.~\ref{Expressions}, we show the final expressions for $L$, $\omega$, and $P_h$
that we use in later sections.

\begin{figure}[!tb]
\def\Size{10.0cm}
\centering
\includegraphics[width=\Size,keepaspectratio,clip]{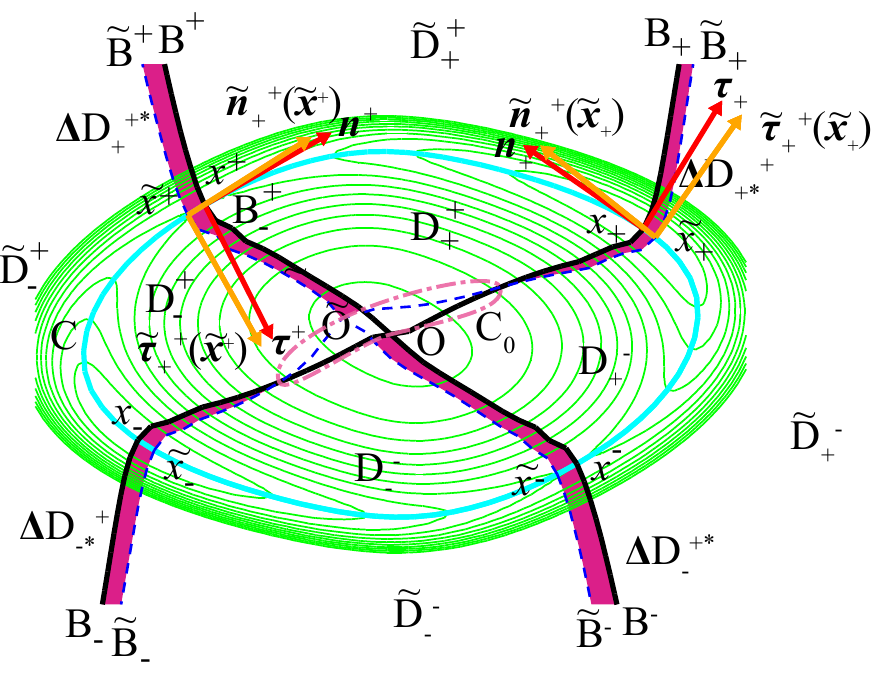}
\caption{
(Color online)
Notation for moving domain boundaries on $V(\boldsymbol{x},t)$.
With $\sigma\in \{-,+\}$ and $\mu\in \{-,+\}$,
$\Tilde{\mathrm{O}}$, $\Tilde{\boldsymbol{x}}_{\sigma}$, and $\Tilde{\boldsymbol{x}}^{\mu}$
represent the local maximum, the local minimum, and the saddle points, respectively, 
of $V(\boldsymbol{x},t)$.
The 2D space is divided into four domains $\Tilde{\mathrm{D}}_{\sigma}^{\mu}$
by the ridge curves
$\Tilde{\mathrm{B}}_{\sigma}$ and $\Tilde{\mathrm{B}}^{\mu}$
of $V(\boldsymbol{x},t)$.
$\Tilde{\boldsymbol{\tau}}_{\sigma}^{\mu}(\Tilde{\boldsymbol{x}}_{\sigma})$ and
$\Tilde{\boldsymbol{\tau}}_{\sigma}^{\mu}(\Tilde{\boldsymbol{x}}^{\mu})$
[$\Tilde{\boldsymbol{n}}_{\sigma}^{\mu}(\Tilde{\boldsymbol{x}}_{\sigma})$ and
$\Tilde{\boldsymbol{n}}_{\sigma}^{\mu}(\Tilde{\boldsymbol{x}}^{\mu})$]
are the tangent (normal) vectors to 
$\Tilde{\mathrm{B}}_{\sigma}$ and $\Tilde{\mathrm{B}}^{\mu}$
at the minimum and the saddle points.
$\mathrm{C}_0$ (dashed--dotted curve) denotes a closed curve
surrounding a central region
of the potential that at least includes $\mathrm{O}$, $\Tilde{\mathrm{O}}'$ and
either a cross point between $\mathrm{B}_{\sigma}$ and $\Tilde{\mathrm{B}}_{\sigma}$,
or another between $\mathrm{B}^{\mu}$ and $\Tilde{\mathrm{B}}^{\mu}$.
$\Delta\mathrm{D}_{\sigma\ast}^{\mu}$ [$\Delta\mathrm{D}_{\sigma}^{\mu \ast}$] (hatched regions)
denotes the region surrounded by $\Tilde{\mathrm{B}}_{\sigma}$ and
$\mathrm{B}^{\mu}$ [$\Tilde{\mathrm{B}}^{\mu}$ and $\mathrm{B}_{\mu}$]
but excluding the interior of $\mathrm{C}_0$.
}
\label{fig:PotShapes}
\end{figure}

\subsection{\label{DefStates} Coarse-grained states and related definitions}

As shown in Fig.~\ref{fig:PotShapes},
we denote $\mathrm{O}$, $\boldsymbol{x}_{\sigma}$, and $\boldsymbol{x}^{\mu}$
($\sigma \in \{+,-\}$ and $\mu \in \{+,-\}$) as the origin,
the local minimum, and the saddle, respectively, determined by
$\partial_{ \boldsymbol{x}} V_{0}(\boldsymbol{x})= \boldsymbol{0}$.
Hereinafter, the signs ``$+$'' and ``$-$'' are identical with
$+1$ and $-1$, thereby $\boldsymbol{x}_{\sigma} = \sigma\boldsymbol{x}_{+}$ and
$\boldsymbol{x}^{\mu} = \mu\boldsymbol{x}^{+}$, where
$\boldsymbol{x}_{+}$ and $\boldsymbol{x}^{+}$ lie
in $x>0$ and $y>0$, respectively.
Furthermore, $\mathrm{B}_{\sigma}$ ($\mathrm{B}^{\mu}$) denotes the ridge curve running from
$\mathrm{O}$  through $\boldsymbol{x}_{\sigma}$ ($\boldsymbol{x}^{\mu}$) outward;
$\mathrm{D}_{\sigma}^{\mu}$ denotes the domain 
surrounded by $\mathrm{B}_{\sigma}$ and $\mathrm{B}^{\mu}$;
$\mathrm{C}$ denotes the potential valley of $V_{0}(\boldsymbol{x})$.

We extend these static ridge curves to temporally varying ridge curves
on the basis of the function
\begin{equation}
 V(\boldsymbol{x}, t) \equiv V_{0}(\boldsymbol{x})
+ V_{h}(\boldsymbol{x},t) +\frac{I}{2\pi} \theta(\boldsymbol{x})
\end{equation}
with the second and third terms in
Eqs.~(\ref{LEQ:Poth}) and (\ref{LEQ:PotI});
$\Tilde{\mathrm{O}}$, $\Tilde{\boldsymbol{x}}_{\sigma}$, and
$\Tilde{\boldsymbol{x}}^{\mu}$ denote
the local maximum, the local minimum, and the saddle (Fig.~\ref{fig:PotShapes}) given by
$\partial_{ \boldsymbol{x}} V(\boldsymbol{x}, t)= \boldsymbol{0}$,
respectively, which move temporally with the external field.
Similarly, $\Tilde{\mathrm{B}}_{\sigma}$ ($\Tilde{\mathrm{B}}^{\mu}$) denotes
the ridge curves running from
$\Tilde{\mathrm{O}}$ through $\Tilde{\boldsymbol{x}}_{\sigma}$ 
($\Tilde{\boldsymbol{x}}^{\mu}$) outward; 
$\Tilde{\mathrm{D}}_{\sigma}^{\mu}$ denotes the domain surrounded by
$\Tilde{\mathrm{B}}_{\sigma}$ and $\Tilde{\mathrm{B}}^{\mu}$;
$\Tilde{\mathrm{C}}$ denotes the potential valley of $V(\boldsymbol{x}, t)$.

Corresponding to 
$\boldsymbol{\tau}(\boldsymbol{x})$ and 
$\boldsymbol{n}(\boldsymbol{x})$ in 
Eqs.~(\ref{eigen_tau}) and (\ref{eigen_n}),
we denote by 
$\Tilde{\boldsymbol{\tau}}_{\sigma}^{\mu}(\boldsymbol{x})$ and
$\Tilde{\boldsymbol{n}}_{\sigma}^{\mu}(\boldsymbol{x})$
the tangent and normal vectors at the point $\boldsymbol{x}$ on
 the boundary of $\Tilde{\mathrm{D}}_{\sigma}^{\mu}$
($\boldsymbol{x}\in\Tilde{\mathrm{B}}_{\sigma}$ or
$\boldsymbol{x}\in\Tilde{\mathrm{B}}^{\mu}$), where
the reference direction of $\Tilde{\boldsymbol{n}}_{\sigma}^{\mu}(\boldsymbol{x})$
lies in $\Tilde{\mathrm{D}}_{\sigma}^{\mu}$, and
$\Tilde{\boldsymbol{\tau}}_{\sigma}^{\mu}(\boldsymbol{x})$ is
oriented in the right-hand direction of 
$\Tilde{\boldsymbol{n}}_{\sigma}^{\mu}(\boldsymbol{x})$ (Fig.~\ref{fig:PotShapes}).
The vectors
$\Tilde{\boldsymbol{\tau}}_{\sigma}^{\mu}(\boldsymbol{x})$
and
$\Tilde{\boldsymbol{n}}_{\sigma}^{\mu}(\boldsymbol{x})$
are the eigenvectors of the Hessian matrix
$
\Hat{G}(\boldsymbol{x})\equiv
\partial_{\boldsymbol{x}}\partial_{\boldsymbol{x}}^{\transpose}
V(\boldsymbol{x},t)
$, i.e.,
\begin{align}
\Hat{G}(\boldsymbol{x})
\Tilde{\boldsymbol{\tau}}_{\sigma}^{\mu}(\boldsymbol{x})
&=\Lambda_{\tau}(\boldsymbol{x})
\Tilde{\boldsymbol{\tau}}_{\sigma}^{\mu}(\boldsymbol{x}),
\label{G_eigen_tau}
\\
\Hat{G}(\boldsymbol{x})
\Tilde{\boldsymbol{n}}_{\sigma}^{\mu}(\boldsymbol{x})
&=\Lambda_{n}(\boldsymbol{x})
\Tilde{\boldsymbol{n}}_{\sigma}^{\mu}(\boldsymbol{x}),
\label{G_eigen_n}
\end{align}
where
$\Lambda_{\tau}(\boldsymbol{x})$ and $\Lambda_{n}(\boldsymbol{x})$
are the corresponding eigenvalues.
In particular, at
$\boldsymbol{x}\in
 \{\Tilde{\boldsymbol{x}}_{\sigma},\Tilde{\boldsymbol{x}}^{\mu}\}$,
$\Lambda_{\tau}(\boldsymbol{x})$ and $\Lambda_{n}(\boldsymbol{x})$
are the curvatures of $V(\boldsymbol{x},t)$ along the ridge curve and the valley, respectively; 
therefore, we have
$\Lambda_{\tau}(\Tilde{\boldsymbol{x}}_{\sigma})>0$,
$\Lambda_{n}(\Tilde{\boldsymbol{x}}_{\sigma})>0$,
$\Lambda_{\tau}(\Tilde{\boldsymbol{x}}^{\mu})>0$, and
$\Lambda_{n}(\Tilde{\boldsymbol{x}}^{\mu})<0$.

\subsection{Master equation for coarse-grained states}
\label{CGK-theory}

The time evolution of
probability density function (PDF) $p(\boldsymbol{x},t)$
for $\boldsymbol{X}=\boldsymbol{x}$ obeys the Fokker--Planck equation as
\begin{gather}
\label{FPE}
\partial_{t} p(\boldsymbol{x},t) = 
 - 
\partial_{\boldsymbol{x}}\cdot
\boldsymbol{J}(\boldsymbol{x},t)
,
\\
\label{FPE:J} 
\boldsymbol{J}(\boldsymbol{x},t)
\equiv \left[
-
\partial_{\boldsymbol{x}}
V(\boldsymbol{x},t)
\right]
p(\boldsymbol{x},t)
-
D\partial_{\boldsymbol{x}}
p(\boldsymbol{x},t),
\end{gather}
where $\partial_t\equiv \frac{\partial}{\partial t}$ and
$\partial_{\boldsymbol{x}}\cdot \boldsymbol{J}(\boldsymbol{x},t)$ means the 2D divergence
of the probability current density.

In terms of $p(\boldsymbol{x},t)$, a probability for an event
$\boldsymbol{X}\in \mathrm{D}_{\sigma}^{\mu}$ is given by
\begin{equation}
P(\sigma,\mu,t)
\equiv 
\int_{\boldsymbol{x}\in \mathrm{D}_{\sigma}^{\mu}} 
\diff\boldsymbol{x}
p(\boldsymbol{x},t).
\label{Markov:Psig_mu}
\end{equation}
Using this, probabilities for events
$\boldsymbol{X}\in
\mathrm{D}_{\sigma}^{+}\cup  \mathrm{D}_{\sigma}^{-}$
and
$\boldsymbol{X}\in  
\mathrm{D}_{+}^{\mu}\cup  \mathrm{D}_{-}^{\mu}$
are represented as
$P(\sigma,t)=\sum_{\mu} P(\sigma,\mu,t)$ and
$Q(\mu,t)=\sum_{\sigma} P(\sigma,\mu,t)$, respectively.
Furthermore, the conditional probabilities,
 the relative probabilities of the event
$\boldsymbol{X}\in \mathrm{D}_{\sigma}^{\mu}$ under the conditions
$\boldsymbol{X}\in
\mathrm{D}_{\sigma}^{+}\cup  \mathrm{D}_{\sigma}^{-}$
and
$\boldsymbol{X}\in  
\mathrm{D}_{+}^{\mu}\cup  \mathrm{D}_{-}^{\mu}$,
are defined respectively as
\begin{equation}
 P(\sigma\mid\mu,t) \equiv \frac{P(\sigma,\mu,t)}{Q(\mu,t)},
\quad
Q(\mu\mid\sigma,t) \equiv \frac{P(\sigma,\mu,t)}{P(\sigma,t)}.
\label{COND_PQ}
\end{equation}

In addition to the assumptions 1) $V_{h} \ll \Delta V$  and 2) $\Omega T_{r} \ll 1$,
we assume that $D$ is so small that $D\ll \Delta V$ hereinafter.
Then, the PDF peaks sharply at
$\Tilde{\boldsymbol{x}}_{\sigma}$ [=$\boldsymbol{x}_{\sigma} + O(h)$],
otherwise almost vanishes in the other region, and 
the trajectories in the transition between two states
$\boldsymbol{X}\in\Tilde{\mathrm{D}}_{+}^{\mu}$ and
$\boldsymbol{X}\in\Tilde{\mathrm{D}}_{-}^{\mu}$ 
concentrate to $\Tilde{\mathrm{C}}$.

From Eqs.~(\ref{FPE}) and (\ref{Markov:Psig_mu}), the time derivative of
$P(\sigma,\mu,t)$ leads to
\begin{equation}
\partial_t P(\sigma,\mu,t)=
\int_{\boldsymbol{x}\in \mathrm{D}_{\sigma}^{\mu}} 
d\boldsymbol{x}
\left[
- \partial_{\boldsymbol{x}}\cdot
\boldsymbol{J}(\boldsymbol{x},t)
\right].  
\end{equation}
We divide the domain of integration $\mathrm D_{\sigma}^{\mu}$ into
$\Tilde{\mathrm D}_{\sigma}^{\mu}$ and
$\Delta \mathrm D_{\sigma}^{\mu} \equiv 
\mathrm D_{\sigma}^{\mu}-\Tilde{\mathrm D}_{\sigma}^{\mu}
$; 
$\Delta\mathrm{D}_{\sigma}^{\mu}$ consists of two domains
$\{\boldsymbol{x}\mid\boldsymbol{x} \in \mathrm D_{\sigma}^{\mu},
\boldsymbol{x} \notin \Tilde{\mathrm D}_{\sigma}^{\mu}\}$
and
$\{\boldsymbol{x}\mid\boldsymbol{x} \in \Tilde{\mathrm D}_{\sigma}^{\mu},
\boldsymbol{x} \notin \mathrm D_{\sigma}^{\mu}\}$. Therefore,
$\Delta\mathrm{D}_{\sigma}^{\mu}$ partly possesses a ``negative domain''
for which the sign of the integral is inverted.
From the assumptions $h \ll \Delta V$ and $D\ll \Delta V$, we can regard 
the PDF as actually vanishing around 
$\Tilde{\mathrm O}$ and $\mathrm{O}$, or the interior of $\mathrm{C}_0$
in Fig.~\ref{fig:PotShapes}.
We can thus consider the region $\Delta\mathrm{D}_{\sigma}^{\mu}$ as a sum of
the part surrounded by $\mathrm{B}_{\sigma}$ and $\Tilde{\mathrm{B}}_{\sigma}$ 
excluding the interior of  $\mathrm{C}_0$, and the other surrounded by
$\mathrm{B}^{\mu}$ and $\Tilde{\mathrm{B}}^{\mu}$ excluding the interior of $\mathrm{C}_0$, 
as indicated by hatched regions in Fig.~\ref{fig:PotShapes}.
Hereinafter, we denote by
$\Delta\mathrm{D}_{\sigma\ast}^{\mu}$ the former region, and by
$\Delta\mathrm{D}_{\sigma}^{\mu\ast}$ the latter one.
Dividing the domain of integration $\mathrm D_{\sigma}^{\mu}$
into $\Tilde{\mathrm D}_{\sigma}^{\mu}$,
$\Delta\mathrm{D}_{\sigma\ast}^{\mu}$, and
$\Delta\mathrm{D}_{\sigma}^{\mu\ast}$, we have
\begin{align}
\partial_t P(\sigma,\mu,t)
\approx &
\int_{\boldsymbol{x}\in \Tilde{\mathrm{D}}_{\sigma}^{\mu}} 
\diff\boldsymbol{x}
\left[
- \partial_{\boldsymbol{x}}
\cdot\boldsymbol{J}(\boldsymbol{x},t)
\right]
\nonumber
\\
& 
-
\partial_t P(\sigma,\mu,t)\bigr|_{Q}
+
\partial_t P(\sigma,\mu,t)\bigr|_{P},
\label{DP:express} 
\end{align}
where
$\partial_t P(\sigma,\mu,t)\bigr|_{Q} \equiv
-\int_{\Delta\mathrm{D}_{\sigma}^{\mu\ast}} \diff\boldsymbol{x}
\left[
- \partial_{\boldsymbol{x}}
\cdot\boldsymbol{J}(\boldsymbol{x},t)
\right]
$ and
$\partial_t P(\sigma,\mu,t)\bigr|_{P} \equiv
\int_{\Delta\mathrm{D}_{\sigma\ast}^{\mu}} \diff\boldsymbol{x}
\left[
- \partial_{\boldsymbol{x}}
\cdot\boldsymbol{J}(\boldsymbol{x},t)
\right]
$.

From the assumptions, 
we can approximate $p(\boldsymbol{x},t)$ with the thermal equilibrium PDF
[$\propto e^{-V(\boldsymbol{x},t)/D}$]
 around
the minima of $V(\boldsymbol{x},t)$, and we assume
$\boldsymbol{J}(\boldsymbol{x},t)= \boldsymbol{0}$ on
$\Tilde{\mathrm{B}}_{\sigma}$.
Applying this to the first term in Eq.~(\ref{DP:express}), we obtain
\begin{align}
\int_{\boldsymbol{x}\in \Tilde{\mathrm{D}}_{\sigma}^{\mu}} 
\diff\boldsymbol{x}
\left[
 - 
\partial_{\boldsymbol{x}}
\cdot
\boldsymbol{J}(\boldsymbol{x},t)
\right]
&\approx
\int_{\boldsymbol{x}\in \Tilde{\mathrm{B}}^{\mu}}
 \diff\boldsymbol{x}
 \,\Tilde{\boldsymbol{n}}_{\sigma}^{\mu}(\boldsymbol{x})\cdot
 \boldsymbol{J}(\boldsymbol{x},t)
\nonumber
\\
&\equiv 
\left(
\delta_{\sigma,-\mu}-
\delta_{\sigma,\mu}
\right)
J^{\mu}(t),
\label{def:J_mu}
\end{align}
where $J^{\mu}(t)$ represents the probability current
from $\Tilde{\mathrm{D}}_{\mu}^{\mu}$ to
$\Tilde{\mathrm{D}}_{-\mu}^{\mu}$.
Terms $\partial_t P(\sigma,\mu,t)\bigr|_{Q}$ and
$\partial_t P(\sigma,\mu,t)\bigr|_{P}$ are considered as follows.
For simplicity, we show them for the case $\sigma=\mu$ as
\begin{align}
\partial_t P(\mu,\mu,t)\bigr|_{Q}
&
=
\int_{\boldsymbol{x}\in\Tilde{\mathrm{B}}^{\mu}} \,\diff\boldsymbol{x}\,
\Tilde{\boldsymbol{n}}_{\mu}^{\mu}(\boldsymbol{x})
\cdot
\boldsymbol{J}(\boldsymbol{x},t)
-
\int_{\boldsymbol{x}\in\mathrm{B}^{\mu}} \,\diff\boldsymbol{x}\,
\boldsymbol{n}^{\mu}(\boldsymbol{x})
\cdot
\boldsymbol{J}(\boldsymbol{x},t)
\nonumber
\\
&\approx
Q(\mu,t)
\int_{\boldsymbol{x}\in\mathrm{B}^{\mu}} \,\diff\boldsymbol{x}\,
\frac{
\boldsymbol{n}^{\mu}(\boldsymbol{x})
\cdot
\left[
\boldsymbol{J}(\Tilde{\boldsymbol{x}}(\boldsymbol{x}),t)
-
\boldsymbol{J}(\boldsymbol{x},t)
\right]}
{Q(\mu,t)}
\label{_Q_RJ}
\\
&\approx Q(\mu,t)\partial_t P(\mu\mid\mu,t)
,
\label{Q_RJ}
\\
\partial_t P(\mu,\mu,t)\bigr|_{P}
&
\approx
P(\mu,t)
\int_{\boldsymbol{x}\in\mathrm{B}_{\mu}} \,\diff\boldsymbol{x}\,
\frac{\boldsymbol{n}_{\mu}(\boldsymbol{x})
\cdot
\boldsymbol{J}(\boldsymbol{x},t)}
{P(\mu,t)}
\label{_P_RJ}
\\
&\approx
P(\mu,t)\partial_t Q(\mu\mid\mu,t)
,
\label{P_RJ}
\end{align}
where $\Tilde{\boldsymbol{x}}(\boldsymbol{x})$ in Eq.~(\ref{_Q_RJ}) represents a map from a point
$\boldsymbol{x}$ on $\mathrm{B}^{\mu}$ to the corresponding nearest point 
 on $\Tilde{\mathrm{B}}^{\mu}$.
An action of relative current density
$\boldsymbol{J}(\Tilde{\boldsymbol{x}}(\boldsymbol{x}),t)-\boldsymbol{J}(\boldsymbol{x},t)$
in the integrand in Eq.~(\ref{_Q_RJ})  
[Eq.~(\ref{_P_RJ}), 
in which $\boldsymbol{J}(\Tilde{\boldsymbol{x}}(\boldsymbol{x}),t)=\boldsymbol{0}$
($\Tilde{\boldsymbol{x}}(\boldsymbol{x})\in \Tilde{\mathrm{B}}_{\mu}$)] 
is regarded as increasing $P(\mu\mid\mu,t)$
[decreasing $Q(\mu\mid\mu,t)$] without varying $Q(\mu,t)$ [$P(\mu,t)$].

In consequence, Eq.~(\ref{DP:express}) becomes
\begin{gather}
\partial_t
 P(\sigma,\mu,t)
\approx
\left(
\delta_{\sigma,-\mu}
-\delta_{\sigma,\mu}
\right)
J^{\mu}(t)+
J_{\sigma}^{\mu}(t),
\label{DP}
\\
J_{\sigma}^{\mu}(t)
\equiv
P(\sigma,t)
\partial_t
Q(\mu\mid\sigma,t)
-
Q(\mu,t)
\partial_t
P(\sigma\mid\mu,t)
.
\label{def:J'}
\end{gather}
Based on reaction rate theory~\cite{RevModPhys.62.251} or
Langer's method~\cite{PhysRevLett.21.973},
we obtain $J^{\mu}(t)$ in Eq.~(\ref{def:J_mu}) as
\begin{gather}
J^{\mu}(t)
\approx
W(-\mu,\mu,t) P(\mu,t)- W(\mu,\mu,t)P(-\mu,t),
\label{J_mu}
\\
W(\sigma,\mu,t)
\equiv
\frac{1}{2\pi}
e^{-[
V(\boldsymbol{x}^{\mu},t) -V(\boldsymbol{x}_{-\sigma},t)]/D}
\sqrt{
\frac{\Lambda_{\tau}(\boldsymbol{x}_{-\sigma})
\Lambda_{n}(\boldsymbol{x}_{-\sigma})\lvert\Lambda_{n}(\Tilde{\boldsymbol{x}}^{\mu})\rvert}
{\Lambda_{\tau}(\Tilde{\boldsymbol{x}}^{\mu})}},
\label{W_sig_mu}
\end{gather}
where $W(-\mu,\mu,t)$ [$W(\mu,\mu,t)$] is the transition probability
from a state $\boldsymbol{X}\in \mathrm{D}_{-\mu}$ to
a state $\boldsymbol{X}\in \mathrm{D}_{\mu}^{\mu}$
[$\boldsymbol{X}\in \mathrm{D}_{\mu}$ to $\boldsymbol{X}\in \mathrm{D}_{-\mu}^{\mu}$];
$\Lambda_{\tau}(\boldsymbol{x})$ and $\Lambda_{n}(\boldsymbol{x})$ are
the eigenvalues of the Hessian matrix $\Hat{G}(\boldsymbol{x})$.
For details, see Appendix~\ref{master}.

From Eq.~(\ref{DP}), the expectation value for the time derivative of a quantity
$A(\boldsymbol{X})\equiv A$ can be approximated with 
the corresponding coarse-grained variable $A(\boldsymbol{x}_{\sigma}) \equiv A_{\sigma}$
as
\begin{align}
\langle \dot{A} \rangle
&\approx
\sum_{\mu}\left(
A_{-\mu} - A_{\mu}
\right)
J^{\mu}(t)
+\sum_{\sigma,\mu}A_{\sigma}J_{\sigma}^{\mu}(t),
\label{PhysObs}
\end{align}
where $\langle A \rangle =\sum_{\sigma,\mu}A(\boldsymbol{x}_{\sigma})P(\sigma,\mu,t)$,
and $A$ is assumed to be a single-valued function of the position.
However, the MAM ($L$) and MAV ($\omega$) cannot be expressed straightforwardly 
as in Eq.~(\ref{PhysObs}),
e.g., it seems that the idea regarding $\omega$ as being
$\sum_{\sigma,\mu}\theta (\boldsymbol{x}_{\sigma})\partial_t P(\sigma,\mu,t)$
fails. This may be because
the angular momentum and angular velocity are classified as
axial vectors that possess information about the rotational direction
as well as their magnitudes.
Here, apart from Eq.~(\ref{PhysObs}),
we directly relate $L$ and $\omega$ with
the currents $J^{\mu}(t)$ and $J_{\sigma}^{\mu}(t)$
on the basis of physical consideration.
For an example with $\omega$, recalling that
$J^{\mu}(t)$, $J_{\mu}^{\mu}(t)$, and $-J_{-\mu}^{\mu}(t)$
express the counterclockwise currents through
$\mathrm{B}^{\mu}$,
$
\left\{
\theta(\boldsymbol{x}_{-\mu}) 
-
\theta(\boldsymbol{x}_{\mu}) 
\right\}
J^{\mu}(t)
$,
$
\left\{
\theta(\boldsymbol{x}_{-\mu}) 
-
\theta(\boldsymbol{x}_{\mu}) 
\right\}
J_{\mu}^{\mu}(t)$,
and 
$
\left\{
\theta(\boldsymbol{x}_{\mu}) 
-
\theta(\boldsymbol{x}_{-\mu})
\right\}
J_{-\mu}^{\mu}(t)$
approximate the phase velocities measured on the pathway from
$\theta(\boldsymbol{x}_{\mu})$ to $\theta(\boldsymbol{x}_{-\mu})$
through $\mathrm{B}^{\mu}$.

We represent $L$ and $\omega$ as a superposition of two parts
as $L= L^{(I)}+ L^{(h)}$ and $\omega = \omega^{(I)}+\omega^{(h)}$,
and express each term as
\begin{align}
L^{(I)}
&\approx
\frac{g_{L}}{2}
\sum_{\mu}
\left[
\boldsymbol{x}^{\mu}\times 
(\boldsymbol{x}_{-\mu}-\boldsymbol{x}_{\mu}) 
\right]_{z} 
\overline{
J^{\mu}(t)
},
\label{LIdef:LI}
\\
L^{(h)}
&\approx
g_{L}'
\sum_{\sigma,\mu}
(\boldsymbol{x}_{\sigma}\times \boldsymbol{x}^{\mu})_{z}
\nonumber
\\
\ &\times
\overline{
\left[
P(\sigma,t)
\partial_t Q(\mu\mid\sigma,t)
-
Q(\mu,t)
\partial_t P(\sigma\mid\mu,t)
\right]
},
\label{L_expect}
\\
\omega^{(I)} &\approx
\;g_{O}
\sum_{\mu}
\left[
\theta(\boldsymbol{x}_{-\mu}) 
-
\theta(\boldsymbol{x}_{\mu}) 
\right]
\overline{J^{\mu}(t)},
\label{Ex_omg|I}
\\
\omega^{(h)}
&\approx
\;g_{O}'
\sum_{\sigma,\mu}
\left[
\theta(\boldsymbol{x}_{-\mu})-
\theta(\boldsymbol{x}_{\mu})
\right]
\left(
\delta_{\sigma,\mu}-\delta_{\sigma,-\mu}
\right)
\nonumber
\\
\ &\times
\overline{
\left[
P(\sigma,t)\partial_t Q(\mu\mid\sigma,t)
-
Q(\mu,t) \partial_t P(\sigma\mid\mu,t)
\right]
},
\label{Ex_omg|h}
\end{align}
where $L^{(I)}$ and $L^{(h)}$, also 
$\omega^{(I)}$ and $\omega^{(h)}$,
come from the two types of current,
$J^{\mu}(t)$ and $J_{\sigma}^{\mu}(t)$.
Since the coarse-grained variables for the position and velocity vectors
are not exact, we employ dimensionless parameters
$g_{L}$, $g_{L}'$, $g_{O}$, and $g_{O}'$
to adjust the approximations to the numerical results;
as shown in Sec.~\ref{NumericalResults}, their actual values are $O(1)$.
Each summand in Eq.~(\ref{LIdef:LI}) represents 
the $z$-component of the angular momentum at $\boldsymbol{x}^{\mu}$
with the position $\boldsymbol{x}^{\mu}$ and the momentum
$\frac12(\boldsymbol{x}_{-\mu}-\boldsymbol{x}_{\mu})J^{\mu}(t)$,
where the latter is the mean of
$(\boldsymbol{x}_{-\mu}-\boldsymbol{x}^{\mu})J^{\mu}(t)$ and
$(\boldsymbol{x}^{\mu}-\boldsymbol{x}_{\mu})J^{\mu}(t)$.
In Eq.~(\ref{L_expect}), we regard the terms
$\boldsymbol{x}_{\mu}\times [(\boldsymbol{x}^{\mu}-\boldsymbol{x}_{\mu}) J_{\mu}^{\mu}(t)]$
($\sigma=\mu$)
and
$\boldsymbol{x}^{\mu}\times [-(\boldsymbol{x}_{-\mu}-\boldsymbol{x}^{\mu}) J_{-\mu}^{\mu}(t)]$
($\sigma=-\mu$)
as the counterclockwise angular momentum.
The interpretation of each summand in Eqs.~(\ref{Ex_omg|I})
and (\ref{Ex_omg|h}) has already been mentioned in the previous paragraph.
Note that
$\theta(\boldsymbol{x}_{-\mu})-
\theta(\boldsymbol{x}_{\mu})=\pi$.

The long time average in Eqs.~(\ref{L_expect}) and (\ref{Ex_omg|h})
reads as
\begin{align*}
&\overline{
P(\sigma,t)\partial_t Q(\mu,t\mid\sigma,t)
-
Q(\mu,t) \partial_t P(\sigma,t\mid\mu,t)
}
\\
=&
\overline{
-
  P(\sigma,\mu,t) \partial_t \ln P(\sigma,t)
+
  P(\sigma,\mu,t) \partial_t \ln Q(\mu,t)
}
\\
=&
\overline{
  \ln \frac{P(\sigma,t)}{Q(\mu,t)}
\partial_t  P(\sigma,\mu,t)
}
\end{align*}
from Eq.~(\ref{COND_PQ}) and the partial integration.
Substituting this into Eqs.~(\ref{L_expect}) and (\ref{Ex_omg|h}),
we obtain
\begin{align}
 L^{(h)}
&=
g_L'
\sum_{\sigma,\mu}
(\boldsymbol{x}_{\sigma}\times \boldsymbol{x}^{\mu})_{z}
\left(
\delta_{\sigma,-\mu}
-\delta_{\sigma,\mu}
\right)
\overline{
J^{\mu}(t)
\ln \frac{P(\sigma,t)}{Q(\mu,t)}
}
\nonumber
\\
&\approx
-
g_L'
\sum_{\mu}
(\boldsymbol{x}_{\mu}\times \boldsymbol{x}^{\mu})_{z}
\overline{
\left[
\ln \frac{P(\mu,t)}{Q(\mu,t)}
+
\ln \frac{P(-\mu,t)}{Q(\mu,t)}
\right]
J^{\mu}(t)
}
,
\label{Lt}
\\
\omega^{(h)}
&\approx
-\pi g_{O}'
\sum_{\mu}
\overline{
\left[
 \ln\frac{P(\mu,t)}{Q(\mu,t)}
+
 \ln\frac{P(-\mu,t)}{Q(\mu,t)}
\right]
J^{\mu}(t)
}
,
\label{App:PI_h}
\end{align}
where we assume that $J_{\sigma}^{\mu}(t)$ is of higher order in $h$
than $J^{\mu}(t)$ [$\sim O(h)$] in Eqs.~(\ref{DP}) and (\ref{def:J'}).

The mean power consumption $P_h$ in Eq.~(\ref{def:P_h}) can be written as
$ 
\overline{h\boldsymbol{N}_t
\cdot \langle\Dot{\boldsymbol{X}}\rangle}$.
Then, $\langle\Dot{\boldsymbol{X}}\rangle$ is estimated
by applying the first term in Eq.~(\ref{PhysObs}) as
\begin{equation}
\langle\dot{\boldsymbol{X}}\rangle \approx
g_{V} \sum_{\mu} (\boldsymbol{x}_{-\mu}-\boldsymbol{x}_{\mu}) 
J^{\mu}(t) 
\label{X_dot:app}
\end{equation}
with an adjustable parameter
$g_V$ neglecting the higher-order terms other than $O(h)$, and
we obtain
\begin{equation}
 P_{h} =
-2g_V h\sum_{\mu}
\overline{
J^{\mu}(t)
\boldsymbol{N}_t} \cdot
\boldsymbol{x}_{\mu}
.
\label{P_h0} 
\end{equation}
Calculations for
$L^{(I)}$, $L^{(h)}$, 
$\omega^{(I)}$, $\omega^{(h)}$, and $P_h$ are shown in Appendix~\ref{LRT}.

\subsection{\label{Expressions}Expressions for $L$, $\omega$ and $P_h$}

From the details given in
Appendix~\ref{App:MAM} [Eqs.~(\ref{LI:1})--(\ref{App:omg})], we obtain
\begin{gather}
L
\approx 
\frac{ g_{L} W_{0}}{2D}
\left(
\boldsymbol{x}_{+}\times\boldsymbol{x}^{+}
\right)_{z}\,
\{I_0(D)-I\}
,
\label{Lt_final}
\\
 \omega
\approx
\frac{\pi g_{O} W_0 }{2D}
\left\{
I_{0}(D)
-I
\right\}
,
\label{omegaEx}  
\end{gather}
where
\begin{align}
W_0
&\equiv
\frac{1}{2\pi}
e^{-[
V_{0}(\boldsymbol{x}^{+}) -V_{0}(\boldsymbol{x}_{+})]/D
}
\sqrt{
\frac{H_{\tau}H_{n}\lvert G_n\rvert}{G_{\tau}}}
,
\label{def:W0}
\\
&
\approx
\frac{K}{2\pi}\frac{
\lvert\partial_{\boldsymbol{x}}v_{0}(\boldsymbol{x}_{+})\rvert}{
\lvert\partial_{\boldsymbol{x}}v_{0}(\boldsymbol{x}^{+})\rvert
}
\sqrt{ g(\boldsymbol{x}_{+})\lvert g(\boldsymbol{x}^{+})\rvert }
e^{-\Delta V/D}
\quad (m \gg 1)
,
\label{W_sig_mug}
\\
I_{0}(D) 
&\equiv
-
\frac{8 g_{L}' h^2 \Omega}{
g_{L}\sqrt{2\pi D H_{n}}
}
\frac{
\boldsymbol{x}_{+}\cdot \boldsymbol{n}_{+}
}{\Omega + 4 W_0},
\label{TorqBalance} 
\\
&\approx
-
\frac{8 g_{L}' h^2 \Omega}{
g_{L}\sqrt{2\pi K D  g(\boldsymbol{x}_{+})}
}
\frac{
\boldsymbol{x}_{+}\cdot \boldsymbol{n}_{v}(\boldsymbol{x}_{+})
}{\Omega + 4 W_0}
\quad (m \gg 1)
.
\label{I0D}
\end{align}
Here, $H_{\tau}\equiv\Lambda_{\tau}(\boldsymbol{x}_{\sigma})$,
$H_{n}\equiv\Lambda_{n}(\boldsymbol{x}_{\sigma})$,
$G_{n}\equiv\Lambda_{n}(\boldsymbol{x}^{\mu})$, and
$G_{\tau}\equiv\Lambda_{\tau}(\boldsymbol{x}^{\mu})$
from Eqs.~(\ref{eigen_tau}) and (\ref{eigen_n});
$g_{L}$, $g_{L}'$, and $g_{O}$ are adjustable parameters of $O(1)$.
Equations (\ref{W_sig_mug}) and (\ref{I0D})
are obtained from Eqs.~(\ref{DV}) and (\ref{eigenvals}).

Equations (\ref{Lt_final}) and (\ref{omegaEx}) suggests that
the stimuli of the RDDF can support positive work and torque for the load
as long as $I < I_{0}(D)$ ($\gamma L$ is regarded as a viscous torque).
Thus, the quantity $\displaystyle\max_{D} {I_{0}(D)}$ indicates the maximal load for
such productive work;
it quantifies the maximal performance of the ratchet.
From Eq.~(\ref{TorqBalance}), it is found that
a higher value of $\displaystyle\max_{D} I_0(D)$ is gained if the value of
$-\boldsymbol{x}_{+}\cdot \boldsymbol{n}_{+}$ is increased.
As shown in Fig.~\ref{fig:show_pot},
the factor $-\boldsymbol{x}_{+}\cdot \boldsymbol{n}_{+}$ characterizes the
asymmetry in the ratchet shape. 
Additionally, one may anticipate another way of increasing $\displaystyle\max_{D} I_0(D)$, namely by decreasing $H_n$.
However, we note that Eq.~(\ref{TorqBalance})
is not always valid for small $H_n$ either because it eventually conflicts with
the prerequisite $\Omega T_{r}\ll 1$ for small $H_n$
or, because of the time-dependent fields, the potential
with small $H_n$ possibly yields temporal minima
other than $\{\boldsymbol{x}_{\sigma}\}$.
Namely, as $H_n$ becomes vanishingly small, the influence
of the time-dependent fields becomes relatively strong, possibly
breaking the local equilibrium condition on which our theory 
crucially depends (see Appendix~\ref{master}).
So, the effect of decreasing $H_n$ may be limited.

From the results in Appendix~\ref{App:Power}, we also obtain $P_h$ as
\begin{equation}
P_{h}
\approx
\frac{2 g_V h^2\left|\boldsymbol{x}_{+}\right|^2 }{D}
\frac{\Omega W_{0}}{\Omega+4W_{0}}
,
\label{Ph_final}
\end{equation}
where $g_V$ is an adjustable parameter.

\section{\label{Optimization} Optimization of ratchet potential}

\subsection{Optimization problem}
\label{sec:OPT0}

We now consider the problem of maximizing
$\omega$ and $L$ through $I_0(D)$ 
by optimizing $V_0(\boldsymbol{x})$
[see Eqs.~(\ref{Lt_final})--(\ref{TorqBalance})].
This also has the appreciable effect of increasing $\eta$
through the numerator $L\omega$ in Eq.~(\ref{def:eta1}), whereas
the optimization of $V_0(\boldsymbol{x})$ does not crucially affect the denominator $P_h$
according to Eq.~(\ref{Ph_final}).

As mentioned in Sec.~\ref{Quantities}, from Eq.~(\ref{TorqBalance}),
we can carry out the maximization of $I_0(D)$
by designing $V_0(\boldsymbol{x})$ so as to maximize the factor
$-\boldsymbol{x}_{+}\cdot \boldsymbol{n}_{+}(\boldsymbol{x}_{+})$,
which can be replaced with the approximation
$-\boldsymbol{x}_{+}\cdot \boldsymbol{n}_{v}(\boldsymbol{x}_{+})$
for $m\gg 1$ from Eq.~(\ref{I0D}).
In addition to this, we may minimize $H_{n}$
[which corresponds to $g(\boldsymbol{x}_{+})$ in Eq.~(\ref{I0D})]
within a valid range for the local equilibrium condition around
the potential minima.
Hereinafter, we assume $m\gg 1$ even in cases in which the essential 2D ratchet characteristics are retained.
We then treat $-\boldsymbol{x}_{+}\cdot \boldsymbol{n}_{v}(\boldsymbol{x}_{+})$
as the main objective function to maximize and,
if necessary, treat $g(\boldsymbol{x}_{+})$ as an optional objective function 
to minimize within some limited range.

Thus, a goal of the optimization is to optimize
$v_0(\boldsymbol{x})$ or $v_1(\boldsymbol{x})$ to maximize
$-\boldsymbol{x}_{+}\cdot \boldsymbol{n}_{v}(\boldsymbol{x}_{+})$.
As shown in Sec.~\ref{TT-Model}, functions
$v_0(\boldsymbol{x})$ and $v_1(\boldsymbol{x})$ 
set up the shape of the potential valley and
the local minima and saddles in it.
Taking these into account, we first optimize
$v_1(\boldsymbol{x})$ because it immediately affects 
$-\boldsymbol{x}_{+}\cdot \boldsymbol{n}_{v}(\boldsymbol{x}_{+})$
through $\boldsymbol{x}_{+}$.
Here, let $p$ be a parameter in $v_1(\boldsymbol{x})$, and
rewrite it as
$v_1(\boldsymbol{x})\equiv v_1(\boldsymbol{x};p)$
to express its dependence on $p$;
$\boldsymbol{x}_{+}$ also depends on $p$.
In Eq.~(\ref{v1:core}), $p$ corresponds to $\alpha$ or $d$.
Then, our problem is to find an optimized value of $p$ ($\equiv p_{\ast}$), i.e.,
\begin{equation}
p_{\ast} \equiv \argmax_{p}
\left\{
-\boldsymbol{x}_{+}\cdot \boldsymbol{n}_{v}(\boldsymbol{x}_{+})
\right\},
\label{opt_prob}
\end{equation}
where $\boldsymbol{x}_{+}$ ($ \in \mathrm{C}_{\infty}$)
 is subject to
$E_{+} =  v_1(\boldsymbol{x}_{+};p)$ and
\begin{equation}
E^{+} \leq
v_1(\boldsymbol{x};p)
\leq E_{+},
\quad \forall\boldsymbol{x}\in \mathrm{C}_{\infty},
\label{G0}
\end{equation}
with $E^{+} =  v_1(\boldsymbol{x}^{+};p)$
($\boldsymbol{x}^{+} \in \mathrm{C}_{\infty}$).

Because this expression is rather complicated for compact wording,
an alternative for practical computation is as follows.
Here, let us consider $v_1(\boldsymbol{x})$ with the specific form
$
v_1(\boldsymbol{x}) \equiv \boldsymbol{x}^{\mathrm T}
\Hat{O}_{\alpha} \Hat{E}_d \Hat{O}_{\alpha}^{\mathrm T}
\boldsymbol{x}$, 
where
\begin{align}
\Hat{O}_{\alpha}
\equiv
\begin{pmatrix}
 \cos\alpha&  -\sin\alpha
 \\
  \sin\alpha&  \cos\alpha
\end{pmatrix}
,
\quad
\Hat{E}_d 
\equiv
\begin{pmatrix}
 d^2&  0
 \\
0& 1
\end{pmatrix}
.
\label{Def:OE}
\end{align}

In the actual procedure, with $\boldsymbol{x}_{+}$ determined in
\begin{equation}
 \mathrm{G}_1:\
\boldsymbol{x}_{+} = \argmax_{\boldsymbol{x} \in \mathrm{C}_{\infty}} 
\left\{
-\boldsymbol{x}\cdot \boldsymbol{n}_{v}(\boldsymbol{x})
\right\},
\label{G1}
\end{equation}
we fix $(\alpha,d)$ through Eq.~(\ref{3th:force_2}) or
\begin{equation}
\mathrm{G}_2:\
 \boldsymbol{n}_{v}^{\mathrm T}(\boldsymbol{x}_{+})
\Hat{O}_{\alpha} \Hat{E}_d \Hat{O}_{\alpha}^{\mathrm T}
\boldsymbol{x}_{+}
=
0
.
\label{force_2b}
\end{equation}
Hereinafter, $\alpha$ and $d$ range as $0 \leq \alpha < \frac{\pi}{2}$ and $d>1$,
which makes the ratchet direction counterclockwise (see Fig.~\ref{fig:show_pot}).
Note that Eq.~(\ref{G0}) is unchanged
under $d\rightarrow \frac{1}{d}$, $\alpha\rightarrow \frac{\pi}{2}+\alpha$,
$E_{+} \rightarrow  \frac{E_{+}}{d^2}$, and $E_{-} \rightarrow  \frac{E_{-}}{d^2}$.
So far, either $\alpha$ or $d$ is a free parameter, but not both.
For example,
using the replacement $d\equiv\tan \delta$ and the matrix $\Hat{A}_{\alpha}$
defined as
\begin{align}
\Hat{O}_{\alpha} \Hat{E}_d \Hat{O}_{\alpha}^{\mathrm T}
 =
 \frac{1+d^2}{2}\Hat{1}-\frac{1-d^2}{2}
\Hat{A}_{\alpha}
,
\quad \Hat{A}_{\alpha}\equiv
\begin{pmatrix}
 \cos 2\alpha&  \sin 2\alpha
 \\
 \sin 2\alpha&  -\cos 2\alpha
\end{pmatrix}
,
\label{Def:ODO}
 \end{align}
Eq.~(\ref{force_2b}) is read as
\begin{equation}
\mathrm{G}_2':\
 \cos 2\delta = \frac{
  - \boldsymbol{n}_{v}(\boldsymbol{x}_{+})\cdot\boldsymbol{x}_{+}}
  {
-  \boldsymbol{n}_{v}(\boldsymbol{x}_{+})^{\mathrm T}
\Hat{A}_{\alpha}
\boldsymbol{x}_{+}
}
\quad
\left(\frac{\pi}{4} <\delta < \frac{\pi}{2}\right).
\label{delta_alpha}
\end{equation}
This is useful when one chooses $\alpha$ as the free parameter, and
determines $\delta$ (also $d$) with $\alpha$.
If $d$ is given instead, $\alpha$ is determined by solving
Eq.~(\ref{force_2b}).

After determining $\boldsymbol{x}_{+}$ and $(\alpha,d)$, if the right inequality in
Eq.~(\ref{G0}) is satisfied for
$E_{+}=\boldsymbol{x}_{+}^{\mathrm T}
\Hat{O}_{\alpha} \Hat{E}_d \Hat{O}_{\alpha}^{\mathrm T}
\boldsymbol{x}_{+}
$, we settle the (elliptic) curve $\mathrm{E}_{+}$
with these values.
Otherwise, if the inequality is unsatisfied, 
we may search for other values
of $\boldsymbol{x}_{+}$ and $(\alpha,d)$, which may be found at the second
extreme point
$\boldsymbol{x} \in \mathrm{C}_{\infty}$ of
$-\boldsymbol{x}\cdot \boldsymbol{n}_{v}(\boldsymbol{x})$, or
may refine $v_0(\boldsymbol{x})$.
This procedure is finalized by finding
$\boldsymbol{x}^{+}$ ($\in \mathrm{C}_{\infty}$), which satisfies
$
\boldsymbol{n}_{v}^{\mathrm T}(\boldsymbol{x}^{+})
\Hat{O}_{\alpha} \Hat{E}_d \Hat{O}_{\alpha}^{\mathrm T}
\boldsymbol{x}^{+}
=
0
$ and the left inequality in
Eq.~(\ref{G0}) for 
$E^{+}=\boldsymbol{x}^{+\mathrm T}
\Hat{O}_{\alpha} \Hat{E}_d \Hat{O}_{\alpha}^{\mathrm T}
\boldsymbol{x}^{+}$.
The curve $\mathrm{E}^{+}$ is also settled with $\boldsymbol{x}^{+}$ and $E^{+}$.

\subsubsection{\label{sec:LamZero} Elliptic case ($\lambda=0$)}

We show analytical results for 
$L$, $P_h$, and $\eta$ maximized by optimizing
$v_1(\boldsymbol{x})$, through
the parameters $\alpha$ and $d$,
with
$\mathrm{G}_1$ [Eq.~(\ref{G1})] and $\mathrm{G}_2$ [Eq.~(\ref{force_2b})]
for the elliptic $\mathrm{C}_{\infty}$ ($\lambda=0$) and $m\gg 1$.
The maximized expressions
for those in Eqs.~(\ref{Lt_final}), (\ref{W_sig_mug}), (\ref{I0D}),
(\ref{Ph_final}), and (\ref{def:eta0}) are obtained as
\begin{gather}
L
\approx 
\frac{ g_{L} ab W_{0}}{2D}
\{I_0(D)-I\}
\label{lamda0:L}
,
\\
P_{h}
\approx
\frac{2 g_V h^2(a^2 + b^2 - ab)}{D}
 \frac{\Omega W_{0}}{\Omega+4W_{0}}
,
\label{lamda0:Ph}
\\
\eta 
\approx
 \frac{
\added[]{2}
\gamma g_Og_L^{'2}
h^2 \{ab(a-b)\}^2
}{g_L g_V  (a^2 + b^2 - ab) \Delta V D^2} 
 \frac{\Omega W_{0}}{\Omega+4W_{0}}
,
\label{lamda0:eta}
\end{gather}
where, for $a>b>0$,
\begin{gather}
W_0
\approx
\frac{\deleted[]{\cancel{2}}\Delta V}{\pi (a^2 + b^2 - ab)}
e^{-\Delta V/D}
,
\\
I_{0}(D) 
\approx
\frac{4 g_{L}' h^2 \Omega}{
g_{L}\sqrt{\deleted[]{\cancel{2}}\pi \Delta V D}
}
\frac{
\sqrt{ab}(a-b)
}{\Omega + 4 W_0}
,
\label{lamda0:I0}
\\
 \Delta V=
\frac{K}{2}\sqrt{ab}(a+b)(d^2-1)\sin(2\alpha)
.
\label{Zero:DE}
\end{gather}
The details of the above process are given in Appendix~\ref{App:LamZero}.
From Eq.~(\ref{omegaEx}), $\omega$ is proportional to $L$.
Corresponding to Eq.~(\ref{force_2b}) or (\ref{delta_alpha}),
$\alpha$ and $d$ ($>1$) are related as
\begin{equation}
\frac{d^2+1}{d^2-1}=
\frac{ \sqrt{ab}}{a+b}
  \sin  2  \alpha
-
\frac{{a}^{2}+{b}^{2}}{a^2-b^2}
  \cos 2 \alpha
.
\label{ellipse:cond2}  
\end{equation}

In the elliptic case, according to Eq.~(\ref{ellipse:cond2}),
we can choose any value for $\alpha$ 
unless
the prerequisite $\Delta V \gg D$ in
the approximation (see Appendix A)
is violated. Furthermore, we do not need to minimize $g(\boldsymbol{x}_{+})$
(or to optimize $v_1(\boldsymbol{x})$ through $\alpha$). Note that,
in the particular case of $\alpha\rightarrow \frac{\pi}{2}$ (or $0$),
Eq.~(\ref{Zero:DE}) leads to $\Delta V \rightarrow 0$, 
and $\Delta V \gg D$  is violated,
where
$\mathrm{E}_{+}$ and $\mathrm{E}^{+}$ coincide with
$\mathrm{C}_{\infty}$ [$d \rightarrow \frac{a}{b}$ (or $\frac{b}{a}$)].

\subsection{\label{sec:LamNonZero} Nonelliptic case ($\lambda\ne 0$)}

Here, as a second optimization, we consider a strategy for minimizing $E_{+}$.
In the case of $\lambda\ne 0$,
the curve $\mathrm{E}_{+}$ never coincides with $\mathrm{C}_{\infty}$
for any $(\alpha,d)$.
When minimizing $E_{+}$ with respect to $(\alpha,d)$, $E_{+} > E^{+}$ is retained, and 
both $\alpha$ and $d$ acquire definitive values.
At the minimized $E_{+}$, the two curves $\mathrm{E}_{+}$ and
$\mathrm{E}^{+}$ tightly enclose $\mathrm{C}_{\infty}$.
This suggests that minimizing $E_{+}$ causes
$g(\boldsymbol{x}_{+})$ (corresponding to  $H_{n}$) to decrease.

In the case of $\lambda \ne 0$, 
in addition to the procedure $\mathrm{G}_1$
in Eq.~(\ref{G1}), firstly, we impose
\begin{equation}
\mathrm{G}_3:\quad
(\alpha_{\ast}, d_{\ast}) = \argmin_{ 0 \leq \alpha < \frac{\pi}{2}, d(\alpha)>1} E_{+},
\label{CondG3}
\end{equation}
where $d(\alpha)$ denotes $d$ as a function of $\alpha$ defined in
Eq.~(\ref{delta_alpha}) [or Eq.~(\ref{force_2b})]; thus,
the essential number of optimization parameters is one.
Specifically, after determining
$\boldsymbol{x}_{+}$
via $\mathrm{G}_1$,
from the set of the pairs $(\alpha, d)$ satisfied in $\mathrm{G}_2$,
$\mathrm{G}_3$ selects $\alpha_{\ast}$ and $d_{\ast}$
such that they minimize $E_{+}$
(this automates the tuning of parameters).
As mentioned above, the procedure $\mathrm{G}_3$
flattens the potential profile along the valley, and narrows
the intersection of the valley.
It is then expected that the fluctuation of the rotor trajectory 
may be suppressed within the valley.
This accords with our intention to improve the rotational efficiency.

Here, we should note that 
$\boldsymbol{x}_{+}$ in Eq.~(\ref{delta_alpha}) has been obtained
in the limit $m\rightarrow \infty$ and in the absence
of the external fields ($h= 0$ and $I= 0$). However,
the actual minimum point deviates from $\boldsymbol{x}_{+}$;
if determining $\boldsymbol{x}_{+}$ with
$
\partial_{ \boldsymbol{x}} V_{0}(\boldsymbol{x})
=h\boldsymbol{N}_{t}+\boldsymbol{f}_{I}(\boldsymbol{x})
$, Eqs.~(\ref{force_1}) and (\ref{force_2}) are modified.
In particular, in the case of $h\ne 0$, $I=0$, and $m \rightarrow \infty$,
Eq.~(\ref{3th:force_2}) is modified as
$
 \boldsymbol{n}_{v}
\cdot
\partial_{ \boldsymbol{x}}
v_1'(\boldsymbol{x},\Phi_t)
=
0
$
($\boldsymbol{x}\in \mathrm{C}_{\infty}$) with
\begin{gather}
v_1'(\boldsymbol{x},\Phi_t)
\equiv
v_1(\boldsymbol{x})
-
\frac{h}{K} \boldsymbol{x}\cdot \boldsymbol{N}_{t}
\label{v1d}
\end{gather}
for minima and saddles.
In this case, a curve 
$\mathrm{E}(\Phi_t):\{\boldsymbol{x}\mid v_1'(\boldsymbol{x},\Phi_t)=E\}$
is the same ellipse as $\mathrm{E}$ except that
the center of $\mathrm{E}(\Phi_t)$ moves around the origin.
Because of this movement,
the minimum point, at which $\mathrm{E}(\Phi_t)$ is circumscribed to $\mathrm{C}_{\infty}$,
also moves along $\mathrm{C}_{\infty}$.
There is a single circumscribed point corresponding to the global minimum
 and a single inscribed point
corresponding to a saddle, which we
denote by  $\boldsymbol{x}_{\ast t}$ and
$\boldsymbol{x}_t^{\ast}$, respectively.
Similarly, corresponding to Eq.~(\ref{G0}), such a minimum and a saddle satisfy
$
E^{\ast}(\Phi_t) \leq
v_1'(\boldsymbol{x},\Phi_t)
\leq E_{\ast}(\Phi_t)
$
for $\forall\boldsymbol{x}\in \mathrm{C}_{\infty}$
with
$E^{\ast}(\Phi_t) \equiv  v_1'(\boldsymbol{x}_t^{\ast},\Phi_t)$ and
$E_{\ast}(\Phi_t) \equiv  v_1'(\boldsymbol{x}_{\ast t},\Phi_t)$.

As the circumscribed ellipse
$\mathrm{E}_{\ast}(\Phi_t):\{\boldsymbol{x}\mid v_1'(\boldsymbol{x},\Phi_t)=E_{\ast}(\Phi_t)\}$
varies with the external field,
$\boldsymbol{x}_{\ast t}$ ($\boldsymbol{x}_t^{\ast}$) is not always
close to either $\boldsymbol{x}_{+}$ or $\boldsymbol{x}_{-}$
($\boldsymbol{x}^{+}$ or $\boldsymbol{x}^{-}$). Rather, it may
sometimes jump to another point on $\mathrm{C}_{\infty}$ away from them,
which creates a temporal minimum.
The occurrence of such events
depends on the parameters $(\alpha, d)$ or the shape of $\mathrm{C}_{\infty}$.
In the experimental observation shown in Sec.~\ref{lambda_nonzero},
the temporal minimum 
is likely to arise when $\mathrm{C}_{\infty}$ (of larger $\lambda$) 
is tightly enclosed by $\mathrm{E}_{+}$ and $\mathrm{E}^{+}$,
as a result of optimizing $(\alpha, d)$ in
$\mathrm{G}_3$.
It is also expected that the temporal minimum may become an obstacle
in the conversion of power to net rotational output, and
may have a negative influence on the efficiency.
Therefore, we moderate $\mathrm{G}_3$
by adding a relaxation such
that the gap between
$\mathrm{E}^{+}$ and $\mathrm{E}_{+}$ becomes wider to a sufficient extent.
Since 
$d$ is minimized to $d_{\ast}$ in $\mathrm{G}_3$,
then, to relax it, we replace $d$ with
\begin{equation}
 d=d_{\ast}+\epsilon\quad (\epsilon > 0),
\label{Det:d}
\end{equation}
where $\epsilon$ is a relaxation parameter.
Again applying this $d$ to
$\mathrm{G}_2$ [Eq.~(\ref{force_2b})],
we obtain a revised $\alpha$. Now, 
with the ratchet potential of this $(\alpha,d)$,
we can expect that the contact point between
the ellipse $\mathrm{E}_{+}(\Phi_t)$ and $\mathrm{C}_{\infty}$ is always
close to either $\boldsymbol{x}_{+}$ or $\boldsymbol{x}_{-}$,
and that the local equilibrium can be retained.

\section{\label{NumericalResults} Numerical results}

\begin{table}[tbh]
\centering
 \begin{tabular}{c|ccccl}
Label  &\multicolumn{2}{c}{Key param.\,vals.} & Fig.~\ref{fig:pot_zero_lamda}
& \multicolumn{2}{c}{$\Delta V$, $K$, and/or $d$.}
\\
\hline
\multirow{3}{*}{A1} &\multirow{3}{*}{$\alpha$} & $0.03\pi$  & (a)
& \multirow{3}{*}{$(K,\Delta V, d)$} & $(0.0284, 0.1500, d_{0.4\pi})$\\
& & $0.4\pi$ &  (b) & &$(0.102, 0.1498, d_{0.4\pi})$ \\ 
& &  $0.48\pi$ &  (c) & & $(0.193, 0.1501, 2.1)$ \\
\hline
\multirow{3}{*}{A2} & \multirow{3}{*}{$(a,b)$} & $(1.8,1)$& (b)&\multirow{3}{*}{$(K,\Delta V)$} & $(0.102, 0.1498)$ \\
& & $(2.7, 1.5)$ & (d)& & $(0.0454, 0.1500)$\\ 
& & $(3.6, 2)$ & (e) &  & $(0.0255, 0.1500)$\\
\hline
\multirow{3}{*}{A3}&\multirow{3}{*}{$m$} &$1$ & (f) &\multirow{3}{*}{$\Delta V$} & $0.1896$  \\ 
& & $2$& (b) & & $0.1498$  \\ 
& & $3$& (g) & & $0.1436$ \\
\hline
\multirow{3}{*}{A4}  & \multirow{3}{*}{$a$} & $1.2$ & (h) & \multirow{3}{*}{$(K, \Delta V)$} & $(0.672, 0.1501)$ \\ 
 & & $1.8$&  (b) & &$(0.102, 0.1498)$ \\ 
& & $2.4$& (i) & & $(0.038, 0.1495)$\\
\hline
 \end{tabular}
\caption{
List of parameter families in the elliptic case ($\lambda=0$).
The families are labeled as in the first column, and
their key parameters are listed in the second and third columns.
The common parameters in each family are as follows:
A1: $(m,a,b) = (2, 1.8, 1)$,
A2: $(m,d,\alpha) = (2, d_{0.4\pi}, 0.4\pi)$,
A3: $(a,b,K,d,\alpha) = (1.8, 1.0, 0.102,  d_{0.4\pi}, 0.4\pi)$,
A4: $(m,b,\alpha) = (2,1,0.4\pi)$.
In A4, $d$ is determined by Eq.~(\ref{ellipse:cond2}) for
each $(a,b,\alpha)$.
$\Delta V \approx 0.15$ is maintained by modifying $K$ (fifth and sixth columns) except for A3.
$d_{0.4\pi}\approx 1.860118$.
}
\label{TAB:A}
\end{table}

We show the numerical results of $L$ (MAM) in Eq.~(\ref{MAM}),
$\omega =
\left\langle\frac{\Theta_{T_{\mathrm{tot}}}}{T_{\mathrm{tot}}}\right\rangle_{\Phi}$ (MAV)
with $\Theta_t$ in Eq.~(\ref{def:theta}),
$P_h$ in Eq.~(\ref{def:P_h}), and
$\eta$ in Eq.~(\ref{def:eta0})
for several parameter families of $V_{0}(\boldsymbol{x})$.
We also discuss the utility of the optimization strategy described in 
Secs.~\ref{sec:OPT0} and \ref{sec:LamNonZero}.
The numerical simulation of Eq.~(\ref{LEQ}) was carried out using
the second-order stochastic Runge--Kutta method
with a time increment of $0.005$ ($m=1,2$) or $0.002$ ($m=3$).
The long time average, $\overline{A(\boldsymbol{X},\Phi_t)}$, was obtained by
averaging $128$ independent trials of the time series of
$T_{\mathrm{tot}}\Omega = 2^{17}$.
Throughout this paper, the parameters of $V_h(\boldsymbol{x}, t)$
in Eq.~(\ref{LEQ:Poth}) are set to $h=0.01$ and $\Omega=0.001$;
no load is applied ($I=0$);
the fitting parameters in Eqs.~(\ref{Lt_final}), (\ref{omegaEx}),
(\ref{TorqBalance}), and (\ref{Ph_final}) are set to
$g_L=\VALgL$, $g_L'=\VALgLr$, $g_O=\VALgO$, and $g_V=\VALgV$.

\begin{figure}[!tb]
%
%
\def\Size{12.3cm}
\centering
\includegraphics[width=\Size,keepaspectratio,clip]{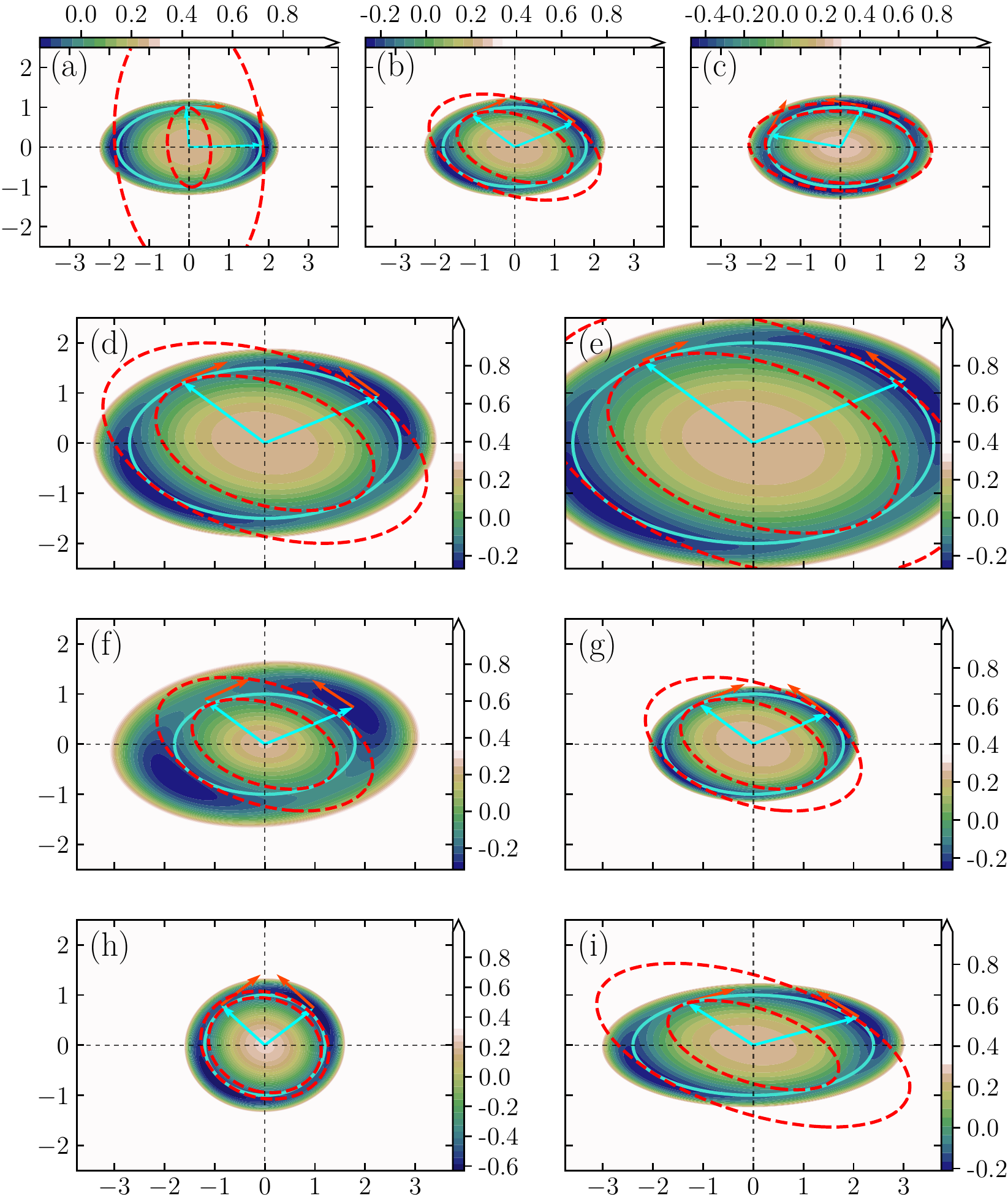}
\caption{
(Color online)
Contour graphs of $V_{0}(\boldsymbol{x})$ 
on $\{\boldsymbol{x}\mid \lvert x\rvert \leq 3.75, \lvert y\rvert\leq 2.5\}$
in
the parameter families A1, A2, A3, and A4 in Table~\ref{TAB:A}.
In (a), (b), and (c), the locations of the local minima differ (A1);
in (b), (d), and (e), the shapes of the elliptic valley have a similarity
with the ratio of diameters as $1:2:3$ (A2);
in (f), (b), and (g), $m = 1$, $2$, and $3$ (A3);
in (h), (b), and (i), the eccentricities differ (A4).
See the fourth column in Table~\ref{TAB:A} for
the correspondences.
The solid and dashed closed curves indicate
$\mathrm{C}_{\infty}$, $\mathrm{E}_{+}$ (ellipse circumscribed to $\mathrm{C}_{\infty}$), and 
$\mathrm{E}^{+}$ (ellipse inscribed to $\mathrm{C}_{\infty}$), respectively.
The arrows starting at the origin and ending
at the minimum and saddle
(near the circumscribed and inscribed points) indicate $\boldsymbol{x}_{+}$ (minimum)
and $\boldsymbol{x}^{+}$ (saddle), respectively.
The arrows tangent to $\mathrm{C}_{\infty}$
at $\boldsymbol{x}_{+}$ and $\boldsymbol{x}^{+}$ indicate
$\boldsymbol{n}_{+}$ and $\boldsymbol{n}^{+}$, respectively.
}
\label{fig:pot_zero_lamda}
\end{figure}

\subsection{\label{LamZero:num} Elliptic case ($\lambda=0$)}

\begin{figure}[hbt]
\def\Size2{7.5cm}
\centering
\includegraphics[height=\Size2,keepaspectratio,clip]{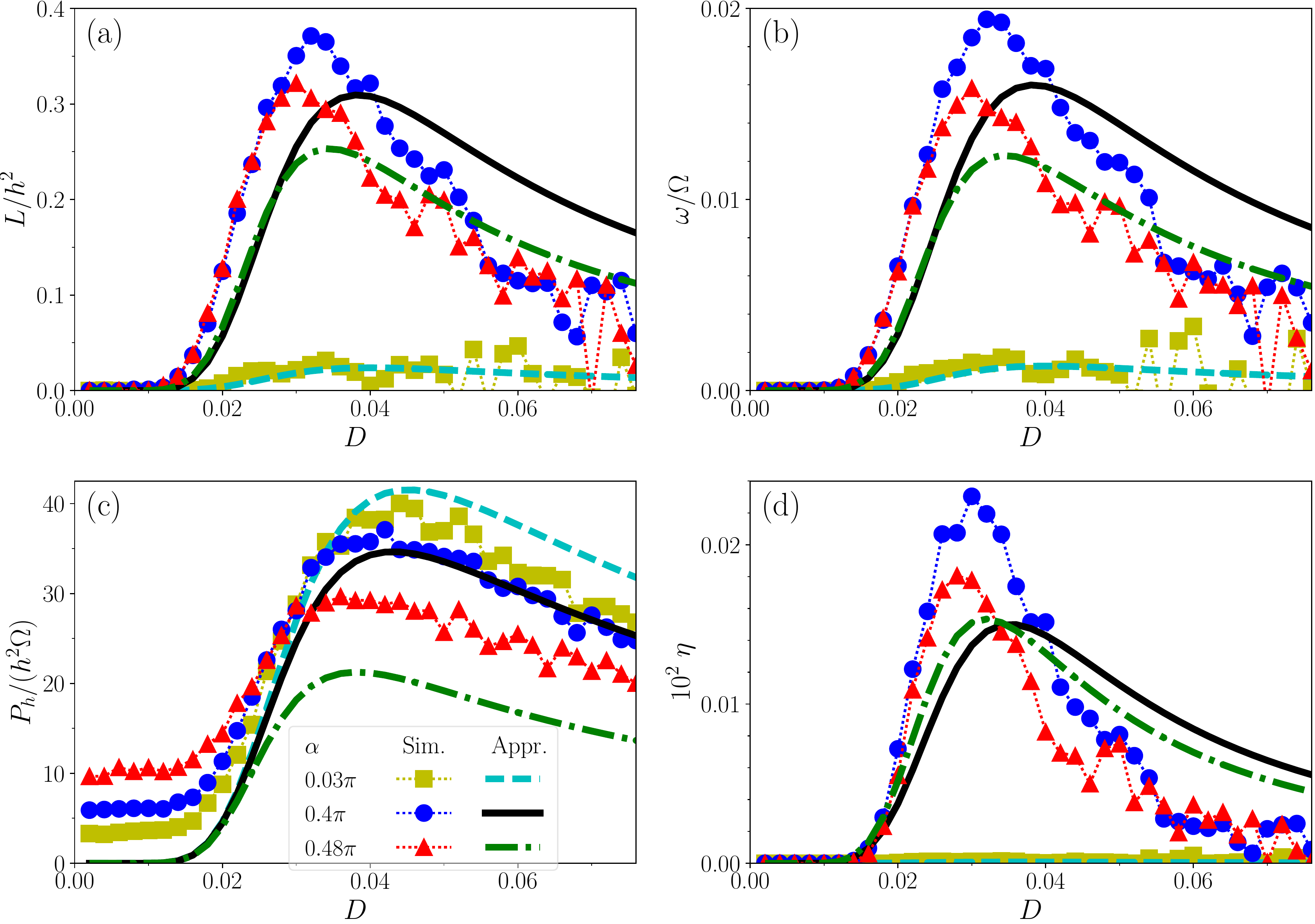}
 \caption{
(Color online)
(a) Scaled mean angular momentum $L/h^2$,
(b) scaled mean angular velocity $\omega/\Omega$,
(c) scaled input power $P_h/(h^2\Omega)$, and (d) efficiency $\eta$
versus noise intensity $D$.
As shown in the legend box in the lower-right panel,
connected symbols ($\Box$, $\bullet$, and $\triangle$) and
(dashed, solid, and dotted) curves represent the numerical (Sim.) and approximation (Appr.)
results under the potentials of parameter family A1, 
which are shown in Fig.~\ref{fig:pot_zero_lamda}(a)--(c).
}
\label{fig:ang}
\end{figure}

We show the outcome of the optimization for
the performance indexes according to the parameter families
A1--A4 in Table~\ref{TAB:A},
and test the results in Eqs.~(\ref{lamda0:L}) and (\ref{lamda0:Ph}).
The contour graphs of $V_{0}(\boldsymbol{x})$ for the parameter sets in Table~\ref{TAB:A}
are displayed in Fig.~\ref{fig:pot_zero_lamda}.

In parameter family A1, it is mainly $\alpha$ that is varied so that
the local minima are positioned near
the $x$ axis ($\alpha = 0.03\pi$) as in Fig.~\ref{fig:pot_zero_lamda}(a),
the optimized position ($\alpha = 0.4\pi$) as in (b),
and near the $y$ axis ($\alpha = 0.48\pi$) as in (c).
In the second case, the factor
$-\boldsymbol{x}_{+}\cdot\boldsymbol{n}_{+}$ in
$I_{0}(D)$ [Eq.~(\ref{TorqBalance})] is maximized with
the optimized position $\boldsymbol{x}_{+}$ in Eq.~(\ref{G1}), and
the parameter $d$ satisfies Eq.~(\ref{ellipse:cond2})
[corresponding to Eq.~(\ref{force_2}) or Eq.~(\ref{force_2b})].
In contrast, in the first and third cases,
$\alpha$ and $d$ do not satisfy Eq.~(\ref{ellipse:cond2}).
As in Fig.~\ref{fig:pot_zero_lamda}(a) and (c),
neither $\mathrm{E}_{+}$ nor $\mathrm{E}^{+}$ are tangent to
$\mathrm{C}_{\infty}$.

Figure~\ref{fig:ang} shows the plots of
$L$, $\omega$, $P_h$, and $\eta$ for $D$ in parameter family A1.
The sets of connected symbols and the (dashed, solid, and dashed--dotted) curves
represent the results of the numerical simulations
(Sim.) and the approximations
(Appr.), i.e.,
Eqs.~(\ref{Lt_final}), (\ref{omegaEx}), (\ref{Ph_final}), and (\ref{def:eta0}),
respectively (see the legend box for the correspondences between the parameters
and the types of symbol or curve).
Each of these curves has a peak with respect to $D$ that can be 
estimated from the relation $\Omega \sim W_{0}$ as the steepest point
of the factor $W_{0}/(\Omega+4W_{0})$ in Eqs.~(\ref{lamda0:L})--(\ref{lamda0:eta}).
Comparing the peaks of $L$ (also $\omega$ and $\eta$) in the series of $\alpha$,
the highest one is found at $\alpha = 0.4\pi$, where $d=d_{0.4\pi}$
[for such comparisons, we attempt to impose consistency on $\Delta V$ 
by modifying $K$ ($\Delta V \approx 0.15$ in Sec.~\ref{LamZero:num})].
This confirms that the optimization for
$v_1(\boldsymbol{x})$ (or $\alpha$ and $d$ in it)
via $\mathrm{G}_1$ [Eq.~(\ref{G1})]
and $\mathrm{G}_2$ [Eq.~(\ref{force_2b}) or $\mathrm{G}_2'$ in Eq.~(\ref{delta_alpha})]
works well.

\begin{figure}[tbh]
\def\Size2{7.5cm}
\centering
\includegraphics[height=\Size2,keepaspectratio,clip]{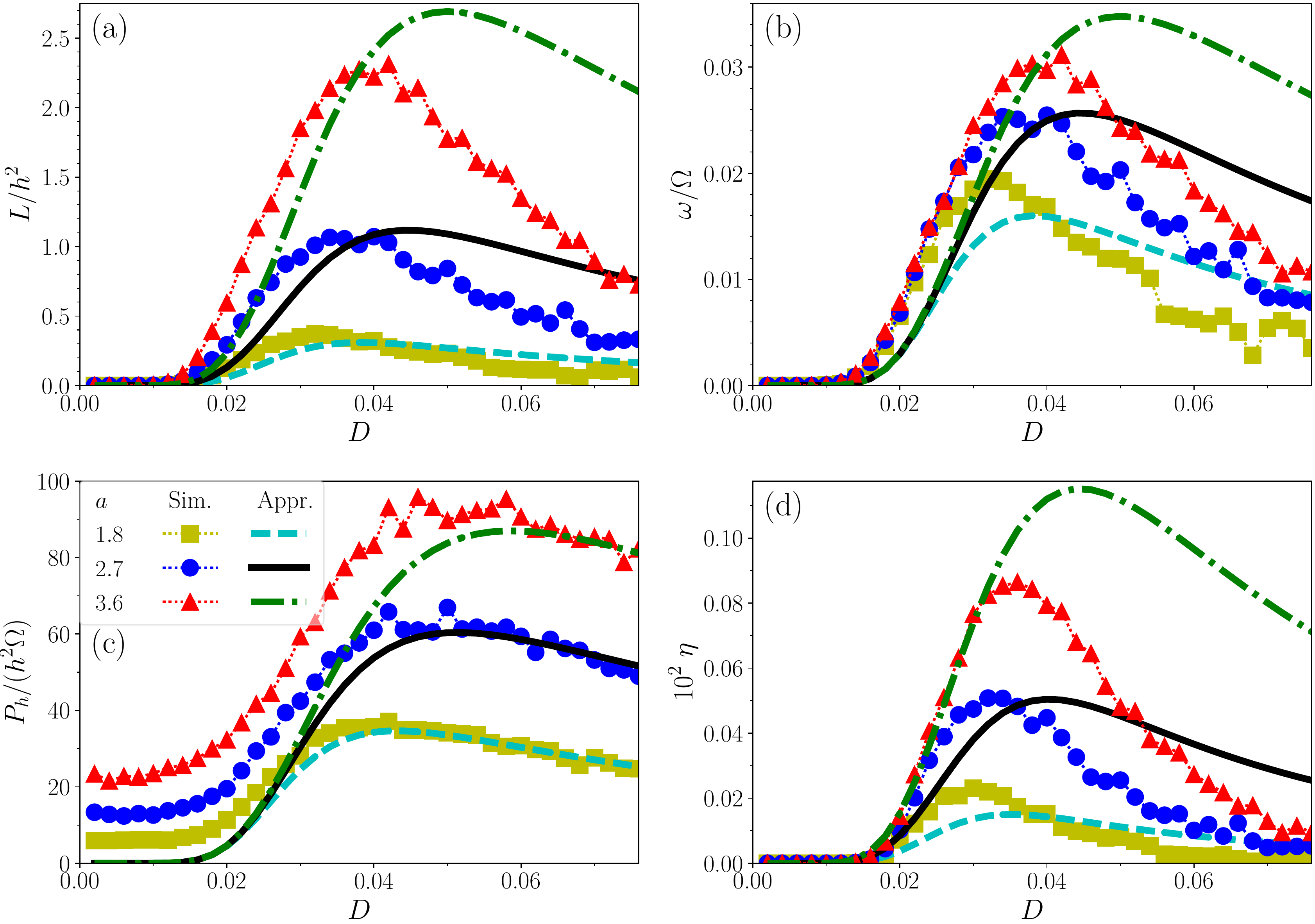}
 \caption{
(Color online)
(a) $L/h^2$, (b) $\omega/\Omega$, (c) $P_h/(h^2\Omega)$, and (d) $\eta$ versus $D$
under the potentials of parameter family A2 
[Fig.~\ref{fig:pot_zero_lamda}(b), (d), and (e)].
}
\label{fig:mag}
\end{figure}

In parameter family A2, the major and minor radii 
of the elliptic pathway of the valley
are varied as $(a,b) = (1.8, 1)$, $(2.7, 1.5)$, and $(3.6, 2)$
while retaining the similarity. Their corresponding potential landscapes
are shown in 
Fig.~\ref{fig:pot_zero_lamda}(b), (d), and (e).
With the common parameters $(m,\alpha,\lambda) = (2,0.4\pi,0)$,
we set $d$ as in Eq.~(\ref{ellipse:cond2}).
Thus, $v_1(\boldsymbol{x})$ is optimized so that 
the factor $-\boldsymbol{x}_{+}\cdot\boldsymbol{n}_{+}$
is maximized.
Figure~\ref{fig:mag} shows that the peaks of $L$, $\omega$, $P_h$, and $\eta$ increase
with the diameter of the elliptic pathway.
These are consistent with Eqs.~(\ref{lamda0:L})--(\ref{Zero:DE}).
Here, it should be noted that
as the diameter of the pathway increases,
the typical magnitude of $V_h(\boldsymbol{x}, t)$ for $\Delta V$ increases.
Then, in order to maintain the local equilibrium condition,
it is necessary to decrease $h$ and $\Omega$ with the diameter.

\begin{figure}[ht]
\def\Size2{7.5cm}
\centering
\includegraphics[height=\Size2,keepaspectratio,clip]{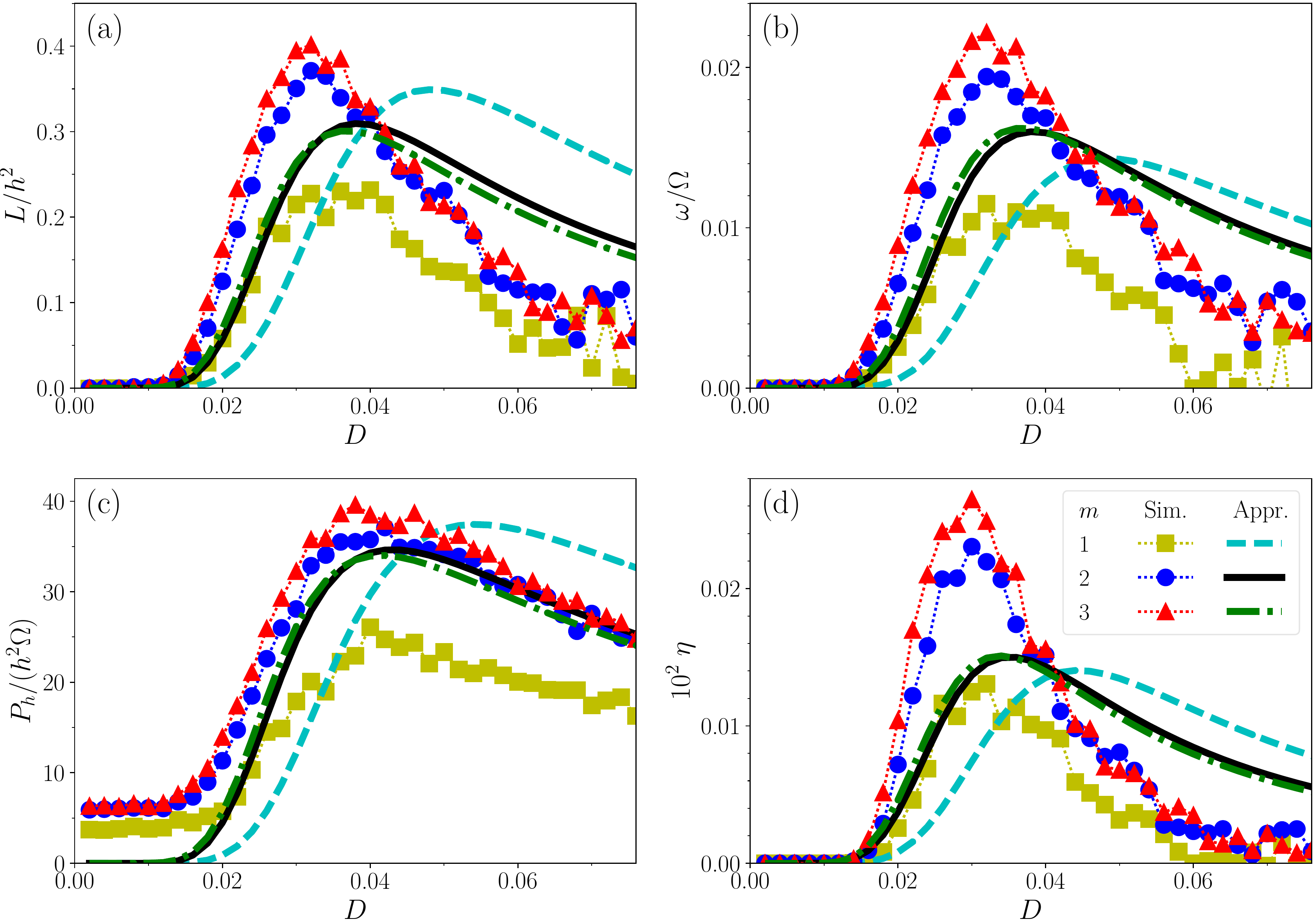}
 \caption{
(Color online)
(a) $L/h^2$, (b) $\omega/\Omega$, (c) $P_h/(h^2\Omega)$, and (d) $\eta$  versus $D$
under the potentials of parameter family
A3 [Fig.~\ref{fig:pot_zero_lamda}(b), (f), and (g)].
}
\label{fig:m}
\end{figure}

In parameter family A3,
only $m$ is increased as $m\in \{1,2,3\}$.
The corresponding potential landscapes are shown in
Fig.~\ref{fig:pot_zero_lamda}(f), (b), and (g), respectively.
In this family,
the intersection of the valley narrows for large $m$, whereas
the diameters of the pathway are nearly equal.
In Fig.~\ref{fig:m}, we can see that
for both numerical and approximation results,
each curve of $L$, $\omega$, $P_h$, and $\eta$ 
is likely to approach a certain curve as $m$ increases.
The approximation result of $m=1$ deviates exceptionally from such an asymptotic
approach. 
For this reason, we consider that
the influence of the external field on the thermal equilibrium condition
is relatively large at $m=1$ because of the
smaller curvature in the intersection of the valley.

\begin{figure}[tbh]
\def\Size2{7.5cm}
\centering
\includegraphics[height=\Size2,keepaspectratio,clip]{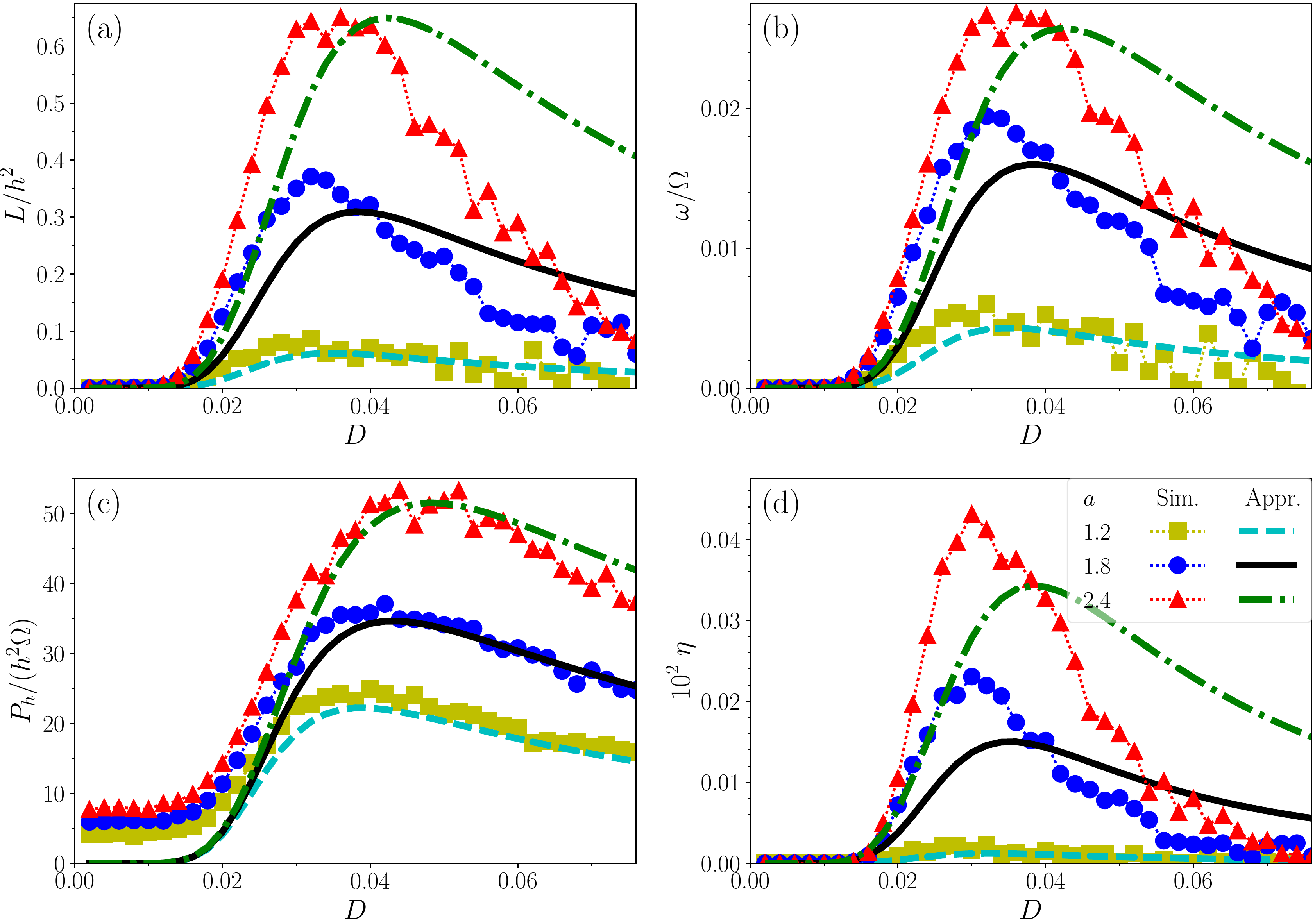}
\caption{
(Color online)
(a) $L/h^2$, (b) $\omega/\Omega$, (c) $P_h/(h^2\Omega)$, and (d) $\eta$
versus $D$ under the potentials of parameter family A4
[Fig.~\ref{fig:pot_zero_lamda}(b), (h), and (i)].
}
\label{fig:elp}
\end{figure}

In parameter family A4, the eccentricity of the elliptic pathway
is increased as $\frac{a}{b} = 1.2$ [Fig.~\ref{fig:pot_zero_lamda}(h)],
$1.8$ [(b)], and $2.4$ [(i)].
Each value of $d$ obeys Eq.~(\ref{ellipse:cond2}), in which case
$-\boldsymbol{x}_{+}\cdot\boldsymbol{n}_{+}$ is maximized.
In Fig.~\ref{fig:elp}, we 
can see that the peaks 
of $L$, $\omega$, $P_h$, and $\eta$ increase with $\frac{a}{b}$.
These are consistent with Eqs.~(\ref{lamda0:L})--(\ref{Zero:DE}).
As mentioned previously, for consistency 
with the local equilibrium condition at larger $\frac{a}{b}$,
it is necessary to keep
$\Omega$ and $h$ sufficiently small.

We make two remarks about the comparison of the 
approximation and simulation results.
Firstly, our approximation has the adjustable parameters
$g_L$, $g_L'$, $g_O$, and $g_V$
for absorbing complexities in the coarse-grained approach,
which we have determined by eye so that the approximations
agree as much as possible with all the simulation results.
Therefore, rather than focusing on the difference in height between the two results for each individual parameter, it is reasonable to compare them
in relation to the similarities among
the plotted curves in a parameter family.
From this respect,
regarding the relationship between the peak heights in Figs.~\ref{fig:ang}--\ref{fig:elp},
the approximation is consistent with the simulation results
except for the case of $m=1$ in Fig.~\ref{fig:m}.
As mentioned above,  if the local equilibrium condition holds well,
our approximation can have such a consistency.
Secondly, it can be
observed that the agreement between the two results
seems better for the lowest curves
in Figs.~\ref{fig:ang} and \ref{fig:elp}.
We consider this to be a visual effect whereby, when observing the upper and lower curves for a couple of parameter sets in a panel in these figures, the difference between the two results for the lower curve is more inconspicuous than that for the upper one.

\subsection{Weakly distorted elliptic case ($\lambda\ne 0$)}
\label{lambda_nonzero}

\begin{table}[bth]
\centering
 \begin{tabular}{c|ccccl}
Label  &\multicolumn{2}{c}{Key param.\,vals.} & Figs.
& \multicolumn{2}{c}{$\alpha$ and/or $d$}
\\
\hline
\multirow{3}{*}{B1}  & \multirow{3}{*}{$d$}  & $1.9000_{\ast}$ & Fig.~\ref{fig:TAB_B}(a) & \multirow{3}{*}{$\alpha$}  & $0.4766\pi_{\ast}$ \\
 & & $2$ & Fig.~\ref{fig:TAB_B}(b) & & $0.4095\pi$ \\ 
 & & $3$ & Fig.~\ref{fig:show_pot} & & $0.3392\pi$ \\
\hline
\multirow{3}{*}{B2} & \multirow{3}{*}{$e$} &$2$& Fig.~\ref{fig:TAB_B}(c) & \multirow{3}{*}{$(\alpha,d)$} & $(0.4824\pi_{\ast}, 1.7384_{\ast})$ \\
& & $3$ & Fig.~\ref{fig:TAB_B}(d)& & $(0.4785\pi_{\ast}, 1.8270_{\ast})$\\ 
& & $8$ & Fig.~\ref{fig:TAB_B}(a) &  & $(0.4766\pi_{\ast}, 1.9000_{\ast})$\\
\hline
\multirow{4}{*}{B3}&\multirow{4}{*}{$\lambda$} &$0.1$ & Fig.~\ref{fig:TAB_B}(e) & \multirow{4}{*}{$(\alpha, d)$} & $(0.4902\pi_{\ast}, 1.8335_{\ast})$\\ 
& &$0.1$ &   & & $(0.4249\pi, 1.85)$\\ 
& & $0.27$& Fig.~\ref{fig:TAB_B}(a) & & $(0.4766\pi_{\ast},1.9000_{\ast})$  \\ 
& & $1.2$& Fig.~\ref{fig:TAB_B}(f) & & $(0.4297\pi_{\ast}, 2.4091_{\ast})$ \\
\hline
\multirow{5}{*}{B4} & \multirow{5}{*}{$\beta$} & $0$ & Fig.~\ref{fig:TAB_B}(g) & \multirow{5}{*}{$(\alpha, d)$}& $(0.4639\pi_{\ast},1.8784_{\ast})$ \\ 
& & $0$ &  & & $(0.4150\pi,1.9)$ \\ 
& & $0.05\pi$& Fig.~\ref{fig:TAB_B}(a) & &$(0.4766\pi_{\ast},1.9000_{\ast})$ \\  
& & $0.05\pi$&  & &$(0.4238\pi,1.95)$ \\  
& & $0.15\pi$& Fig.~\ref{fig:TAB_B}(h) & & $(0.4980\pi_{\ast}, 1.8133_{\ast})$
\\
\hline
 \end{tabular}
\caption{
List of parameter families in weakly distorted elliptic case.
The first and second columns consist of labels and key parameters, respectively.
For values of $\alpha$ and $d$ in the sixth column,
those with an asterisk ``$*$'' were
determined through $\mathrm{G}_3$ [Eq.~(\ref{CondG3})]; without an asterisk, 
$d$ is modified as $d=d_{\ast}+\epsilon$ with $\epsilon > 0$, 
and then $\alpha$ is determined through $\mathrm{G}_2$.
The common parameters for each family are as follows:
B1:
$(m,a, b, e, f, \lambda, \beta) = (2, 1.8, 1, 8, 1, 0.27, 0.05\pi)$;
B2:
$(m,a, b, f, \lambda, \beta) = (2, 1.8, 1, 1, 0.27, 0.05\pi)$;
B3:
$(m, a, b, e, f, \beta) = (2, 1.8, 1, 8, 1, 0.05\pi)$;
B4:
$(m, a, b, e, f, \lambda) = (2, 1.8, 1, 8, 1, 0.27)$.
$\Delta V = 0.15$ is maintained by modifying $K$,
which is more precise than the elliptic case.
}
\label{TAB:B}
\end{table}

\begin{figure}[!t]
%
\def\Size{14.6cm}
\centering
\includegraphics[height=\Size,keepaspectratio,clip]
{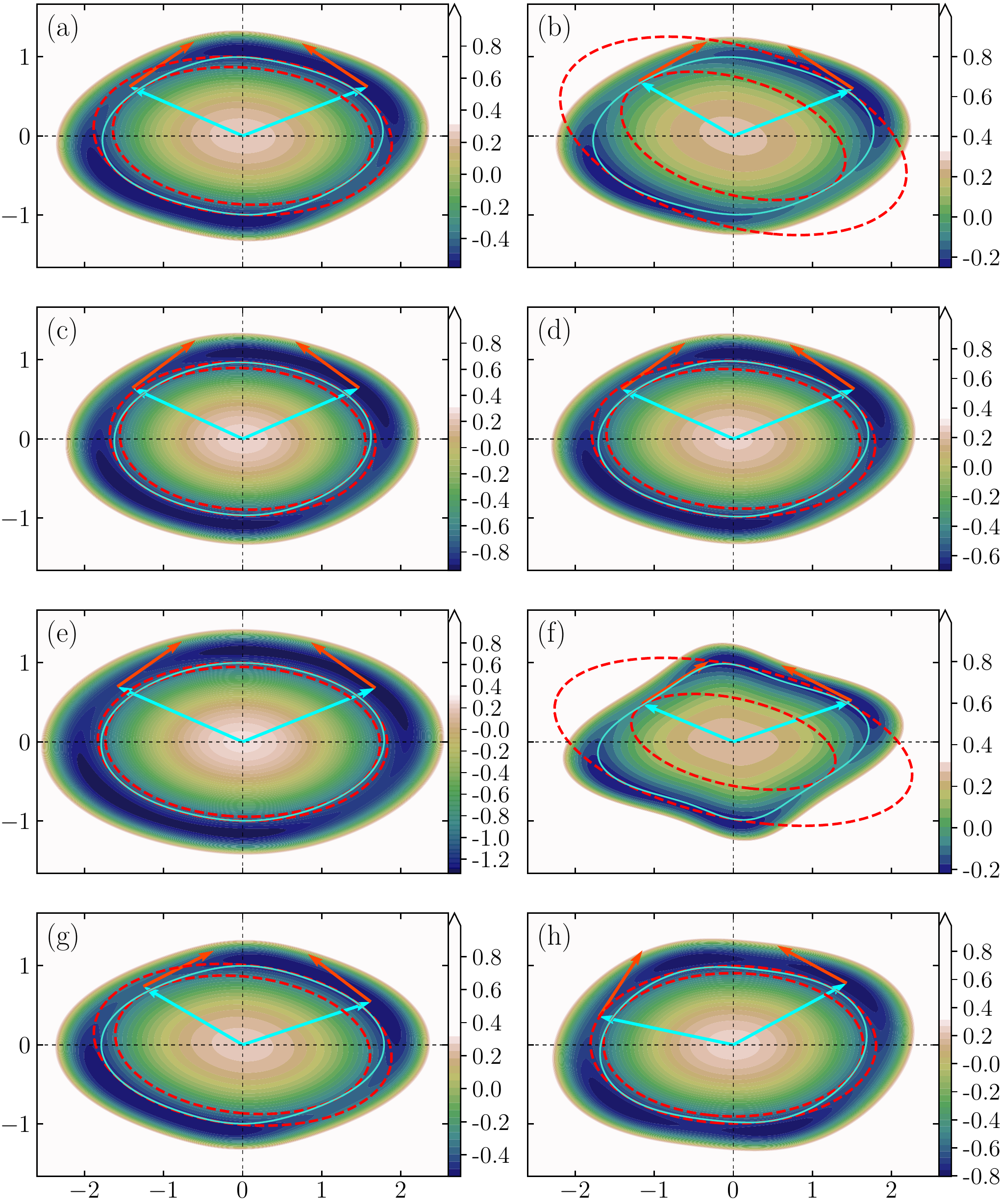} 
\caption{
(Color online)
Contour graphs of $V_{0}(\boldsymbol{x})$
on $\{\boldsymbol{x}\mid \lvert x\rvert \leq 2.6, \lvert y\rvert\leq 1.664\}$
in the parameter families B1, B2, B3, and B4 in Table~\ref{TAB:B}, and
the curves $\mathrm{C}_{\infty}$, $\mathrm{E}_{+}$, and $\mathrm{E}^{+}$.
In (a) and (b), $d=d_{\ast}$ and $d_{\ast}+\epsilon$ (B1);
in (c) and (d), modulation of the four-fold symmetry ($\frac{e}{f}$) differs (B2);
in (e) and (f), $\lambda$ differs (B3);
in (g) and (h), $\beta$ differs (B4).
See the fourth column in Table~\ref{TAB:B} for
the correspondences.
}
\label{fig:TAB_B}
\end{figure}

\begin{figure}[hbt]
%
\def\Size2{7.5cm}
\centering
\includegraphics[height=\Size2,keepaspectratio,clip]{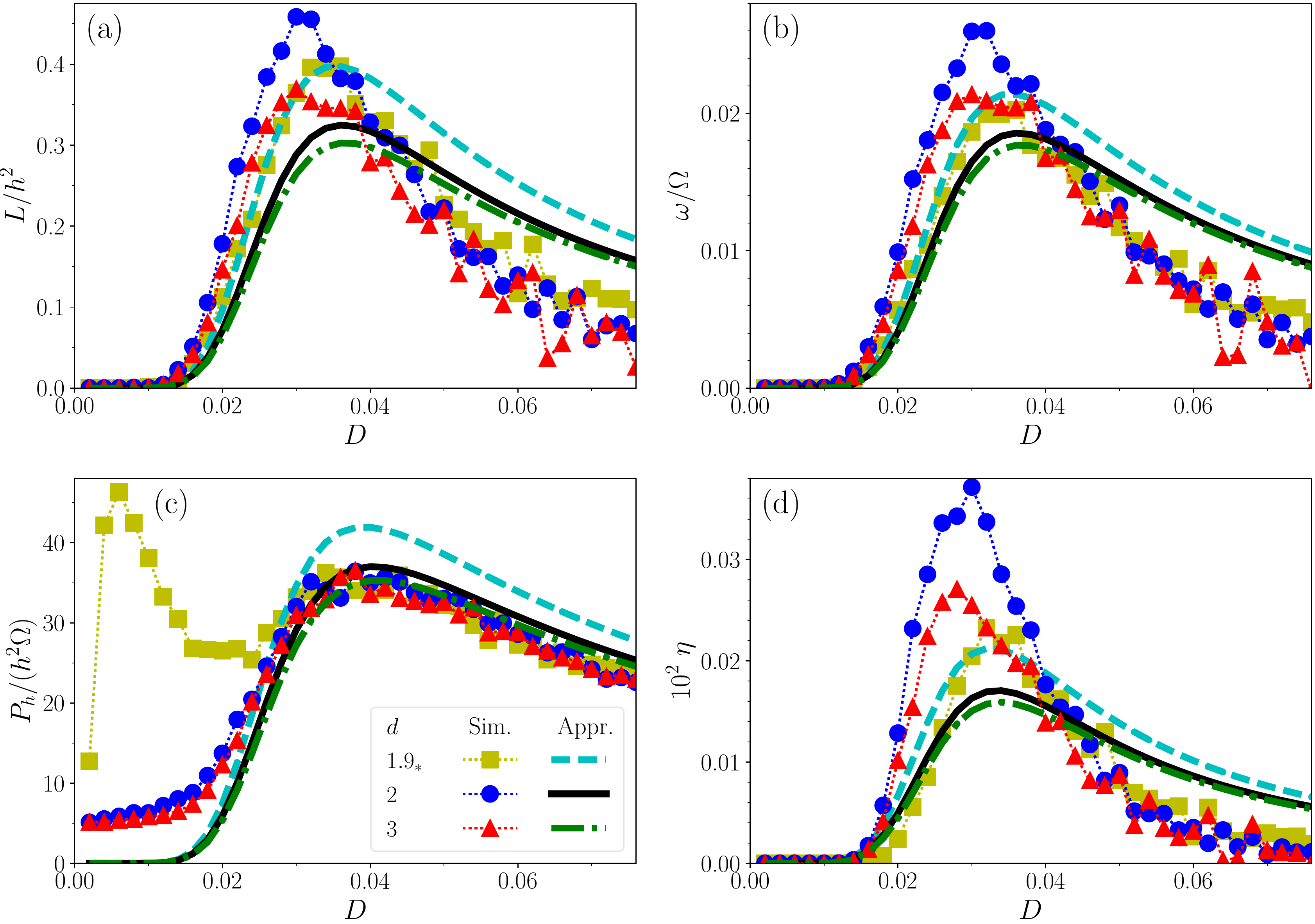}
 \caption{
(Color online)
(a) $L/h^2$, (b) $\omega/\Omega$, (c) $P_h/ (h^2\Omega)$, and (d) $\eta$
versus $D$ under the potentials of parameter family B1
[Figs.~\ref{fig:TAB_B}(a), \ref{fig:TAB_B}(b), and \ref{fig:show_pot}].
}
\label{fig:opt}
\end{figure}

\begin{figure}[tbh]
\def\Size2{7.5cm}
\centering
\includegraphics[height=\Size2,keepaspectratio,clip]{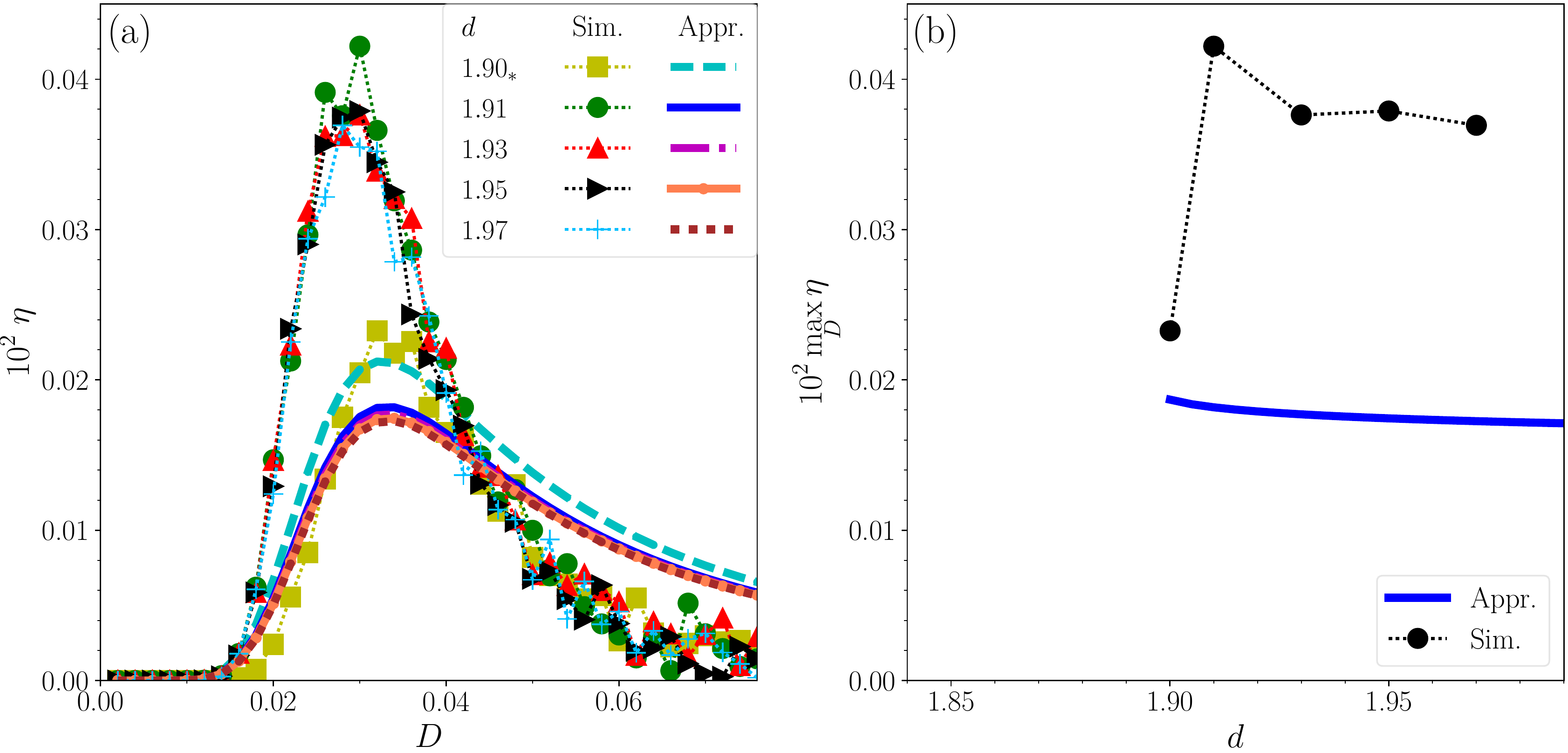}
 \caption{
(Color online)
(a)
$\eta$ versus $D$ for a series of $d$.
For each of $d \in \{1.90_{\ast}, 1.91, 1.93, 1.95, 1.97\}$,
retaining $\Delta V = 0.15$ and
$(m,a, b, e, f, \lambda, \beta) = (2, 1.8, 1, 8, 1, 0.27, 0.05\pi)$,
$\alpha$ is optimized via
 $\mathrm{G}_2$; particularly,
in the case of $d = 1.90_{\ast}$, which being
in the parameter family B1 (the first line in Table.~\ref{TAB:B}),
$(\alpha,d)$ is optimized via $\mathrm{G}_3$.
(b)
$\displaystyle\max_{D}\eta$ as a function of $d$ 
for $d \geq 1.90$.
As $d$ varies, $\alpha$ is optimized simultaneously via
$\mathrm{G}_2$ with the other parameters being the same as those in (a).
For $1\leq d < 1.90$, there is no optimized value of $\alpha$, and the curve
is not drawn.
The correspondences between the parameters
and the types of symbol (numerical simulation results) and
curve (approximation results) are shown in the legend boxes.
}
\label{fig:psi.so.d}
\end{figure}

\begin{figure}[tbh]
\def\Size2{7.5cm}
\centering
\includegraphics[height=\Size2,keepaspectratio,clip]{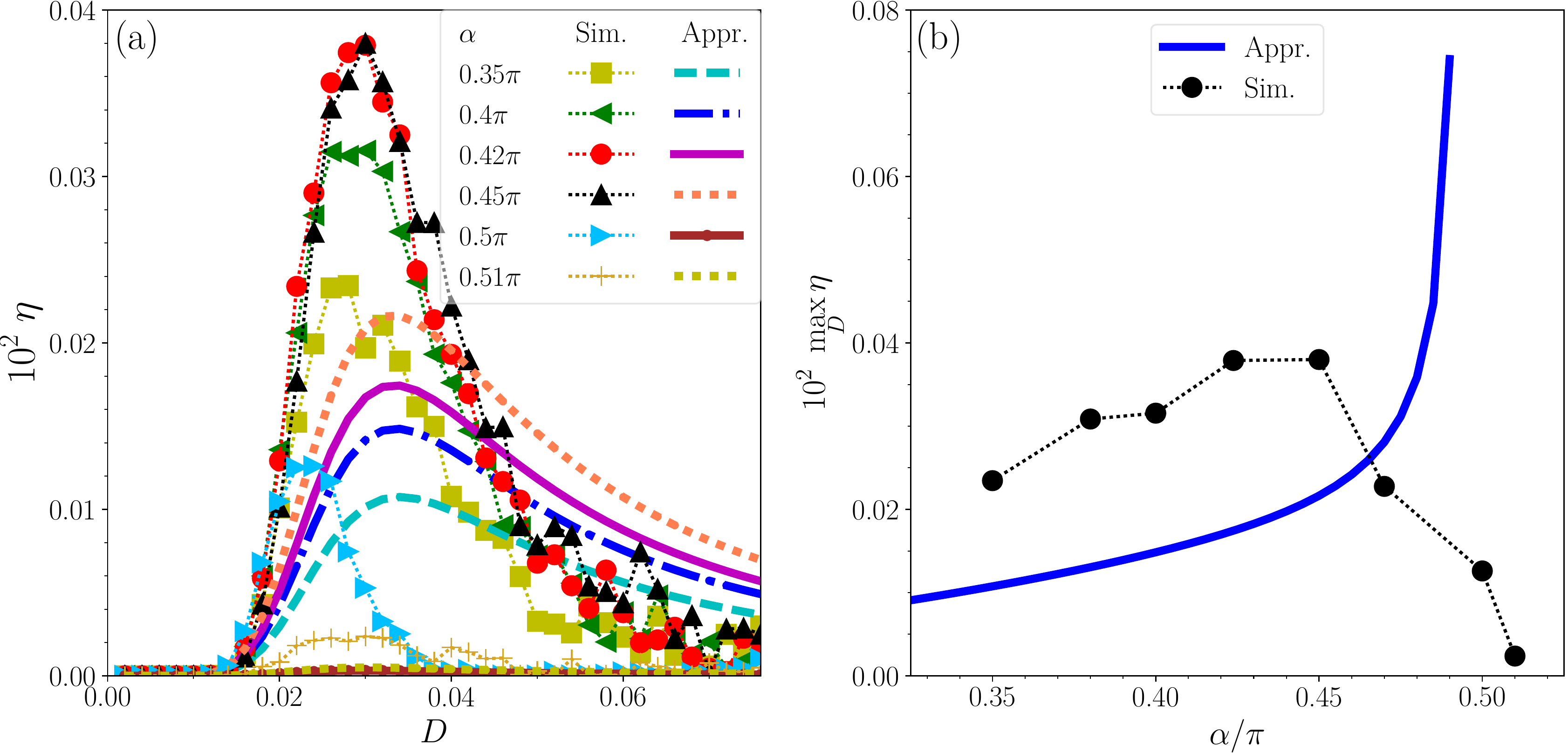}
\caption{
(Color online)
(a)
$\eta$ versus $D$ for a series of $\alpha$ from $0.35\pi$ to $0.51\pi$.
(b)
$\displaystyle\max_{D}\eta$ as a function of $\alpha$
over the range treated in (a).
While $\alpha$ varies, $d$ is fixed at $1.95$ (i.e.,
$v_1(\boldsymbol{x})$ is not optimized),
 $\Delta V=0.15$ is retained, and
$(m,a, b, e, f, \lambda, \beta) = (2, 1.8, 1, 8, 1, 0.27, 0.05\pi)$.
}
\label{fig:psi.so.alp}
\end{figure}

\begin{figure}[hbt]
\def\Size2{7.5cm}
\centering
\includegraphics[height=\Size2,keepaspectratio,clip]
{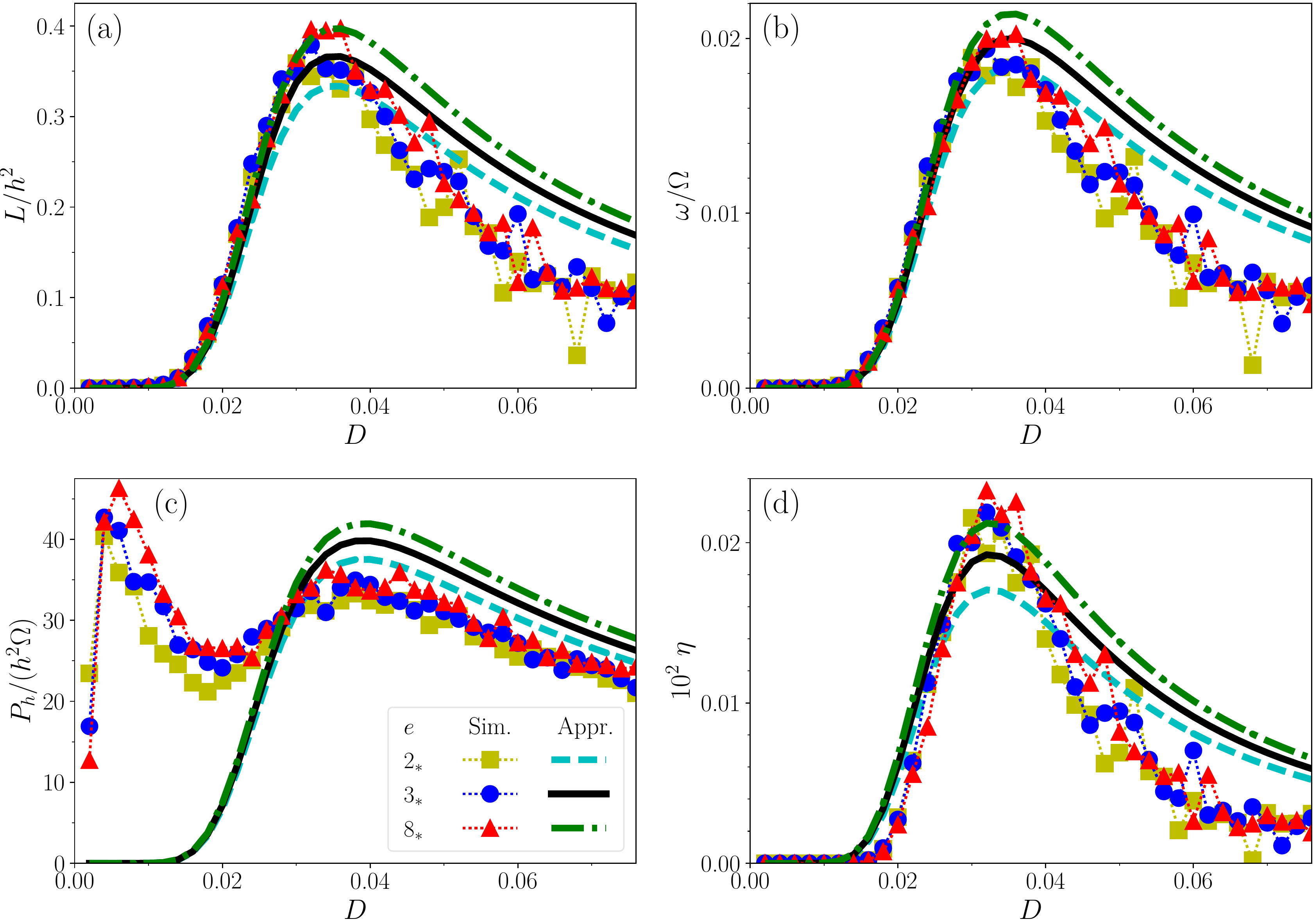}
\caption{
(Color online)
(a) $L/h^2$, (b) $\omega/\Omega$, (c) $P_h/ (h^2\Omega)$, and (d) $\eta$
versus $D$ under the potentials of parameter family B2
[Fig.~\ref{fig:TAB_B}(c), (d), and (a)].
}
\label{fig:ef}
\end{figure}

Outcomes of the optimization described in Sec.~\ref{sec:LamNonZero}
for $V_{0}(\boldsymbol{x})$ of nonelliptic pathway ($\mathrm{C}_{\infty}$)
are shown with the results of the performance indexes
according to the parameter families B1--B4 in Table~\ref{TAB:B}.
Firstly, let us observe the effect of the relaxation for $d$
in Eq.~(\ref{Det:d}).
In parameter family B1, 
$d$ is varied as $d_{\ast}$, $2$, and $3$, i.e.,
the first one is determined by $\mathrm{G}_3$ [Eq.~(\ref{CondG3})] together with
$\alpha_{\ast}$, and the second and third ones are increased from $d_{\ast}$ in accordance with
the moderation procedure [Eq.~(\ref{Det:d})]
followed by readjustment of $\alpha$ through $\mathrm{G}_2$.
To see the curves
$\mathrm{C}_{\infty}$, $\mathrm{E}_{+}$, and $\mathrm{E}^{+}$
in Figs.~\ref{fig:TAB_B}(a), \ref{fig:TAB_B}(b), and \ref{fig:show_pot},
$\mathrm{E}_{+}$ and $\mathrm{E}^{+}$ closely contact to
$\mathrm{C}_{\infty}$ for $d=d_{\ast}$ [Fig.~\ref{fig:TAB_B}(a)] 
and, as $d$ is increased, the space between $\mathrm{E}_{+}$ and $\mathrm{E}^{+}$ 
becomes wider [Figs.~\ref{fig:TAB_B}(b) and \ref{fig:show_pot}].

The simulation results of $L$, $\omega$, and $\eta$ in Fig.~\ref{fig:opt}
demonstrate that the curves of $d=2$ are higher than those of 
$d = d_{\ast}\approx 1.900$
around the peak region.
Turning to the plot of $P_h$, the curve of $d=d_{\ast}$
has another peak around $D\approx 0.006$, while the others have 
only a single peak.
A reason for this new peak in $P_h$ is, as mentioned in Sec.~\ref{sec:LamNonZero},
as follows.
In the presence of time-dependent fields,
instead of the curves $\mathrm{E}_{+}$ and $\mathrm{E}^{+}$,
which are defined for $m\rightarrow \infty$ and $h=I=0$,
we should consider the temporally moving curves $E^{\ast}(\Phi_t)$ and $E_{\ast}(\Phi_t)$
with $v_1'(\boldsymbol{x},\Phi_t)$ in Eq.~(\ref{v1d}).
The motion of the circumscribed point of $E_{\ast}(\Phi_t)$ may
temporally create another minimum at a point distant from both $\boldsymbol{x}_{+}$ and
$\boldsymbol{x}_{-}$, and then may induce a jump of state.
Such a jump motion may expend power associated with a small amount of thermal activation.
We can thus relate such a power consumption to the new peak in $P_h$.
This also suggests that the input power is not applied efficiently 
to the rotation while employing $v_1(\boldsymbol{x})$ such
that $\mathrm{E}_{+}$ and $\mathrm{E}^{+}$ 
enclose $\mathrm{C}_{\infty}$ without sufficient room.
In contrast, when making a suitably loose gap between 
$\mathrm{E}_{+}$ and $\mathrm{E}^{+}$ with
$\epsilon$ in Eq.~(\ref{Det:d}), the movement of the minimum can be restricted near
either $\boldsymbol{x}_{+}$ or $\boldsymbol{x}_{-}$, in which case the local equilibrium
is maintained.
We then expect that
incorporating the moderation brings a better efficiency.
This is consistent with the numerical results for $\eta$ in Fig.~\ref{fig:opt}.

We should also note that the presented approximation cannot predict the extra peak
of $P_h$.
This is because we have assumed that the local equilibrium always holds around the minima of $V_{0}(\boldsymbol{x})$, and have
ignored any temporally induced current due to the creation of a temporal minimum.
Thus, for the case of $V_{0}(\boldsymbol{x})$ optimized with the moderation,
we can assume a local equilibrium, and basically regard the approximation to be
consistent with the results of numerical simulation.

We give a more detailed view on the marginal behaviors of $\eta$ in
the optimization for $v_1(\boldsymbol{x})$ 
under the procedure $\mathrm{G}_3$ followed by the moderation Eq.~(\ref{Det:d}).
Figure~\ref{fig:psi.so.d}(a) shows the graphs of
$\eta$ versus $D$ for a series of $d$ from $1.90$
(the case of $d=1.90_{\ast}\equiv d_{\ast}$ and $\alpha=0.48\pi_{\ast}\equiv \alpha_{\ast}$ in the parameter family B1 
in Table~\ref{TAB:B}) to $1.97$,
where, for each $d$, $\alpha$ is simultaneously readjusted 
in accordance with $\mathrm{G}_2$, i.e.,
$\alpha = \displaystyle\argmax_{ 0 \leq \alpha < \frac{\pi}{2}}
(-\boldsymbol{x}_{+}\cdot \boldsymbol{n}_{v})$,
and $\Delta V =  0.15$ is retained by modulating $K$.
These curves indicate that the peak is higher
 as $d$ is closer to $d_{\ast}$, but drops at $d=d_{\ast}$.
Figure~\ref{fig:psi.so.d}(b) shows
the dependence of the peak height on $d$
in the aforementioned settings of parameters.
The solid curve thus may approximate
$\displaystyle\max_{D,\alpha}\eta$ for $d > d_{\ast}$,
whereas it is not defined for $1 \leq d < d_{\ast}$, in which
no optimized value of $\alpha$ 
satisfying $\mathrm{G}_2$ exists.
One can see that the numerical results (symbols) follow the solid curve,
except for the difference in their heights.
Figure~\ref{fig:psi.so.alp} shows (a) the graphs of
$\eta$ versus $D$ as only $\alpha$ varies around $\alpha \approx 0.42\pi$
with $d=1.95$, $\Delta V = 0.15$ and
$(m,a, b, e, f, \lambda, \beta) = (2, 1.8, 1, 8, 1, 0.27, 0.05\pi)$,
and (b) $\displaystyle\max_{D}\eta$ over the range
of $\alpha$ treated in the panel (a).
Recalling that the referenced parameters $\alpha \approx 0.42\pi$ and $d=1.95$
(filled circles or solid curve)
are obtained in the moderation
procedure for the case of $\alpha_{\ast} > 0.42\pi$ and $d_{\ast}$, there is a possibility of raising
the peak of $\eta$, i.e., $\displaystyle\max_{D}\eta$, by increasing $\alpha$ from
$\alpha \approx 0.42\pi$. However, in the numerical results (symbols),
as $\alpha$ is increased, $\displaystyle\max_{D}\eta$
soon plateaus and goes down for $\alpha \geq 0.45\pi$.
For $\alpha < 0.42\pi$, the peak diminishes monotonically; 
this implies that $\alpha$ moves away from the optimized point on $d =1.95$.
The solid curve for $\displaystyle\max_{D}\eta$ in Fig.~\ref{fig:psi.so.alp}(b)
 has a discontinuity at $\alpha\approx 0.49\pi$,
where the original two minima of $V_0(\boldsymbol{x})$
 switch to another two minima
(The number of minima of $V_0(\boldsymbol{x})$
changes as two, four, and two 
for $\alpha < 0.47\pi$, $0.47\pi \leq \alpha \leq 0.49\pi$, and $0.49\pi < \alpha$,
respectively), therefore, the curve is drawn only for
the domain lower than the singular point ($\alpha\approx 0.49\pi$).
Around that point,
it is expected that the local equilibrium assumption breaks,
the rotational performance drops as mentioned above,
and also our approximation becomes inconsistent
with the original assumptions such that the potential always has two minima.
Consequently, these results reveal that
the moderation procedure works well with a
small relaxation parameter.

In parameter family B2, $e$ is increased;
with $\frac{e}{f}$ [see Eq.~(\ref{v0:core})], we can enhance the fourth-order circular harmonic distortion
of the shape of the pathway along the potential valley.
It is deformed gradually from an ellipse as $\frac{e}{f}$ differs from one.
In Fig.~\ref{fig:TAB_B}, we see the shapes
of the pathway for $e = 2$ (c), $3$ (d), and $8$ (a).
In Fig.~\ref{fig:ef}, the approximation curves indicate that
the indexes rise as $e$ increases, and the numerical results seem
to follow such a tendency, although it is not as clear.
The emergence of the peak at $D\approx 0.006$ in $P_h$ is, as mentioned above,
because of the fact that $(\alpha, d)$ is determined by $\mathrm{G}_3$ without
the moderation.
As in the figure legends, we add an asterisk ``$\ast$'' to the parameter
value(s) for which
$(\alpha, d)$ is determined in $\mathrm{G}_3$ (see Table~\ref{TAB:B}).

\begin{figure}[hbt]
\def\Size2{7.5cm}
\centering
\includegraphics[height=\Size2,keepaspectratio,clip]{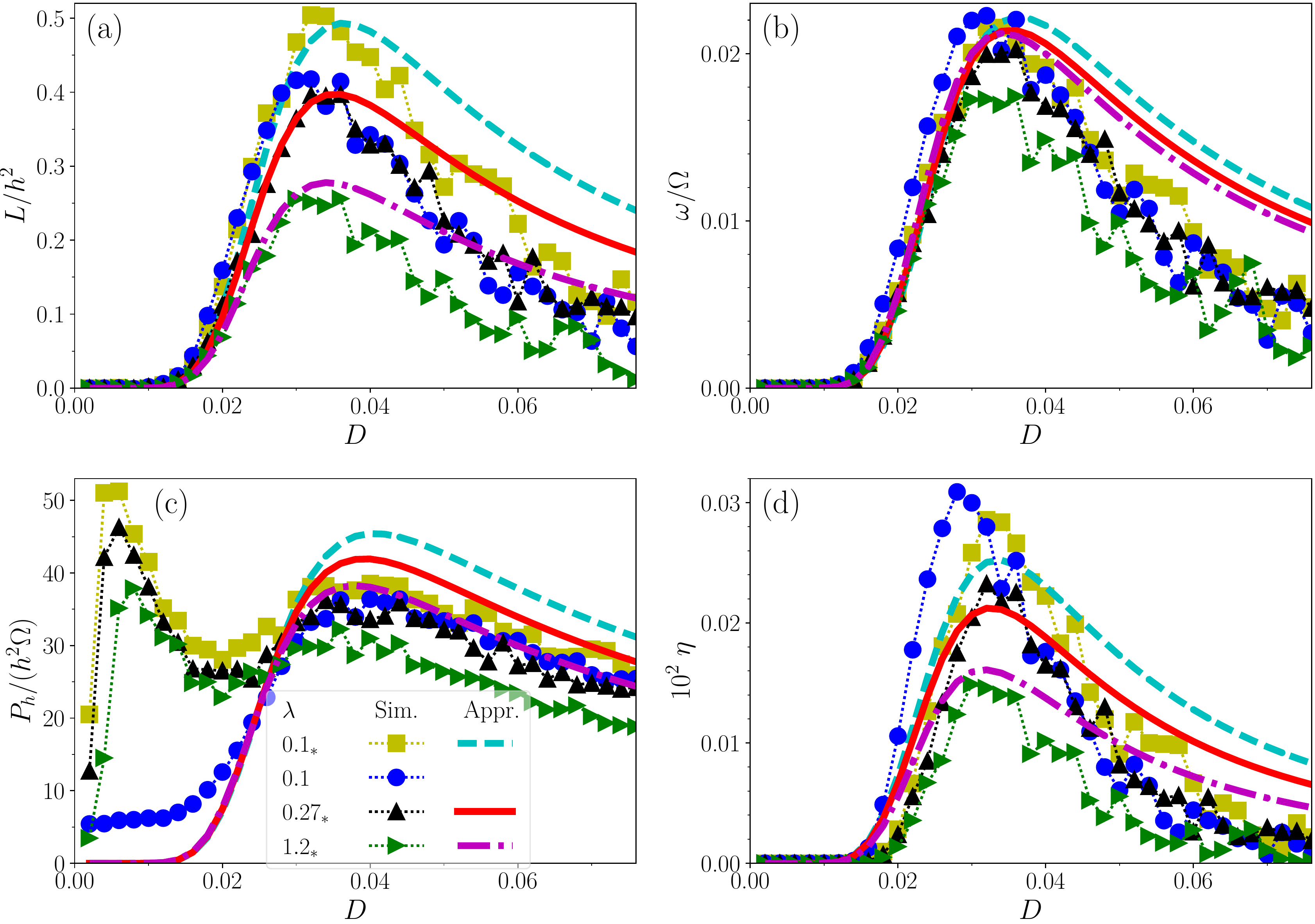}
 \caption{
(Color online)
(a) $L/h^2$, (b) $\omega/\Omega$, (c) $P_h/ (h^2\Omega)$, and (d) $\eta$
versus $D$ under the potentials of parameter family B3
[Fig.~\ref{fig:TAB_B}(e), (a), and (f)].
}
\label{fig:lam}
\end{figure}

In parameter family B3, $\lambda$ in Eq.~(\ref{v0:core}) is increased as
$0.1$, $0.27$, and $1.2$.
As shown in Fig.~\ref{fig:TAB_B}(e), (a), and (f)
for $\lambda = 0.1$, $0.27$, and $1.2$,
the four-fold symmetric modulation on the pathway is conspicuous with $\lambda$.
In Fig.~\ref{fig:lam},
we see that the peaks of $L$, $\omega$, $P_h$, and $\eta$ decrease
with $\lambda$, except for 
the case $\lambda = 0.1$ (filled circles) in which
$(\alpha, d)$ is optimized with the modulation.
A characteristic of this decrease is that
as $\lambda$ is increased, 
the factor $-\boldsymbol{x}_{+}\cdot\boldsymbol{n}_{+}$
increases; however, the other factor
$H_n$ increases simultaneously, in which case all the performance indexes decrease.

\begin{figure}[tbh]
\def\Size2{7.5cm}
\centering
\includegraphics[height=\Size2,keepaspectratio,clip]{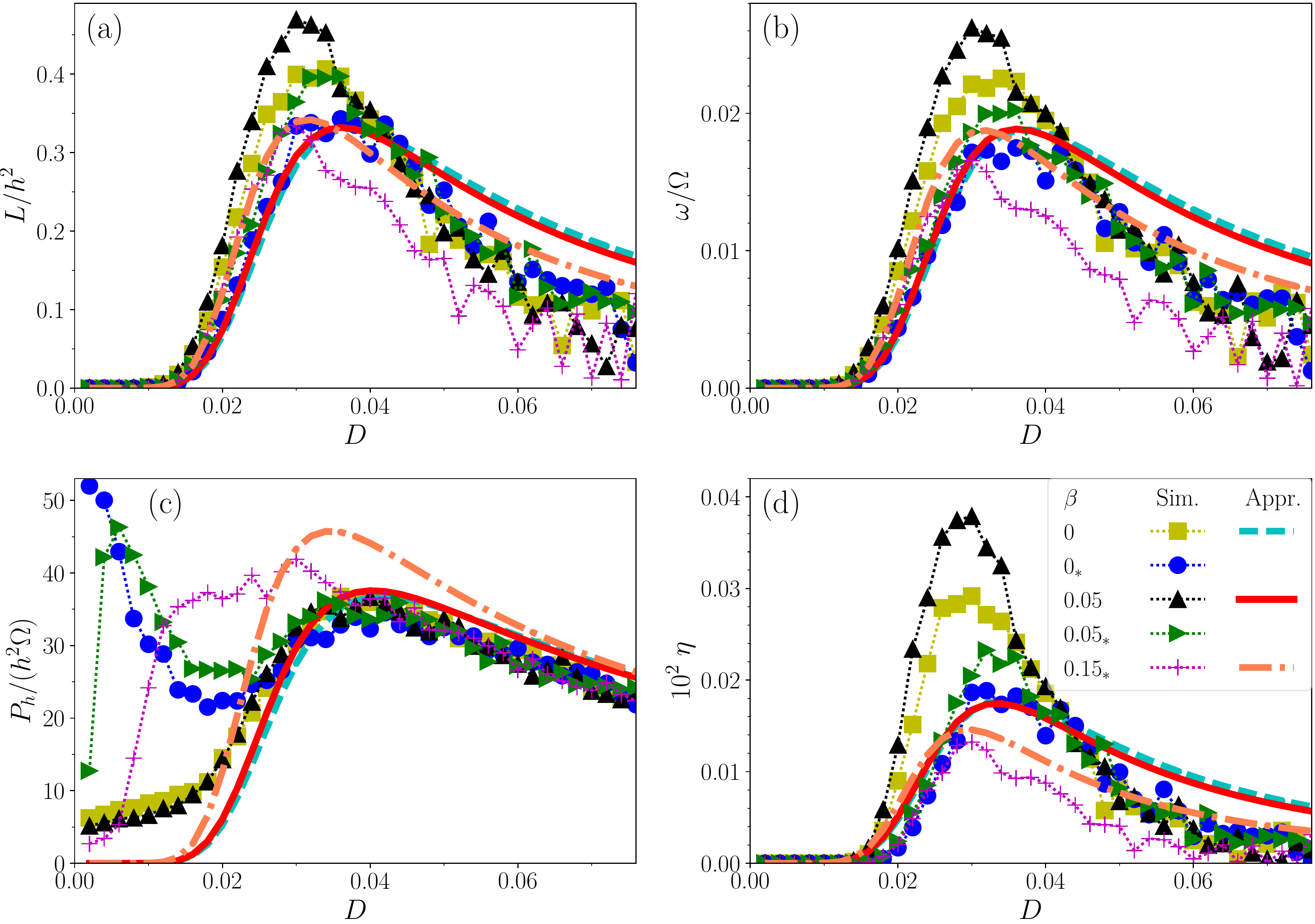}
 \caption{
(Color online)
(a) $L/h^2$, (b) $\omega/\Omega$, (c) $P_h/ (h^2\Omega)$, and (d) $\eta$
versus $D$ under the potentials of parameter family B4
[Fig.~\ref{fig:TAB_B}(g), (a), and (h)].
}
\label{fig:psi}
\end{figure}

In parameter family B4, $\beta$ is varied as
$\beta = 0$, $0.05\pi$, and $0.15\pi$;
with $\beta$, the axis of the fourth-order harmonic distortion rotates.
In Fig.~\ref{fig:TAB_B}(g), (a), and (h) for $\beta = 0$, $0.05\pi$, 
and $0.15\pi$, respectively, we can see such a rotation.
Figure~\ref{fig:psi} shows that
$L$, $\omega$, and $\eta$ have higher peaks
for $\beta=0.05\pi$ as
$v_1(\boldsymbol{x})$ is optimized with the moderation.
Finally, let us compare
the best result in the elliptic case ($\lambda=0$) in Sec.~\ref{LamZero:num}
with that in the parameter families B1--B4
under the same conditions of $(m,a,b)$ 
with respect to the peak of $\eta$.
For the former, see the case of $(m,a,b)=(2,1.8,1)$, i.e.,
the curve of $a=1.8$ in Fig.~\ref{fig:elp} (or that of $m=2$ in
Fig.~\ref{fig:m} or that of $a=1.8$ in Fig.~\ref{fig:mag}).
We can see that $\eta$ for $\beta=0.05\pi$ in Fig.~\ref{fig:psi}
has a higher peak, $\eta\approx 0.038\times 10^{-2}$, than the best one, 
$\eta\approx 0.023\times 10^{-2}$,
in the elliptic case.
This result suggests that the term $\lambda$ can contribute to a better efficiency.
It also implies that the efficiency could be improved by designing
$v_0(\boldsymbol{x})$ and $v_1(\boldsymbol{x})$
more carefully.

So far, 
maximizing the performance indexes under the RDDF [Eq.~(\ref{LEQ:Poth})]
has been considered by optimizing $V_0(\boldsymbol{x})$; however, the value
of $\eta$ is very small. Finally, let us discuss the reason for such small
efficiency, and a possible way of remodeling to improve it.
In the present model,
for a small $h$, the field $h\boldsymbol{N}_t$ 
has a role in modulating the ratchet (saw-tooth) profile along
the valley by varying the positions of
 the minima and saddle (or ridge curves) of $V(\boldsymbol{x},t)$
and the slopes around the minima.
This eventually causes net rotational motion because of the circular ratchet structure
of $V_0(\boldsymbol{x})$.
However, 
because the primary action of the field is
to cause a linear displacement of the minima and saddles,
not all the power of the field
is applied to the unidirectional rotational motion; 
instead, a great deal of the power is scattered to other motions
(i.e., rocking motions without bias in the rotational and radial directions)
\cite{doi:10.7566/JPSJ.84.044004}.
Thus, we can conclude that
the main reason for the small efficiency
lies in the form of the field.
The problem of improving the efficiency within the non-biased fields
can therefore be recast into a problem of designing 
the time-dependent part of the potential, $V_h(\boldsymbol{x},t)$,
or external fields to maximize
its power conversion efficiency.
Exploiting an idea from one-dimensional ratchet models
that incorporate a mechanism for avoiding such a rocking motion
with saw-tooth type potentials that are shifted randomly back or forth
by an appropriate distance
\cite{0295-5075-27-6-002,PhysRevE.60.2127,PhysRevE.69.021102},
 we may consider a form of the field as
$\boldsymbol{f}_h(\boldsymbol{x},t) = 
h q(\boldsymbol{x},t) \partial_{\boldsymbol{x}}\theta(\boldsymbol{x})$.
This represents a circular field around the origin, the
direction of which varies randomly with the spatial dependency of
$q(\boldsymbol{x},t)$.
We expect that
this can reduce the rocking motion in the radial direction, and
may also suppress such diffusive motion in the rotational direction if
we appropriately design
the spatial and temporal variations of $q(\boldsymbol{x},t)$ in accordance with 
$V_0(\boldsymbol{x})$ imposing a constraint that
the spatial average of $\boldsymbol{f}_h(\boldsymbol{x},t)$
has no bias.

\section{\label{Discuss} Direction For Three-tooth Ratchet Model}

So far, we have dealt with optimizing
the two-tooth ratchet potential in
Eqs.~(\ref{LEQ:Pot0})--(\ref{v1:core}).
However, our approach could be applied to more general ratchet potentials.
Here, we show how a similar approach holds for a three-tooth ratchet potential
of the same form as $V_0(\boldsymbol{x})$ in Eq.~(\ref{LEQ:Pot0}).

It is necessary that
$v_{0}(\boldsymbol{x})$ and $v_{1}(\boldsymbol{x})$ 
have three-fold symmetry.
For $m\gg 1$, the curve
$\mathrm{C}_{\infty}: \{\boldsymbol{x}|v_{0}(\boldsymbol{x})=1 \}$
corresponds to a potential valley, and the region of
$v_{0}(\boldsymbol{x})<1$ must be a simply connected space.
Therefore, a simple expression is proposed as
\begin{equation}
v_{0}(\boldsymbol{x})\equiv
\lvert\boldsymbol{x}\rvert^2
+a\lvert\boldsymbol{x}\rvert^4
+b
(\boldsymbol{e}_{0} \cdot \boldsymbol{x})
(\Hat{g}_1  \boldsymbol{e}_{0} \cdot \boldsymbol{x})
(\Hat{g}_2  \boldsymbol{e}_{0} \cdot\boldsymbol{x})
,
\label{ThreeTooth:v0}
\end{equation}
where $a$ is positive, so that we have
$v_{0}(\boldsymbol{x})\rightarrow\infty$ for
$\lvert\boldsymbol{x}\rvert \rightarrow\infty$,
and $\lvert b\rvert$ is sufficiently small
for such $\mathrm{C}_{\infty}$ of a simply connected curve.
Term $\Hat{g}_{1}$ ($\Hat{g}_{2}$) represents a matrix for
a rotation of angle $+\frac{2\pi}{3}$ ($-\frac{2\pi}{3}$):
\begin{align}
\Hat g_1 \equiv
\frac12
\begin{pmatrix}
-1 & -\sqrt{3}\\
\sqrt{3} & -1
\end{pmatrix},
\quad
\Hat g_2 \equiv
\frac12
\begin{pmatrix}
-1 & \sqrt{3}\\
-\sqrt{3} & -1
\end{pmatrix}
.
\end{align}
The third term adds a third circular harmonic in $\mathrm{C}_{\infty}$;
$\boldsymbol{e}_{0}$ is a reference axis on the azimuthal angle
about the origin.
Note that as $\boldsymbol{e}_{0}$ rotates, $\mathrm{C}_{\infty}$ rotates
by the same angle about the origin.
Without loss of generality, we have $b > 0$ and $\boldsymbol{e}_{0}=(1,0)^{\mathrm{T}}$.
Similarly, $v_{1}(\boldsymbol{x})$ is given as
\begin{equation}
v_{1}(\boldsymbol{x})\equiv
\lvert\boldsymbol{x}\rvert^2
+c\lvert\boldsymbol{x}\rvert^4
+d
(\boldsymbol{e}_{1} \cdot \boldsymbol{x})
(\Hat{g}_1  \boldsymbol{e}_{1} \cdot \boldsymbol{x})
(\Hat{g}_2  \boldsymbol{e}_{1} \cdot\boldsymbol{x})
\label{eq:v1_3th}
\end{equation}
with a reference axis
$\boldsymbol{e}_{1}\equiv (\cos\alpha,\sin\alpha)^{\mathrm{T}}$ and
positive values $c$ and $d$.

\begin{figure}[t]
\def\Size{7.3cm}
\centering
\includegraphics[height=\Size,keepaspectratio,clip]{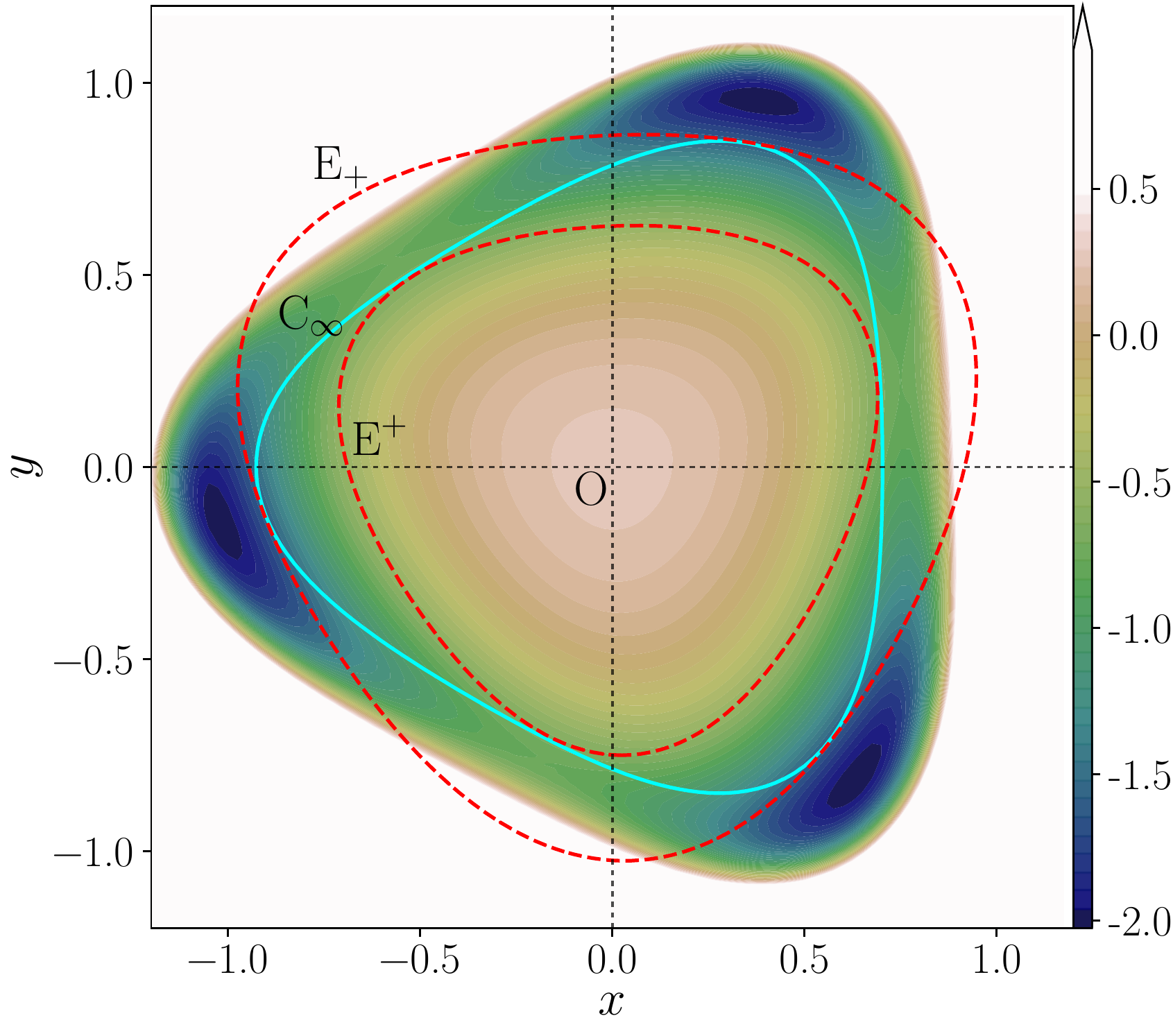}
 \caption{
(Color online)
Contour plot of a three-tooth ratchet potential with skeletons
of $\mathrm{C}_{\infty}: \{v_0(\boldsymbol{x})=1\}$,
 $\mathrm{E}_{+}$, and $\mathrm{E}^{+}$.
The parameters of $V_0(\boldsymbol{x})$ of
Eqs.~(\ref{LEQ:Pot0}), (\ref{ThreeTooth:v0})--(\ref{eq:v1_3th}) are
$(m,a,b,c,d,K,\alpha)=(2,1,3,1,2,1,0.52\pi)$.
$\mathrm{E}_{+}$ and $\mathrm{E}^{+}$ correspond to 
the curves $\{v_1(\boldsymbol{x})=E\}$ for 
$E=1.62$ (externally tangent case) and $0.67$ (internally tangent case),
respectively.
}
\label{fig:3th}
\end{figure}

Figure~\ref{fig:3th} shows a contour graph of the three-tooth ratchet potential
of Eqs.~(\ref{LEQ:Pot0}), (\ref{ThreeTooth:v0})--(\ref{eq:v1_3th}).
The curves
$\mathrm{E}_{+}$ and $\mathrm{E}^{+}$ on the graph
represent the circumscribed and inscribed curves
of $\mathrm{E}:\{\boldsymbol{x}\mid v_1(\boldsymbol{x})=E\}$
to $\mathrm{C}_{\infty}$ with $E=1.62$ and $0.67$, respectively.
The externally (internally) tangent points correspond to
the local minima (saddles) of $V_0(\boldsymbol{x})$.
For $m \rightarrow \infty$, these minima and saddles
satisfy Eq.~(\ref{3th:force_2}).

The optimization of $L$, $\omega$, and $\eta$ can be carried out
through the maximization of a factor such as
$I_0(D)$ [Eq.~(\ref{TorqBalance})], which can be obtained
by following the procedure in Appendix~\ref{App:LamZero}.
Similarly, let us assume that
the factor $-\boldsymbol{x}_{\ast}\cdot \boldsymbol{n}(\boldsymbol{x}_{\ast})$
at a local minimum point $\boldsymbol{x}_{\ast}$ affects
the maximization of $I_0(D)$ more than it does $H_{n}$.
We then employ
the strategy to maximize 
$-\boldsymbol{x}\cdot \boldsymbol{n}_{v}(\boldsymbol{x})$
using $v_{0}(\boldsymbol{x})$ and $v_{1}(\boldsymbol{x})$
with the assumption that $m\gg 1$.
In particular, letting $p$ be a target parameter in $v_1(\boldsymbol{x})$ for the optimization,
the problem is to solve Eqs.~(\ref{opt_prob}) and (\ref{G0});
the actual procedure follows
Eq.~(\ref{G1}) in Sec.~\ref{sec:OPT0} as
\begin{equation}
\boldsymbol{x}_{\ast}=
\argmax_{\boldsymbol{x} \in \mathrm{C}_{\infty}} 
\left\{
-\boldsymbol{x}\cdot \boldsymbol{n}_{v}(\boldsymbol{x})
\right\},
\end{equation}
and then, with this
$\boldsymbol{x}_{\ast}$, we find such $p$ as satisfies
Eqs.~(\ref{3th:force_2}) and (\ref{G0}) [replace $\boldsymbol{x}_{\ast}$ with $\boldsymbol{x}_{+}$].

As described in Sec.~\ref{sec:LamNonZero}, if we choose
$v_1(\boldsymbol{x})$ to be a different functional form from $v_0(\boldsymbol{x})$,
we can further arrange the
values of target parameters in $v_1(\boldsymbol{x})$ to
decrease $g(\boldsymbol{x})$ within a suitable range.
For example, we can consider a $v_0(\boldsymbol{x})$ that has
the sixth circular harmonic deformation.
In such a case, as in Sec.~\ref{sec:LamNonZero}, 
letting $\alpha$ and $d$ in $v_1(\boldsymbol{x})$ be target parameters,  we first determine
$\alpha_{\ast}$ and $d_{\ast}$ as
\begin{equation}
(\alpha_{\ast}, d_{\ast}) = \argmin_{ 0 \leq \alpha < \frac{\pi}{3}, d(\alpha)} E_{+},
\end{equation}
where $d(\alpha)$ means a function that relates $\alpha$ to $d$
through Eq.~(\ref{3th:force_2}).
Next, to prevent the creation of temporal minima,
we moderate the above minimization by replacing $d$ with
$d=d_{\ast}+\epsilon$ ($\epsilon >0$), and revise $\alpha$
to satisfy Eq.~(\ref{3th:force_2}) with this $d$.
We expect this procedure to bring about a robust local equilibrium for external fields
and to reduce the power consumption for rotation.
By observing the numerical result for $P_h$,
we can confirm whether the local equilibrium has been retained.

\section{\label{summary} Summary}

The underlying themes in this study have been to elucidate the types of ratchet model (as combinations of the 2D ratchet potential and
the unbiased randomly varying field) that produce a robust net rotation, 
and to determine how to maximize the rotational output and efficiency.
In this paper, we have shown that the proposed ratchet model, consisting of
a 2D two-tooth ratchet potential and an RDDF,
generates a net rotation in the direction of the ratchet potential, i.e., the chirality.
The 2D three-tooth ratchet model also possesses
such a property \cite{PhysRevE.87.022144,doi:10.7566/JPSJ.84.044004}.

The mechanism of net rotation is not so obvious
because
the deformation along the valley in the 2D ratchet model can be composed of
various types of deformation.
The mathematical origin of the net rotation can be found in
Eq.~(\ref{Lt_2}), i.e.,
$ L,\omega \propto \overline{ \ln\left[
\frac{P(\mu,t) P(-\mu,t)}{Q(\mu,t)^2} \right] J^{\mu}(t)}$, 
in which $J^{\mu}(t)$, the barrier-crossing current,
and the multiplied factor,  the entropy-like measure for 
the deviation of the positional distribution from the equilibrium one,
are correlated as a result of the rectification effect due to the chirality, 
and the average of these products remains a bias [Eq.~(\ref{lnPQJ})].

Another explanation uses a ratchet
exposed to an external field made of superimposed
uni-axially polarized fields within the same 2D plane.
The mechanism for the net rotation of a two-tooth ratchet
under a uni-axially polarized randomly varying field can be
explained using the mechanism for the propeller rotation of a ``gee-haw whammy diddle''
 or ``propeller stick'' \cite{wiki:whammydiddle} (cf. \cite{PhysRevE.84.061119}).
Employing $M$ copies of such a uni-axially polarized field,
we orient their angles of polarization to $\phi_k = \frac{2\pi k}{M}$ ($0\leq k < M$), respectively,
whereby the ratchet is exposed to the field
$\sum_k h_k(t)\boldsymbol{N}_k$ (cf. Eq.~(\ref{LEQ:Poth})), where 
$\boldsymbol{N}_k=(\cos\phi_k, \sin\phi_k)$ and
$h_k(t)$ is a unbiased dichotomic noise, independent of the others and
varying between $-\frac{h}{\sqrt{M}}$ and $\frac{h}{\sqrt{M}}$
with mean frequency $\Omega$. Thus, this field mimics the RDDF.
Then, as a total of the propeller-stick-like responses to the
individual fields, we can expect this ratchet to yield a net rotation
in the direction determined by its chirality.

The optimization of the 2D ratchet potential has been considered by employing
the redesigned form of the ratchet potential in Eq.~(\ref{LEQ:Pot0}).
In the proposed potential, the parameter $m$ controls the sharpness of the valley;
thereby, for $m\gg 1$, the two curves with
$\mathrm{C}_{\infty}: \{\boldsymbol{x}\mid v_0(\boldsymbol{x})=1\}$
 and $\mathrm{E}: \{\boldsymbol{x}\mid v_1(\boldsymbol{x})=E\}$
determine a skeleton of the 2D ratchet potential, and
the eigenvalues of the Hessian matrix are expressed approximately
in terms of the quantities derived from
$\mathrm{C}_{\infty}$ and $\mathrm{E}$ (Sec.~\ref{TT-Model}).
These enable us to easily design a strategy for maximizing
the performance indexes [$L$ (MAM), $\omega$ (MAV), and $\eta$ (efficiency)].

From the analytic expressions for $L$ and $\omega$ 
(Secs.~\ref{Expressions}),
we have specified the factor $-\boldsymbol{x}_{+}\cdot\boldsymbol{n}_{+}$ 
as the main objective function to maximize,
and $H_n$ as the optional one to minimize within
the appropriate range for the local equilibrium condition.
Quantities $-\boldsymbol{x}_{+}\cdot\boldsymbol{n}_{+}$ and $H_n$ are
relevant to the asymmetry of the potential profile along the pathway
and the curvature at the potential minimum, respectively.
Through the optimization of $v_1(\boldsymbol{x})$,
the procedure to maximize the main factor
$-\boldsymbol{x}_{+}\cdot\boldsymbol{n}_{+}$ consists of
$\mathrm{G}_1$ [Eq.~(\ref{G1})] and $\mathrm{G}_2$ [Eq.~(\ref{force_2b})]
(Sec.~\ref{sec:OPT0}), and
the one to minimize $H_n$ consists of $\mathrm{G}_3$ [Eq.~(\ref{CondG3})] and its moderation
[Eq.~(\ref{Det:d})] (Sec.~\ref{sec:LamNonZero}).
The moderation of $\mathrm{G}_3$ is required to prevent the creation of temporal minima.
We reason that such temporal minima cause extra dissipation that is observed
as another peak in $P_h$; the relaxation parameter $\epsilon$
[Eq.~(\ref{Det:d})] is determined
so that $P_h$ has no additional peak in the plot for the noise intensity $D$.
Although the proposed optimization method has 
been implemented on the basis of the two-tooth ratchet model,
it is applicable to three-tooth or other similar ratchet models (Sec.~\ref{Discuss})
if $\mathrm{G}_2$ is generalized as in Eq.~(\ref{3th:force_2}).

The outcomes of the optimization have been shown 
in Secs.~\ref{sec:LamZero} and \ref{NumericalResults} 
for the cases of $\mathrm{C}_{\infty}$ given by elliptic or nonelliptic
curves.
The analytical expressions for the maximized $L$, $\omega$, and $\eta$
are shown in the elliptic case (Secs.~\ref{sec:LamZero} and \ref{App:LamZero}).
Consistent with the numerical simulation results in Sec.~\ref{LamZero:num},
these suggest that the peaks of
$L$, $\omega$, and $\eta$ increase as the diameter or eccentricity of
the ellipse becomes larger. A note for applying such larger values of diameter
or eccentricity is that $h$ and $\Omega$ must be sufficiently small to retain
the local equilibrium.
In the nonelliptic case (Sec.~\ref{sec:LamNonZero}),
the optimization
procedure $\mathrm{G}_3$ with the moderation is useful; compared with
no moderation, it improves the efficiency with a suitable choice of 
the relaxation parameter.
In comparing the efficiency between the elliptic ($\lambda=0$) and
nonelliptic ($\lambda\ne 0$) cases under the same condition of $(m,a,b)$,
we have seen that the best result in the latter case exhibits a higher peak than the best one
in the former.
This suggests that a more sophisticated design of $v_0(\boldsymbol{x})$,
incorporating higher-order harmonic deformations,
could improve efficiency.

\begin{acknowledgments}
We used the supercomputer of the Academic Center for Computing and Media Studies (ACCMS) at Kyoto University in this research.
\end{acknowledgments}

\appendix

\section{correlation matrix of randomly directed force}
\label{OnNoise}

We consider the time correlation matrix for
$\boldsymbol{N}_t=(\cos\Phi_t,\sin\Phi_t)^{\transpose}$, which changes
its direction randomly at the rate $\Omega$ independent of the
current direction. The angle $\Phi_t \in [0,2\pi)$ is a stationary Markov jump
process, whose conditional probability density
for the transition from $\Phi_{t}=\phi'$ during an infinitesimal interval $\Delta t >0$
obeys
$p_{\Phi}(\phi,t+\Delta t\mid\phi',t) = (1- \Delta t \Omega )\delta(\phi-\phi')+
\Delta t \Omega p_{\Phi}(\phi) + o(\Delta t)$ 
with non-negative $p_{\Phi}(\phi)$ satisfying
$\int_0^{2\pi}p_{\Phi}(\phi)d\phi=1$ and
the Dirac's delta function $\delta(\cdot)$.
This leads to the master equation for $p_{\Phi}(\phi,t\mid\phi',t')$ ($t\ge t'$) as
\begin{equation}
\partial_t  p_{\Phi}(\phi,t\mid\phi',t')= - \Omega
 p_{\Phi}(\phi,t\mid\phi',t')
+\Omega p_{\Phi}(\phi).
\end{equation}
It is obvious that the
stationary probability density of $\Phi_t$ coincides with
$p_{\Phi}(\phi)$.

The master equation is solved as
\begin{equation}
p_{\Phi}(\phi,t\mid\phi',t')= p_{\Phi}(\phi)+e^{-\Omega (t-t')} 
\left\{\delta(\phi-\phi') -  p_{\Phi}(\phi)\right\} .
\end{equation}
For $A_t\equiv A(\Phi_t)$ and $B_t\equiv B(\Phi_t)$, where $A(\phi)$
and $B(\phi)$ are any functions of $\phi$, the statistical average of
$A_tB_0$ ($t\ge0$) with respect to $\{\Phi_t\}$ reads as
\begin{align}
\left\langle A_t B_0\right\rangle_{\Phi}
 &=
\int_{0}^{2\pi} d\phi
\int_{0}^{2\pi} d\phi'
 A(\phi) B(\phi') p_{\Phi}(\phi,t\mid\phi',0)p_{\Phi}(\phi')
\nonumber
\\
 &=
\int_{0}^{2\pi} d\phi
\int_{0}^{2\pi} d\phi'
 A(\phi) B(\phi') 
\left[
p_{\Phi}(\phi)+e^{-\Omega t} 
\left\{\delta(\phi-\phi') - p_{\Phi}(\phi)\right\}
\right]p_{\Phi}(\phi')
\nonumber
\\
 &=
(1-e^{-\Omega t})\langle A_0\rangle_{\Phi}
\langle B_0\rangle_{\Phi}
+
\langle A_0B_0\rangle_{\Phi} e^{-\Omega t} ,
\end{align}
which leads to the time correlation function for $t\ge0$:
\begin{equation}
\label{eq:s-corr-ab}
\left\langle A_t B_0\right\rangle_{\Phi}
-\langle A_0\rangle_{\Phi}
\langle B_0\rangle_{\Phi}
=
 e^{-\Omega t}\{\langle A_0B_0\rangle_{\Phi}
-\langle A_0\rangle_{\Phi}
\langle B_0\rangle_{\Phi}\}.
\end{equation}
The rotational symmetry for $\boldsymbol{N}_t$, i.e.,
$p_{\Phi}(\phi)=\frac{1}{2\pi}$, is further assumed in the paper, which
leads to $\left\langle \cos \Phi_0 \sin \Phi_0\right\rangle_{\Phi}
=\left\langle \sin \Phi_0 \cos \Phi_0\right\rangle_{\Phi} =0$ and
$\left\langle \cos \Phi_0 \cos \Phi_0\right\rangle_{\Phi}
=\left\langle \sin \Phi_0 \sin \Phi_0\right\rangle_{\Phi} =1/2$, and
thus, by Eq.~(\ref{eq:s-corr-ab}),
$\left\langle\boldsymbol{N}_{t}\boldsymbol{N}_{0}^{\transpose}
\right\rangle_{\Phi} = (e^{-\Omega t}/2)\Hat{1}$
(Eq.~(\ref{cor_NN})).

\section{\label{master} Transition Rates}

The transition rate $W(\sigma,\mu,t)$ [$\sigma,\mu\in\{+,-\}$]
in Eq.~(\ref{W_sig_mu})
is derived based on Langer's
method \cite{PhysRevLett.21.973,RevModPhys.62.251}.
Let $\Tilde{\mathrm{B}}_{\epsilon}^{\mu}$ be
a narrow band region with thickness $2\epsilon$ inside which the ridge curve
$\Tilde{\mathrm{B}}^{\mu}$ is contained
(see Fig.~\ref{fig:PotShapes}, or
Fig.~6 in \cite{doi:10.7566/JPSJ.84.044004}). In the regions
$\Tilde{\mathrm{B}}_{\epsilon}^{\mu}$, the current density
$\boldsymbol{J}(\boldsymbol{x},t)$ is concentrated by bottleneck
structures, whereas, in the central region 
of $\mathrm{D}_{\epsilon}^{\mu}$, $\boldsymbol{J}(\boldsymbol{x},t)$
 can be regarded approximately as vanishing.
Thus, we may specify locally non-equilibrium or
equilibrium regions either inside or outside
$\Tilde{\mathrm{B}}_{\epsilon}^{\mu}$. On each region, we
assume $\boldsymbol{J}(\boldsymbol{x},t)$ as follows \cite{doi:10.7566/JPSJ.84.044004}.
\begin{enumerate}
\item[{\bf A.}]
In the domain $\mathrm{D}_{\sigma}^{\mu}$ complementary to
$\Tilde{\mathrm{B}}_{\epsilon}^{\mu}$, i.e.,
$\mathrm{D}_{\sigma}^{\mu} \setminus
\Tilde{\mathrm{B}}_{\epsilon}^{\mu}$,
we assume $\boldsymbol{J}(\boldsymbol{x},t)\approx \boldsymbol{0}$, i.e.,
$p(\boldsymbol{x},t)$ approximately obeys 
the thermal equilibrium probability density
function.
Then, we have
$p(\boldsymbol{x},t)\approx
e^{-V(\boldsymbol{x},t)/D}e^{V(\boldsymbol{y},t)/D}p(\boldsymbol{y},t)
$
for
$\boldsymbol{x},\boldsymbol{y} \in
\mathrm{D}_{\sigma}^{\mu} \setminus \Tilde{\mathrm{B}}_{\epsilon}^{\mu}$.
From Eq.~(\ref{Markov:Psig_mu}),
this leads to
\begin{equation}
P(\sigma,\mu,t)
\approx
\int_{\boldsymbol{x}\in \mathrm{D}_{\sigma}^{\mu}} 
d\boldsymbol{x}\,
 e^{-V(\boldsymbol{x},t)/D}e^{V(\boldsymbol{y},t)/D}
 p(\boldsymbol{y},t),
\label{Mathrm:Psig_mu}
\end{equation}
where we assume $p(\boldsymbol{x},t)\approx 0$
for $\boldsymbol{x}\in\Tilde{\mathrm{B}}_{\epsilon}^{\mu}$.
Also, we have
\begin{equation}
P(\sigma,t)
\approx
\int_{\boldsymbol{x}\in \mathrm{D}_{\sigma}}
d\boldsymbol{x}\,
e^{-V(\boldsymbol{x},t)/D}e^{V(\boldsymbol{y},t)/D}
p(\boldsymbol{y},t).
\label{Mathrm:Psig}
\end{equation}
 \item[{\bf B.}] 
Consider a family of curves that are parallel to
the curve $\Tilde{\mathrm{B}}^{\mu}$ in
$\Tilde{\mathrm{B}}_{\epsilon}^{\mu}$, and unit vectors
$\Tilde{\boldsymbol{\tau}}_{\sigma}^{\mu}(\boldsymbol{x})$ and
$\Tilde{\boldsymbol{n}}_{\sigma}^{\mu}(\boldsymbol{x})$
that are tangent and normal, respectively, to such a curve passing through
a point $\boldsymbol{x}\in \Tilde{\mathrm{B}}_{\epsilon}^{\mu}$.
Then, we assume that a current can arise along the vector field
$\Tilde{\boldsymbol{n}}_{\sigma}^{\mu}(\boldsymbol{x})$,
while an equilibrium condition is retained along the direction
$\Tilde{\boldsymbol{\tau}}_{\sigma}^{\mu}(\boldsymbol{x})$.
Namely, we have
$\Tilde{\boldsymbol{\tau}}_{\sigma}^{\mu}(\boldsymbol{x})\cdot
\boldsymbol{J}(\boldsymbol{x},t)=0$ and
$\Tilde J_{\mu}\equiv\Tilde{\boldsymbol{n}}_{\sigma}^{\mu}(\boldsymbol{x})\cdot
\boldsymbol{J}(\boldsymbol{x},t)$ in which
$\Tilde J_{\mu}$ is a constant on a curve perpendicularly crossing
the family of the curves parallel to $\Tilde{\mathrm{B}}^{\mu}$
($\Tilde J_{\mu}$ depends on the coordinate on $\Tilde{\mathrm{B}}^{\mu}$).
Therefore, $J^{\mu}(t)$ in Eq.~(\ref{def:J_mu}) reads as
\begin{equation}
J^{\mu}(t) \approx
\left(
\delta_{\sigma,-\mu}-
\delta_{\sigma,\mu}
\right)
\int_{\boldsymbol{x}\in \Tilde{\mathrm{B}}^{\mu}} 
d\boldsymbol{x}\,
\Tilde J_{\mu}.
\label{def:Jhat}
\end{equation}
\end{enumerate}

To estimate the integration in Eq.~(\ref{Mathrm:Psig_mu}),
let us define a local coordinate system
$\boldsymbol{x} = \boldsymbol{x}_{\sigma}+
\sigma\mu ( \xi\boldsymbol{\tau}_{\sigma}
+ \eta\boldsymbol{n}_{\sigma}) $ near
$\boldsymbol{x}_{\sigma}$ with the unit tangential and normal
vectors to $\mathrm{B}_{\sigma}$, $\boldsymbol{\tau}_{\sigma}$, and
$\boldsymbol{n}_{\sigma}$, at $\boldsymbol{x}=\boldsymbol{x}_{\sigma}$,
as eigenvectors of $\Hat{G}(\boldsymbol{x}_{\sigma})
=\partial_{\boldsymbol{x}}\partial_{\boldsymbol{x}}^{\transpose}
V(\boldsymbol{x}_{\sigma},t)$.
Here, the values of $\sigma$ and $\mu$,  ``$+$'' and ``$-$'',
are mapped to the numbers $+1$ and $-1$, respectively.
Then, we expand $V(\boldsymbol{x},t)$ as
\begin{gather}
V(\boldsymbol{x},t)\approx
V(\boldsymbol{x}_{\sigma},t)
 -\mu\boldsymbol{f}_{\sigma}\cdot\left(
\xi\boldsymbol{\tau}_{+}
+\eta\boldsymbol{n}_{+} \right)+ 
\frac{1}{2}
\Lambda_{\tau}(\boldsymbol{x}_{\sigma})\xi^{2}
+\frac{1}{2}
\Lambda_{n}(\boldsymbol{x}_{\sigma})
\eta^{2}
,
\end{gather}
where
$\boldsymbol{f}_{\sigma}\equiv\boldsymbol{f}_{I}(\boldsymbol{x}_{\sigma})
+h\boldsymbol{N}_{t}$.
Note that
the eigenvalues of $\Hat{G}(\boldsymbol{x}_{\sigma})$,
$\Lambda_{\tau}(\boldsymbol{x}_{\sigma})$,
and
$\Lambda_{n}(\boldsymbol{x}_{\sigma})$
depend on $I$.
Since $h$ and $D$ are assumed to be small,
neglecting the terms of $O(h^2)$, we estimate the integration
in Eq.~(\ref{Mathrm:Psig_mu}) as
\begin{align}
\int_{\boldsymbol{x}\in \mathrm{D}_{\sigma}^{\mu}} 
d\boldsymbol{x}
 e^{- \frac{V(\boldsymbol{x},t)}{D}}
&\approx
 e^{-\frac{V(\boldsymbol{x}_{\sigma},t)}{D}}
\int_{0}^{\infty}
d\eta
\int_{-\infty}^{\infty}
d\xi
 e^{-
\frac{
\Lambda_{\tau}(\boldsymbol{x}_{\sigma})\xi^{2}+
\Lambda_{n}(\boldsymbol{x}_{\sigma})\eta^{2}
}{2D}}
\left(
1+
\frac{\mu
\boldsymbol{f}_{\sigma}\cdot\boldsymbol{n}_{+}
}{D}\eta
\right)
\nonumber
\\
&
\approx
\frac{e^{-\frac{V(\boldsymbol{x}_{\sigma},t)}{D}}}{2}
\sqrt{\frac{2\pi D}{\Lambda_{\tau}(\boldsymbol{x}_{\sigma})}}
\left(
\sqrt{\frac{2\pi D}{\Lambda_{n}(\boldsymbol{x}_{\sigma})}}
+
\frac{2\mu\boldsymbol{f}_{\sigma}\cdot\boldsymbol{n}_{+}}
{\Lambda_{n}(\boldsymbol{x}_{\sigma})} 
\right),
\label{App:IntegraEV}
\end{align}
where we have used the Gaussian integral approximation
by the replacement
$
\int_{\boldsymbol{x}\in \mathrm{D}_{\sigma}^{\mu}}
d\boldsymbol{x}
\rightarrow
\int_{-\infty}^{\infty}
\int_{0}^{\infty}
d\xi d\eta
$.

Substituting Eq.~(\ref{App:IntegraEV}) to
Eqs.~(\ref{Mathrm:Psig_mu}) and (\ref{Mathrm:Psig}),
we obtain
\begin{gather}
P(\sigma,t)
\approx
\frac{2\pi D}{\sqrt{\Lambda_{\tau}(\boldsymbol{x}_{\sigma})
\Lambda_{n}(\boldsymbol{x}_{\sigma})}}
 e^{-V(\boldsymbol{x}_{\sigma},t)/D}
e^{V(\boldsymbol{y},t)/D}
p(\boldsymbol{y},t),
\label{Mathrm:Psig_mu2}
\\
P(\sigma,\mu,t)
\approx
Q(\mu\mid\sigma,t)
P(\sigma,t),
\\
Q(\mu\mid\sigma,t)
\approx
\frac{1}{2}
\left(
1+
\frac{2\mu\boldsymbol{f}_{\sigma}\cdot\boldsymbol{n}_{+}}
{\sqrt{2\pi D \Lambda_{n}(\boldsymbol{x}_{\sigma}) }}
\right).
\label{App:Q_sig_mu}
\end{gather}

Similarly, on the local coordinate system near 
$\Tilde{\boldsymbol{x}}^{\mu} \in \Tilde{\mathrm{B}}^{\mu}$,
$\boldsymbol{x} = \Tilde{\boldsymbol{x}}^{\mu}+
 \xi\Tilde{\boldsymbol{\tau}}^{\mu}_{\sigma}
+\eta\Tilde{\boldsymbol{n}}^{\mu}_{\sigma}
$, where $\Tilde{\boldsymbol{\tau}}_{\sigma}^{\mu}\equiv
\Tilde{\boldsymbol{\tau}}_{\sigma}^{\mu}(\Tilde{\boldsymbol{x}}^{\mu})$ and
$\Tilde{\boldsymbol{n}}_{\sigma}^{\mu}\equiv
\Tilde{\boldsymbol{n}}_{\sigma}^{\mu}(\Tilde{\boldsymbol{x}}^{\mu})$
(see Sec.~\ref{DefStates}),
we expand $V(\boldsymbol{x},t)$ as
\begin{gather}
V(\boldsymbol{x},t) \approx V(\Tilde{\boldsymbol{x}}^{\mu},t)
+V_{\tau}(\xi,t)+V_{n}(\eta,t),
\\
V_{\tau}(\xi,t)\equiv
\frac{1}{2}
\Lambda_{\tau}(\Tilde{\boldsymbol{x}}^{\mu})
\xi^{2},
\quad
V_{n}(\eta,t) \equiv \frac{1}{2}
\Lambda_{n}(\Tilde{\boldsymbol{x}}^{\mu})
\eta^{2}.
\end{gather}
Because $\Tilde{\boldsymbol{\tau}}_{\sigma}^{\mu}\cdot
\boldsymbol{J}(\boldsymbol{x},t)=0$, or
\begin{equation}
0= \left[
-
\partial_{\xi}
V(\boldsymbol{x},t)
\right]
p(\boldsymbol{x},t)
-
D
\partial_{\xi}
p(\boldsymbol{x},t),
\end{equation}
then by separation of variables, we have
$
p(\boldsymbol{x},t)\equiv
\exp\left[- \frac{V_{\tau}(\xi,t)}{D}\right]
p_{n}(\eta,t)
$
for $\boldsymbol{x}\in\Tilde{\mathrm{B}}_{\epsilon}^{\mu}$.

Multiplying $\Tilde J_{\mu}=\Tilde{\boldsymbol{n}}_{\sigma}^{\mu}(\boldsymbol{x})\cdot
\boldsymbol{J}(\boldsymbol{x},t)$ by $e^{V_{n}(\eta,t)/D}$, and 
integrating over $\eta$ in the range $[-\epsilon,\epsilon]$, we obtain
\begin{equation}
\int_{-\epsilon}^{\epsilon}d\eta\, e^{V_{n}(\eta,t)/D}
\Tilde J_{\mu}
= 
\int_{-\epsilon}^{\epsilon}d\eta\, e^{V_{n}(\eta,t)/D}
\left\{
\left[
-
\partial_{\eta}
V(\boldsymbol{x},t)
\right]
p(\boldsymbol{x},t)
-
D
\partial_{\eta}
p(\boldsymbol{x},t)
\right\}.
\end{equation}
From the assumption for $\Tilde J_{\mu}$, this leads to
\begin{equation}
\Tilde J_{\mu}=
\frac{D}
{
\int_{-\epsilon}^{\epsilon}dy\, e^{V_{n}(y,t)/D}
}
\left.
 \exp\left[
\frac{
-
V_{\tau}(\xi,t)
+
V_{n}(\eta,t)
}{D}
\right]
p_{n}(\eta,t)
\right\rvert_{\eta=\epsilon}^{\eta=-\epsilon}.
\label{App:J_mu}
\end{equation}
From Eq.~(\ref{Mathrm:Psig_mu2}), we have
\begin{align}
e^{V(\boldsymbol{y},t)/D}
 p(\boldsymbol{y},t)
\bigr\rvert_{\boldsymbol{y}=\Tilde{\boldsymbol{x}}^{\mu}+\epsilon\mu\Tilde{\boldsymbol{n}}_{\sigma}}
&\approx
\frac{\sqrt{
\Lambda_{\tau}(\boldsymbol{x}_{\mu\sigma})
\Lambda_{n}(\boldsymbol{x}_{\mu\sigma})
}}{2\pi D
 e^{-V(\boldsymbol{x}_{\mu\sigma},t)/D}
}
P(\mu\sigma,t).
\end{align}
Applying this to
$
 e^{V_{n}(\eta,t)/D} p_{n}(\eta,t)
\bigr\rvert_{\eta=\epsilon}^{\eta=-\epsilon}
=
e^{\{V(\boldsymbol{y},t)-
V(\Tilde{\boldsymbol{x}}^{\mu},t)\}/D}
 p(\boldsymbol{y},t)
\bigr\rvert_{\boldsymbol{y}=
\Tilde{\boldsymbol{x}}^{\mu}+\epsilon \mu\Tilde{\boldsymbol{n}}_{\sigma}}
^{\boldsymbol{y}=
\Tilde{\boldsymbol{x}}^{\mu}-\epsilon \mu\Tilde{\boldsymbol{n}}_{\sigma}}
$ in Eq.~(\ref{App:J_mu}), we obtain
\begin{align}
\Tilde J^{\mu}=
W_{\xi}(\sigma,\mu,t)P(-\sigma,t)
-
W_{\xi}(-\sigma,\mu,t)P(\sigma,t)
\end{align}
with
\begin{align}
W_{\xi}(\sigma,\mu,t)
\equiv
\frac{1}{2\pi}
\frac{
\sqrt{
\Lambda_{\tau}(\boldsymbol{x}_{-\sigma})
\Lambda_{n}(\boldsymbol{x}_{-\sigma})
}
}{
\int_{-\epsilon}^{\epsilon}dy\, e^{V_{n}(y,t)/D}
}
\exp\left\{
\frac{
V(\boldsymbol{x}_{-\sigma},t)
-
V_{\tau}(\xi,t)
-
V(\Tilde{\boldsymbol{x}}^{\mu},t)
}{D}
\right\}.
\label{App:W0}
\end{align}

From Eqs.~(\ref{def:Jhat}) and (\ref{App:W0}), the transition rate
$W(\sigma,\mu,t)$ in Eq.~(\ref{J_mu}) is found to be
\begin{equation}
W(\sigma,\mu,t)
\approx
\frac{1}{2\pi}
e^{
-\left[
V(\boldsymbol{x}^{\mu},t) -V(\boldsymbol{x}_{-\sigma},t)
\right]/D
}
\sqrt{
\frac{
\Lambda_{\tau}(\boldsymbol{x}_{-\sigma})
\Lambda_{n}(\boldsymbol{x}_{-\sigma})
\lvert
\Lambda_{n}(\Tilde{\boldsymbol{x}}^{\mu})
\rvert}{
\Lambda_{\tau}(\Tilde{\boldsymbol{x}}^{\mu})
}}.
\label{App:W_sig_mu}
\end{equation}
Here, we have approximated
$
\int_{-\epsilon}^{\epsilon}dy\, e^{V_{n}(y,t)/D}
$ and
$
\int_{\boldsymbol{x}\in \mathrm{B}^{\mu}} d\xi\,
e^{-V_{\tau}(\xi,t)/D}
$
with the Gaussian integrals
$
\int_{-\infty}^{\infty}d\eta\, e^{
\Lambda_{n}(\Tilde{\boldsymbol{x}}^{\mu})\eta^{2}/(2D)}
=
\sqrt{\frac{2\pi D}{\lvert
\Lambda_{n}(\Tilde{\boldsymbol{x}}^{\mu})\rvert}}
$ and
$
\int_{-\infty}^{\infty}d\xi\, e^{-
\Lambda_{\tau}(\Tilde{\boldsymbol{x}}^{\mu})
\xi^{2}/(2D)}
=
\sqrt{\frac{2\pi D}{
\Lambda_{\tau}(\Tilde{\boldsymbol{x}}^{\mu})}
}
$, 
respectively.
We have also replaced
$V(\Tilde{\boldsymbol{x}}^{\mu},t)$ with
$V(\boldsymbol{x}^{\mu},t)$, because, from
$\Tilde{\boldsymbol{x}}^{\mu}-\boldsymbol{x}^{\mu}\sim O(h)$,
$V(\Tilde{\boldsymbol{x}}^{\mu},t)
=V(\boldsymbol{x}^{\mu},t)+O(h^2)$.
Then
we obtain Eq.~(\ref{W_sig_mu}).

\section{Linear response approximations}
\label{LRT}

In this section, $J^{\mu}(t)$, $P(\sigma,t)$, and $Q(\mu,t)$,
which are required in the calculations for
$L$, $\omega$, and $P_h$,
are estimated within a linear response approximation for small $h$ and $I$.
For those estimations in $O(h)$ and $O(I)$, we employ
\begin{equation}
\partial_t P(\sigma,\mu,t)
\approx
\delta_{\sigma,-\mu}
J^{\mu}(t) -
\delta_{\sigma,\mu}
J^{\mu}(t),
\label{P_sig_mu}
\end{equation}
assuming $J_{\sigma}^{\mu}(t) \sim O(h^2)$ [which is confirmed later in Eq.~(\ref{lnPQJ})]
in Eqs.~(\ref{DP}) and (\ref{def:J'}).
We expand $P(\sigma,t)$ and $W(\sigma,\mu,t)$
in Eqs.~(\ref{J_mu}) and (\ref{W_sig_mu}) as
\begin{gather}
P(\sigma,t) \approx P_0(\sigma) + P_{1}(\sigma,t),
\label{P_div}
\\
W(\sigma,\mu,t)
\approx
W_{0}
\left[
1
+
\frac{h}{D}
\boldsymbol{N}_{t}\cdot 
(\boldsymbol{x}^{\mu}-\boldsymbol{x}_{-\sigma})
-I
\frac{
\theta(\boldsymbol{x}^{\mu}) -\theta(\boldsymbol{x}_{-\sigma}) 
}{2\pi D}
\right],
\label{W_sig_mu2}
\end{gather}
where the first and second 
[and the third in Eq.~(\ref{W_sig_mu2})] terms
are of zeroth- and first-order in $h$ and $I$, respectively;
normalizations
$\sum_{\sigma} P_0(\sigma) = 1$ and
$\sum_{\sigma} P_1(\sigma,t) = 0$ are assumed.
Term $W_0$, defined in Eq.~(\ref{def:W0}), represents the rate of barrier-crossing events
under the thermal activation in the absence
of the load and the external field.
In the expansion for Eq.~(\ref{W_sig_mu2}), the eigenvalues of
$\Hat{G}(\boldsymbol{x})$ in Eq.~(\ref{W_sig_mu})
are replaced with those of
$
\partial_{\boldsymbol{x}}\partial_{\boldsymbol{x}}^{\transpose}
V_{0}(\boldsymbol{x})
$, for simplicity.

Substituting Eqs.~(\ref{P_div}) and (\ref{W_sig_mu2})
into Eq.~(\ref{J_mu}), we obtain
$P_{0}(\sigma) = 1/2$ from the zeroth-order equality, 
and, up to $O(h)$ and $O(I)$,
\begin{align}
J^{\mu}(t)
\approx
 W_{0} 
\biggl[
&
P_{1}(\mu,t)
-
P_{1}(-\mu,t)
-
\frac{\mu h}{D}
\boldsymbol{N}_{t}\cdot\boldsymbol{x}_{+}
-
\frac{
  I
}{4D}
\biggr].
\label{eq:J_mu_1}
\end{align}
Note that
we have
$\boldsymbol{x}_{-\sigma}=-\boldsymbol{x}_{\sigma}$
from the two-fold symmetry, and, since
$\theta(\boldsymbol{x}^{\mu})-\theta(\boldsymbol{x}_{-\sigma}) 
= \angle  \boldsymbol{x}_{-\sigma} \mathrm{O} \boldsymbol{x}^{\mu} 
$ denoting the angle from $\boldsymbol{x}_{-\sigma}$ to
$\boldsymbol{x}^{\mu}$,
$\angle \boldsymbol{x}_{\mu}\mathrm{O}
\boldsymbol{x}^{\mu} > 0$ and
$\angle \boldsymbol{x}_{-\mu}\mathrm{O}
\boldsymbol{x}^{\mu} < 0$,
we have
$
\theta(\boldsymbol{x}^{\mu})-\theta(\boldsymbol{x}_{\mu}) 
-[
\theta(\boldsymbol{x}^{\mu})-\theta(\boldsymbol{x}_{-\mu})
] =\pi$.

Applying this to
$\partial_t P_1(\sigma,t) \approx
J^{-\sigma}(t)-J^{\sigma}(t)$
from Eq.~(\ref{P_sig_mu}),
we find 
\begin{gather}
P_1(\sigma,t)=\sigma\int_{-\infty}^{t} ds K(t-s) F_s,
\label{P1_sig}
\end{gather}
where 
$K(t) =  e^{-4W_0 t}$ ($t\geq 0$) and
$F_t = \frac{2 h W_{0}}{D}  \boldsymbol{N}_{t}\cdot \boldsymbol{x}_{+}$.
Hence, we obtain
\begin{align}
P(\sigma,t)&\approx
\frac{1}{2}
\left[
1+2\sigma\int_{-\infty}^{t} ds K(t-s)F_s
\right]
.
\label{P_sig_app}
\end{align}
Assuming the local equilibrium around the potential minima,
$P(\sigma,\mu,t)$ and $Q(\mu\mid\sigma,t)$ are found as
\begin{gather}
P(\sigma,\mu,t)
\approx
Q(\mu\mid\sigma,t)
P(\sigma,t),
\\
Q(\mu\mid\sigma,t)\approx\frac{1}{2}
\left(
1
+
\frac{2\mu\boldsymbol{f}_{\sigma}\cdot\boldsymbol{n}_{+}}
{\sqrt{2\pi D \Lambda_{n}(\boldsymbol{x}_{\sigma})}}
\right),
\label{eq:Q_sig_mu}
\end{gather}
where $\boldsymbol{f}_{\sigma}\equiv 
\boldsymbol{f}_{I}(\boldsymbol{x}_{\sigma})
+h\boldsymbol{N}_t$ (Eq.~(\ref{App:Q_sig_mu}) in Appendix~\ref{master}).
Therefore, substituting
 Eqs.~(\ref{P_sig_app}) and (\ref{eq:Q_sig_mu}) into
$Q(\mu,t)=\sum_{\sigma\in\{\mu,-\mu\}} Q(\mu\mid\sigma,t)P(\sigma,t)$,
we find
\begin{equation}
Q(\mu,t) \approx
\frac{1}{2}
\biggl(
1+ 
\frac{
2 \mu h\boldsymbol{N}_t\cdot\boldsymbol{n}_{+}
}{\sqrt{2\pi D H_{n}}}
\biggr).
\label{Q_mu_app}
\end{equation}
Substituting Eq.~(\ref{P_sig_app}) into
Eq.~(\ref{eq:J_mu_1}), we obtain
\begin{align}
J^{\mu}(t)
\approx &
2\mu W_{0}\int_{-\infty}^{t} ds K(t-s)F_s
-\frac{\mu}{2} F_t - \frac{IW_{0}}{4D}
.
\label{J_mu_app}
\end{align}

\subsection{Calculations of MAM ($L$) and MAV ($\omega$)}
\label{App:MAM}

Firstly, $L^{(I)}$ and $\omega^{(I)}$ in
Eqs.~(\ref{LIdef:LI}) and (\ref{Ex_omg|I}) are calculated as follows.
From Eqs.~(\ref{cor_NN}) and (\ref{J_mu_app}), we have
$\overline{J^{\mu}(t)}=\langle J^{\mu}(t) \rangle_{\Phi}=-\frac{IW_0}{4D}$, and
\begin{align}
L^{(I)}
&\approx 
-
\frac{ g_{L} W_{0}I}{2D}
\left(
\boldsymbol{x}_{+}\times\boldsymbol{x}^{+}
\right)_{z},
\label{LI:1}
\\ 
\omega^{(I)}
&\approx
-
\frac{ \pi g_{O}  W_{0}I}{2D}.
\label{omega:1}
\end{align}
Note that 
$\left(\boldsymbol{x}_{+}\times \boldsymbol{x}^{+}\right)_{z} > 0$.

Terms $L^{(h)}$ and $\omega^{(h)}$ are 
 approximated up to $O(h^2)$ as follows.
From Eqs.~(\ref{P_sig_app}) and (\ref{Q_mu_app}), up to $O(h)$,
$\ln \frac{P(\sigma,t)}{Q(\mu,t)}$ in
Eqs.~(\ref{Lt}) and (\ref{App:PI_h}) reads as
\begin{align}
\ln\frac{P(\sigma,t)}{ Q(\mu,t)} 
\approx &
2\sigma\int_{-\infty}^{t} ds K(t-s)F_s
-
\frac{2\mu h
\boldsymbol{N}_t\cdot\boldsymbol{n}_{+}
}{\sqrt{2\pi D  H_{n}}}
.
\label{Lt_1}
\end{align}
Thus, we have
\begin{equation}
\overline{
\left[
 \ln\frac{P(\mu,t)}{Q(\mu,t)}
+
 \ln\frac{P(-\mu,t)}{Q(\mu,t)}
\right]
J^{\mu}(t)
}
=
-
\frac{4\mu h}{\sqrt{2\pi D  H_{n}}}
\overline{
\left(
\boldsymbol{N}_t\cdot\boldsymbol{n}_{+}
\right)
J^{\mu}(t) 
}
,
\label{Lt_2}
\end{equation}
and, from Eq.~(\ref{J_mu_app}),
\begin{align}
\mu
\overline{
\left(
\boldsymbol{N}_t\cdot\boldsymbol{n}_{+}
\right)
J^{\mu}(t) 
}
&=
\mu
\overline{
\left(
\boldsymbol{N}_t\cdot\boldsymbol{n}_{+}
\right)
\left[
2\mu W_{0}\int_{-\infty}^{t} ds K(t-s)F_s
-\frac{\mu}{2} F_t
\right]
}
\nonumber
\\
&=
2 W_{0}\int_{-\infty}^{t} ds e^{-4W_0(t-s)}
\left\langle
F_s
\left(
\boldsymbol{N}_t\cdot\boldsymbol{n}_{+}
\right)
 \right\rangle_{\Phi}
-
\frac{1}{2}
\left\langle
 F_t
\left(
\boldsymbol{N}_t\cdot\boldsymbol{n}_{+}
\right)
 \right\rangle_{\Phi}.
\label{App:<>_p}
\end{align}
From Eqs.~(\ref{cor_NN}) and (\ref{P1_sig}), we also have
\begin{align*}
\left\langle
 F_s
\left(
\boldsymbol{N}_t\cdot\boldsymbol{n}_{+}
\right) 
\right\rangle_{\Phi}
&=
\frac{2hW_{0}}{D}
\left\langle
\left(
\boldsymbol{N}_s\cdot\boldsymbol{x}_{+}
\right)
\left(
\boldsymbol{N}_t\cdot\boldsymbol{n}_{+}
\right)
\right\rangle_{\Phi}
\\
&=
\frac{hW_{0}}{D}
\left(
\boldsymbol{x}_{+}\cdot\boldsymbol{n}_{+}
\right)
e^{-\Omega (t-s)}
,
\end{align*}
and
\begin{align*}
\left\langle
 F_t
\left(
\boldsymbol{N}_t\cdot\boldsymbol{n}_{+}
\right)
\right\rangle_{\Phi}
&=
\frac{hW_{0}}{D}
\boldsymbol{x}_{+}\cdot\boldsymbol{n}_{+}
.
\end{align*}
Substituting these into Eq.~(\ref{App:<>_p}),
we find
\begin{align*}
\mu
\overline{
\left(
\boldsymbol{N}_t\cdot\boldsymbol{n}_{+}
\right)
J^{\mu}(t) 
}
&=
\frac{2h W_{0}^{2}}{D}
\left(\boldsymbol{x}_{+}\cdot\boldsymbol{n}_{+}\right)
\int_{-\infty}^{t} ds e^{-4W_{0}(t-s)}
 e^{-\Omega(t-s)}
-
\frac{h W_{0}}{2D}
\left(\boldsymbol{x}_{+}\cdot\boldsymbol{n}_{+}\right)
\\
\nonumber
&=
-
\frac{h W_{0}}{2D}
\frac{\Omega}{\Omega+4W_{0}}
\left(\boldsymbol{x}_{+}\cdot\boldsymbol{n}_{+}\right)
.
\end{align*}
Thus, Eq.~(\ref{Lt_2}) reads as
\begin{equation}
\overline{
\left[
 \ln\frac{P(\mu,t)P(-\mu,t)}{Q(\mu,t)^2}
\right]
J^{\mu}(t)
}
=
\frac{2 h^2}{D\sqrt{2\pi D  H_{n}}}
\frac{\Omega W_{0} }{\Omega+4W_{0}}
\left(\boldsymbol{x}_{+}\cdot\boldsymbol{n}_{+}\right)
.
\label{lnPQJ}
\end{equation}
Substituting this into
Eqs.~(\ref{Lt}) and (\ref{App:PI_h}), we obtain
\begin{gather}
L^{(h)}
\approx
-
\frac{4g_L' h^2}{D\sqrt{2\pi D  H_{n}}}
 \frac{\Omega W_{0}}{\Omega+4W_{0}}
(\boldsymbol{x}_{+}\times \boldsymbol{x}^{+})_{z}
\left(\boldsymbol{x}_{+}\cdot\boldsymbol{n}_{+}\right)
,
\label{Lh_final}
\\
\omega^{(h)}
\approx
-\frac{4\pi g_{O}' h^2}{
D\sqrt{2\pi D H_{n}}
}
\frac{\Omega W_0}{\Omega + 4W_0}
\boldsymbol{x}_{+}\cdot \boldsymbol{n}_{+}
.
\label{App:omg}
\end{gather}
Combining Eqs.~(\ref{LI:1}) and (\ref{Lh_final}), 
also Eqs.~(\ref{omega:1}) and (\ref{App:omg}),
we obtain
Eqs.~(\ref{Lt_final})--(\ref{TorqBalance}).
Here, $\frac{g_{O}'}{g_{O}} =\frac{g_{L}'}{g_{L}}$
is assumed so that $\omega$ is proportional to $L$.

\subsection{\label{App:Power} Power}

Applying Eqs.~(\ref{cor_NN}) and (\ref{J_mu_app}) to
Eq.~(\ref{P_h0}), within the approximation of
$O(h^2)$, we obtain
\begin{align}
 P_{h}
&\approx
 -g_V h\sum_{\mu}
\overline{
\left[
4 W_{0}\int_{-\infty}^{t}\! ds K(t-s)F_s
- F_t
\right]
\left(
\mu
\boldsymbol{N}_t \cdot
\boldsymbol{x}_{\mu}
\right)
}
\nonumber
\\
&=
 -
\frac{4g_V h^2 W_{0}}{D}
\left[
4 W_{0}\int_{-\infty}^{t} ds 
e^{-4W_{0}(t-s)}
\left\langle
\left(
\boldsymbol{N}_s \cdot
\boldsymbol{x}_{+}
\right)
\left(
\boldsymbol{N}_t \cdot
\boldsymbol{x}_{+}
\right)
\right\rangle_{\Phi}
-
\left\langle
\left(
\boldsymbol{N}_t \cdot
\boldsymbol{x}_{+}
\right)^2
\right\rangle_{\Phi}
\right]
\nonumber
\\
&=
 -
\frac{2g_V h^2 \left\lvert\boldsymbol{x}_{+}\right\rvert^2 W_{0}}{D}
\left[
4 W_{0}\int_{-\infty}^{t}\! ds 
e^{-4W_{0}(t-s)-\Omega(t-s)}
-
1
\right]
\nonumber
\\
&=
\frac{2g_V h^2 \left\lvert\boldsymbol{x}_{+}\right\rvert^2 W_{0}}{D}
\frac{\Omega}{\Omega + 4W_0}
.
\end{align}
Therefore, we find Eq.~(\ref{Ph_final}).

\subsection{Check of $\overline{L_t'}\approx L$ and $\overline{\omega_t'}\approx \omega_t$}
\label{App:XxX}

From Eq.~(\ref{def:L'}), we have
\begin{equation}
 \overline{L_t'} = L - 
\overline{
 \left(\langle\boldsymbol{X}\rangle 
\times \langle\dot{\boldsymbol{X}}\rangle \right)_z
}.
\label{App:L_t'}
\end{equation}
Substituting Eqs.~(\ref{P_sig_app}) and (\ref{J_mu_app}) into
$
 \langle\boldsymbol{X}\rangle \approx \sum_{\sigma} \boldsymbol{x}_{\sigma}
P(\sigma,t)
$ and
$
\langle\dot{\boldsymbol{X}}\rangle \approx
g_{V} \sum_{\mu} \boldsymbol{x}_{\mu} \left\{
J^{-\mu}(t)-J^{\mu}(t)
\right\}
$ [Eq.~(\ref{X_dot:app})],
we obtain
$
 \langle\boldsymbol{X}\rangle \propto \sum_{\sigma} 
\sigma\boldsymbol{x}_{\sigma}
$
and
$
\langle\dot{\boldsymbol{X}}\rangle \propto
- \sum_{\mu} \mu\boldsymbol{x}_{\mu}$
omitting the proportional coefficients.
Therefore, the second term in Eq.~(\ref{App:L_t'}) reads as
\begin{equation}
\overline{
 \left(\langle\boldsymbol{X}\rangle 
\times \langle\dot{\boldsymbol{X}}\rangle \right)_z}
\propto 
-\sum_{\sigma,\mu}\sigma\mu
\left(
\boldsymbol{x}_{\sigma}\times\boldsymbol{x}_{\mu}
\right)_z = 0.
\end{equation}
Since we have neglected the terms of $O(h^2)$ in
$\langle\boldsymbol{X}\rangle$ and
$\langle\dot{\boldsymbol{X}}\rangle$,
we can regard Eq.~(\ref{App:L_t'}) as
$\overline{L_t'} = L + o(h^2)$.
Similarly, $\omega_t'$ in Eq.~(\ref{def:omega'}) reads as
\begin{equation}
 \overline{\omega_t'} = \omega_t - 
\frac{1}{\lvert\boldsymbol{x}\rvert^2}
\overline{
 \left(\langle\boldsymbol{X}\rangle 
\times \langle\dot{\boldsymbol{X}}\rangle \right)_z
} + o(h^2).
\label{App:omega_t'}
\end{equation}
Therefore, neglecting the terms of $o(h^2)$, we have
$\overline{\omega_t'} \approx \omega_t$.

\section{\label{App:LamZero} Detailed analysis in the elliptic two-tooth ratchet case}

For the elliptic curve $\mathrm{C}_{\infty}=
\left\{\boldsymbol{x}\mid v_0(\boldsymbol{x})=\frac{x^2}{a^2}+\frac{y^2}{b^2}=1\right\}$
($a>b$), 
its trajectory, as well as the normal and tangential vectors along it, are parameterized
with the angular variable $\theta \in [0,2\pi)$ as
$\boldsymbol{x}\equiv (a\cos\theta, b\sin\theta)^{\mathrm T}$ and
\begin{equation}
 \boldsymbol{\tau}_{v}=  \frac{1}{N_{v}}
   \begin{pmatrix}
    b\cos\theta\\ a\sin\theta
   \end{pmatrix}
   ,
   \quad
   \boldsymbol{n}_{v}=  \frac{1}{N_{v}}
   \begin{pmatrix}
    -a\sin\theta\\ b\cos\theta
   \end{pmatrix}
,
\label{App:tau_n}
\end{equation}
respectively, where $N_{v}=\sqrt{a^2\sin^2\theta + b^2\cos^2\theta}$.
Letting $\theta_{+}$ be the angle corresponding to the local minimum
$\boldsymbol{x}_{+}$, it is determined by
$\mathrm{G}_1$:
$\theta_{+}=
\argmax_{\theta}
\{-\boldsymbol{n}_{v}\cdot \boldsymbol{x}\}$. 
We therefore have
\begin{equation}
 \cos 2\theta_{+} = \frac{a-b}{a+b},
  \quad
  \sin 2\theta_{+} = \frac{2\sqrt{ab}}{a+b},
\label{ellipse:theta+}
\end{equation}
and $\max_{\theta}\{-\boldsymbol{n}_{v}\cdot \boldsymbol{x}_{+}\}=a-b$.
Also, in the same parameterization,
$v_1(\boldsymbol{x})$ is represented as
\begin{align}
\boldsymbol{x}^{\mathrm T}
\Hat{O}_{\alpha} \Hat{E}_d \Hat{O}_{\alpha}^{\mathrm T}
 \boldsymbol{x}
 \equiv  E_0 + \frac{\epsilon}{2}\cos(2\theta - 2\theta_{+})
\equiv
E(\theta),
\label{app:E_theta}
\end{align}
where
\begin{gather}
4 E_0 \equiv
 \left( {d}^{2}-1 \right)  \left( {a}^{2}-{b}^{2} \right) \cos
 \left( 2\,\alpha \right) + \left( {d}^{2}+1 \right)  \left( {a}^
{2}+{b}^{2} \right),
\nonumber
\\
2\epsilon \cos 2\theta_{+} \equiv
 \left( {d}^{2}-1 \right)  \left( {a}^{2}+{b}^{2} \right) \cos
 \left( 2\,\alpha \right) + \left( {d}^{2}+1 \right)  \left( {a}^
{2}-{b}^{2} \right),
\label{DEcos}
\\
\epsilon \sin 2\theta_{+} \equiv
ab \left( d^2-1 \right)
 \sin \left( 2\,\alpha \right)
.
\label{DEsin}
\end{gather}
From Eq.~(\ref{app:E_theta}), assuming $m \rightarrow \infty$
with $\epsilon \geq 0$, $d > 1$, and $0 \leq \alpha < \frac{\pi}{2}$,
the local minimum and the saddle on $\mathrm{C}_{\infty}$ correspond to
$\theta = \theta_{+}$ and $\frac{\pi}{2}+ \theta_{+}$, and
we have $E_{+} = E_0  + \frac{\epsilon}{2}$ and $E^{+} = E_0 -\frac{\epsilon}{2}$
[$K\epsilon = \Delta V$ from Eq.~(\ref{DV})]
for the circumscribed and inscribed ellipses $\mathrm{E}_{+}$ and $\mathrm{E}^{+}$,
respectively.

We obtain Eqs.~(\ref{Zero:DE}) and (\ref{ellipse:cond2}) as follows:
from Eq.~(\ref{ellipse:theta+}),
$
\epsilon =\left.
\epsilon \cos(2\theta - 2\theta_{+})
\right|_{\theta=\theta_{+}}
$ and $\left.\frac{dE(\theta)}{d\theta}\right|_{\theta=\theta_{+}}=0$
(which corresponds to $\mathrm{G}_2$), we have
\begin{gather}
\epsilon =
\left[
\epsilon \cos 2\theta_{+} \cos 2\theta
+
\epsilon \sin 2\theta_{+} \sin 2\theta
\right]_{\theta=\theta_{+}}
\nonumber\\
 =
\left(
\epsilon \cos 2\theta_{+} \right) \frac{a-b}{a+b}
+
\left(\epsilon \sin 2\theta_{+} \right) \frac{2\sqrt{ab}}{a+b}
,
\label{DE0}
\\
0 = \left[
\epsilon \sin 2\theta_{+} \cos 2\theta
-
\epsilon \cos 2\theta_{+} \sin 2\theta
\right]_{\theta=\theta_{+}}
\nonumber\\
= \left(
\epsilon \sin 2\theta_{+}
\right)  \frac{a-b}{a+b}
-\left(
\epsilon \cos 2\theta_{+}
\right)\frac{2\sqrt{ab}}{a+b}
,
\label{ellipse:cond}
\end{gather}
then, substituting Eqs.~(\ref{DEcos}) and (\ref{DEsin}) into
Eq.~(\ref{ellipse:cond}), we find Eq.~(\ref{ellipse:cond2}),
and also 
substituting Eq.~(\ref{DEsin}) and
$
2\epsilon \cos 2\theta_{+}=
\sqrt{ab}(a-b) \left( d^2-1 \right)
 \sin \left( 2\,\alpha \right)
$ [from  Eqs.~(\ref{ellipse:cond2}) and (\ref{DEcos})]
to Eq.~(\ref{DE0}),
we find Eq.~(\ref{Zero:DE}).

Furthermore, based on Eqs.~(\ref{Hessian:G}) and (\ref{Hessian:g})
in Sec.~\ref{subsec:Hesse},
we obtain the eigenvalues of the Hessian matrix
at the local minimum and the saddle in $m\gg 1$ as follows.
Since, from Eqs.~(\ref{3th:force_2}) and
(\ref{App:tau_n})--(\ref{app:E_theta}), we have
\begin{gather}
 \partial_{\boldsymbol{x}}v_{0}(\boldsymbol{x})
  =\frac{2}{ab}
  \begin{pmatrix}
   b\cos\theta\\
      a\sin\theta
  \end{pmatrix},
\quad
 \partial_{\boldsymbol{x}}
   \partial_{\boldsymbol{x}}^{\mathrm T}v_{0}(\boldsymbol{x})
  =2
  \begin{pmatrix}
   \frac{1}{a^2}&0\\ 0&   \frac{1}{b^2}
  \end{pmatrix},
\nonumber\\
\boldsymbol{\tau}_{v}\cdot
\partial_{ \boldsymbol{x}}
v_1(\boldsymbol{x}) 
=2\boldsymbol{\tau}_{v}^{\mathrm T}
\Hat{O}_{\alpha} \Hat{E}_d \Hat{O}_{\alpha}^{\mathrm T}
 \boldsymbol{x}
= \frac{2E(\theta)}{\boldsymbol{x}\cdot  \boldsymbol{\tau}_{v}}
= 2N_v \frac{E(\theta)}{ab}
,
\nonumber\\
N_v^2
\boldsymbol{n}_{v}^{\mathrm T} 
\left\{
\partial_{ \boldsymbol{x}}
\partial_{ \boldsymbol{x}}^{\mathrm{T}}
v_1(\boldsymbol{x})
\right\}
\boldsymbol{n}_{v}
= 
\left.
\boldsymbol{x}^{\mathrm T} 
\partial_{ \boldsymbol{x}}
\partial_{ \boldsymbol{x}}^{\mathrm{T}}
v_1(\boldsymbol{x})
\boldsymbol{x}
\right|_{\theta \rightarrow \theta + \frac{\pi}{2}} 
=2E\left(\theta+\frac{\pi}{2}\right),
\nonumber
\end{gather}
noting that $\boldsymbol{x}= (\boldsymbol{x}\cdot  \boldsymbol{\tau}_{v})
 \boldsymbol{\tau}_{v} + (\boldsymbol{x}\cdot  \boldsymbol{n}_{v})\boldsymbol{n}_{v}$
and Eq.~(\ref{3th:force_2}) in the second line,
we find the diagonal components
of $\Hat G_{0}(\boldsymbol{x})$ in Eqs.~(\ref{Hessian:G}) and (\ref{Hessian:g}) as
\begin{gather}
 \frac{m^2}{2}
\lvert\partial_{\boldsymbol{x}}v_{0}(\boldsymbol{x})\rvert^2
=
\frac{m^2} {a^2b^2}
\left\{a^2 + b^2 - (a^2-b^2)\cos 2\theta\right\}
,
\\
 g(\boldsymbol{x})=
2 \frac{
E(\theta )-E(\theta+\frac{\pi}{2})
}{a^2\sin^2\theta + b^2\cos^2\theta
}
=\frac{4\epsilon\cos(2\theta - 2\theta_{+})}
{a^2 + b^2 - (a^2-b^2)\cos2\theta}
.
\label{App:g}
\end{gather}
Then, at $\boldsymbol{x}= \boldsymbol{x}_{+}$ ($\theta =\theta_{+}$) and 
$\boldsymbol{x}= \boldsymbol{x}^{+}$ ($\theta =\frac{\pi}{2}+\theta_{+}$), 
Eq.~(\ref{Hessian:G}) reads as
\begin{align}
\Hat G_0(\boldsymbol{x}_{+})
&\approx \frac{2m^2}{ab}
\boldsymbol{\tau}_{v}
\boldsymbol{\tau}_{v}^{\mathrm T}
+\frac{2\Delta V}{ab}
 \boldsymbol{n}_{v}
 \boldsymbol{n}_{v}^{\mathrm T}
,
\label{app:G_+}
\\
\Hat G_0(\boldsymbol{x}^{+})
&\approx
 \frac{2m^2}{a^2b^2}
(a^2 + b^2 - ab)
\boldsymbol{\tau}_{v}
\boldsymbol{\tau}_{v}^{\mathrm T}
-\frac{2\Delta V}{a^2 + b^2 - ab}
 \boldsymbol{n}_{v}
 \boldsymbol{n}_{v}^{\mathrm T}.
\label{app:G^+}
\end{align}
The diagonal components of $\Hat G_0(\boldsymbol{x}_{+})$
[$\Hat G_0(\boldsymbol{x}^{+})$] in Eq.~(\ref{app:G_+}) [Eq.~(\ref{app:G^+})]
correspond to $H_{\tau}$ and $H_{n}$ ($G_{\tau}$ and $G_n$)
in Eqs.~(\ref{def:W0}) and (\ref{TorqBalance}), respectively.

Substituting these results into
Eqs.~(\ref{Lt_final}), (\ref{W_sig_mug}), (\ref{I0D}),
(\ref{Ph_final}), and (\ref{def:eta0}), we find
Eqs.~(\ref{lamda0:L})--(\ref{lamda0:I0}).


\bibliography{dynamics,DPT,tutu}
\end{document}